%% file: main.tex
\documentclass[11pt,a4paper]{article}
\pdfoutput=1
\usepackage{jheppub}
\usepackage[T1]{fontenc}
\usepackage{lmodern}
\usepackage{booktabs}
\usepackage{mathtools}
\usepackage{changepage}
\usepackage{microtype}
\usepackage[section]{placeins}
\usepackage{tikz}
\usetikzlibrary{arrows.meta,decorations.markings}
\graphicspath{{figures/}}
\allowdisplaybreaks[2]
\newcommand{\dd}{\mathrm d}
\newcommand{\ii}{\mathrm i}
\newcommand{\Tr}{\operatorname{Tr}}
\newcommand{\ON}{O(N)}
\newcommand{\GN}{\mathrm{GNY}}
\newcommand{\TJJ}{\langle TJJ\rangle}
\newcommand{\TTT}{\langle TTT\rangle}
\newcommand{\nsJ}{n_{\mathrm s}^{TJJ}}
\newcommand{\nfJ}{n_{\mathrm f}^{TJJ}}
\newcommand{\nsT}{n_{\mathrm s}^{TTT}}
\newcommand{\nfT}{n_{\mathrm f}^{TTT}}
\newcommand{\SJ}{\mathcal S_{TJJ}}
\newcommand{\FJ}{\mathcal F_{TJJ}}
\newcommand{\HJ}{\mathcal H_{TJJ}}
\newcommand{\ST}{\mathcal S_{TTT}}
\newcommand{\FT}{\mathcal F_{TTT}}
\newcommand{\HT}{\mathcal H_{TTT}}
\newcommand{\intk}{\int\!\frac{\dd^3k}{(2\pi)^3}}

\numberwithin{equation}{section}

\title{\boldmath \texorpdfstring{$1/N$}{1/N} corrections to $\langle TJJ\rangle$ and $\langle TTT\rangle$ in the critical
\texorpdfstring{$O(N)$}{O(N)} vector and Gross--Neveu--Yukawa models}
\author[a,b]{Yongwei Guo}
\author[a,b]{and Zhijin Li}
\affiliation[a]{School of Physics, Southeast University, 2 Sipailow Street, Nanjing 210096, China}
\affiliation[b]{Shing-Tung Yau Center, Southeast University, 2 Sipailow Street, Nanjing 210096, China}
\emailAdd{guoyw33@seu.edu.cn}
\emailAdd{zhijin\_li@seu.edu.cn}
\abstract{We compute the subleading corrections in the $1/N$ expansion of the
three-point functions $\langle TJJ\rangle$ and $\langle TTT\rangle$ in the
three-dimensional critical $O(N)$ vector and Gross--Neveu--Yukawa models.
We determine the coefficients of the two parity-even conformal structures
and the corresponding conformal collider parameters. The calculation uses
an expansion of momentum-space Feynman diagrams in a soft external
momentum, together with the contact terms required by background gauge
and metric sources. The sums of the structure coefficients reproduce the
known current and stress-tensor two-point normalizations. We also compute these coefficients independently using the metric and mixed metric--gauge
derivatives of the sphere free energy. Extrapolating the large $N$ results to the $O(2)$ vector model yields
reasonably good agreement with recent numerical bootstrap bounds. 
At $N=2$, our $\gamma$ and $t_4$ predictions lie within the joint bootstrap region under weaker spectral assumptions; their deviations from the more restrictive interval midpoints are only $3.28\%$ and $4.44\%$, respectively.
We discuss the implications for energy flux,
high-frequency conductivity, and shear response near quantum criticality.}

\tikzset{
 gny fermion/.style={line width=.75pt,postaction={decorate},
 decoration={markings,mark=at position #1 with {\arrow{Stealth[length=2mm,width=2mm]}}}},
 gny fermion/.default=.55,
 gny sigma/.style={dashed,line width=.75pt},
 gny contact/.style={rectangle,fill,inner sep=2pt}
}
\newcommand{\GNYTriangle}[5]{%
  \coordinate (T) at (0,0);
  \coordinate (A) at (1.5,1);
  \coordinate (B) at (1.5,-1);
  \draw[gny fermion=#1] (T) -- (A);
  \draw[gny fermion=#2] (A) -- (B);
  \draw[gny fermion=#3] (B) -- (T);
  \foreach \v in {T,A,B} {\fill (\v) circle (2pt);}
  \node[left] at (T) {$T(q)$};
  \node[right] at (A) {$#4$};
  \node[right] at (B) {$#5$};
}
\newcommand{\GNYThreePointFigure}[3]{%
\resizebox{\textwidth}{!}{%
\begin{tikzpicture}[line width=.75pt,font=\normalsize]
\begin{scope}[shift={(0.5,-0.0)}]
  \GNYTriangle{0.55}{0.55}{0.55}{#1}{#2}
  \node at (0.75,-1.5) {$D_0$};
\end{scope}
\begin{scope}[shift={(6.3,-0.0)}]
  \GNYTriangle{0.55}{0.55}{0.55}{#1}{#2}
  \path (T) -- coordinate[pos=0.25] (u) coordinate[pos=0.75] (v) (A);
  \draw[gny sigma] (u) to[bend right=80,looseness=1.3] (v);
  \foreach \w in {u,v} {\fill (\w) circle (1.6pt);}
  \node at (0.75,-1.5) {$D_1$};
\end{scope}
\begin{scope}[shift={(12.1,-0.0)}]
  \GNYTriangle{0.55}{0.55}{0.55}{#1}{#2}
  \path (A) -- coordinate[pos=0.25] (u) coordinate[pos=0.75] (v) (B);
  \draw[gny sigma] (u) to[bend right=80,looseness=1.3] (v);
  \foreach \w in {u,v} {\fill (\w) circle (1.6pt);}
  \node at (0.75,-1.5) {$D_2$};
\end{scope}
\begin{scope}[shift={(0.5,-3.2)}]
  \GNYTriangle{0.55}{0.55}{0.55}{#1}{#2}
  \path (B) -- coordinate[pos=0.25] (u) coordinate[pos=0.75] (v) (T);
  \draw[gny sigma] (u) to[bend right=80,looseness=1.3] (v);
  \foreach \w in {u,v} {\fill (\w) circle (1.6pt);}
  \node at (0.75,-1.5) {$D_3$};
\end{scope}
\begin{scope}[shift={(6.3,-3.2)}]
  \GNYTriangle{0.75}{0.55}{0.35}{#1}{#2}
  \path (T) -- coordinate[pos=0.4] (u) (A);
  \path (T) -- coordinate[pos=0.4] (v) (B);
  \draw[gny sigma] (u) -- (v);
  \foreach \w in {u,v} {\fill (\w) circle (1.6pt);}
  \node at (0.75,-1.5) {$D_4$};
\end{scope}
\begin{scope}[shift={(12.1,-3.2)}]
  \GNYTriangle{0.35}{0.75}{0.55}{#1}{#2}
  \path (A) -- coordinate[pos=0.4] (u) (T);
  \path (A) -- coordinate[pos=0.4] (v) (B);
  \draw[gny sigma] (u) -- (v);
  \foreach \w in {u,v} {\fill (\w) circle (1.6pt);}
  \node at (0.75,-1.5) {$D_5$};
\end{scope}
\begin{scope}[shift={(0.5,-6.4)}]
  \GNYTriangle{0.55}{0.35}{0.75}{#1}{#2}
  \path (B) -- coordinate[pos=0.4] (u) (T);
  \path (B) -- coordinate[pos=0.4] (v) (A);
  \draw[gny sigma] (u) -- (v);
  \foreach \w in {u,v} {\fill (\w) circle (1.6pt);}
  \node at (0.75,-1.5) {$D_6$};
\end{scope}
\begin{scope}[shift={(5.8,-6.4)}]
  \coordinate (T) at (0,0);
  \coordinate (u) at (0.8,0.65);
  \coordinate (v) at (0.8,-0.65);
  \coordinate (s) at (1.8,0.65);
  \coordinate (t) at (1.8,-0.65);
  \coordinate (A) at (2.5,1);
  \coordinate (B) at (2.5,-1);
  \draw[gny fermion] (T) -- (u);
  \draw[gny fermion] (u) -- (v);
  \draw[gny fermion] (v) -- (T);
  \draw[gny fermion] (s) -- (A);
  \draw[gny fermion] (A) -- (B);
  \draw[gny fermion] (B) -- (t);
  \draw[gny fermion] (t) -- (s);
  \draw[gny sigma] (u) -- (s);
  \draw[gny sigma] (v) -- (t);
  \foreach \w in {u,v,s,t} {\fill (\w) circle (1.6pt);}
  \foreach \w in {T,A,B} {\fill (\w) circle (2pt);}
  \node[left] at (T) {$T(q)$};
  \node[right] at (A) {$#1$};
  \node[right] at (B) {$#2$};
  \node at (1.25,-1.5) {$D_7$};
\end{scope}
\begin{scope}[shift={(11.6,-6.4)}]
  \coordinate (T) at (0,0);
  \coordinate (u) at (0.9,0.6);
  \coordinate (v) at (0.9,-0.6);
  \coordinate (s) at (1.9,0.6);
  \coordinate (t) at (2.5,-0.6);
  \coordinate (A) at (2.5,1.1);
  \coordinate (B) at (1.9,-1.1);
  \draw[gny fermion] (T) -- (u);
  \draw[gny fermion] (u) -- (v);
  \draw[gny fermion] (v) -- (T);
  \draw[gny sigma] (u) -- (s);
  \draw[gny sigma] (v) -- (t);
  \draw[gny fermion] (s) -- (A);
  \draw[gny fermion] (A) -- (t);
  \draw[gny fermion] (t) -- (B);
  \draw[gny fermion,preaction={draw,white,line width=5pt}] (B) -- (s);
  \foreach \w in {u,v,s,t} {\fill (\w) circle (1.6pt);}
  \foreach \w in {T,A,B} {\fill (\w) circle (2pt);}
  \node[left] at (T) {$T(q)$};
  \node[right] at (A) {$#1$};
  \node[right,yshift=-5pt] at (B) {$#2$};
  \node at (1.25,-1.5) {$D_8$};
\end{scope}
\ifnum#3=1
\begin{scope}[shift={(5.8,-10.0)}]
  \coordinate (T) at (0,0);
  \coordinate (u) at (0.7,0.6);
  \coordinate (v) at (0.7,-0.6);
  \coordinate (s) at (1.6,1.2);
  \coordinate (A) at (2.5,1.5);
  \coordinate (t) at (2.3,0.5);
  \coordinate (x) at (2.3,-0.5);
  \coordinate (B) at (2.5,-1.5);
  \coordinate (y) at (1.6,-1.2);
  \draw[gny fermion] (T) -- (u);
  \draw[gny fermion] (u) -- (v);
  \draw[gny fermion] (v) -- (T);
  \draw[gny fermion] (s) -- (A);
  \draw[gny fermion] (A) -- (t);
  \draw[gny fermion] (t) -- (s);
  \draw[gny fermion] (x) -- (B);
  \draw[gny fermion] (B) -- (y);
  \draw[gny fermion] (y) -- (x);
  \draw[gny sigma] (u) -- (s);
  \draw[gny sigma] (t) -- (x);
  \draw[gny sigma] (y) -- (v);
  \foreach \w in {u,v,s,t,x,y} {\fill (\w) circle (1.6pt);}
  \foreach \w in {T,A,B} {\fill (\w) circle (2pt);}
  \node[left] at (T) {$T(q)$};
  \node[right] at (A) {$#1$};
  \node[right] at (B) {$#2$};
  \node at (1.25,-2.1) {$D_{12}$};
\end{scope}
\fi
\end{tikzpicture}%
}%
}
\newcommand{\GNYContactPanel}[4]{%
\begin{tikzpicture}[line width=.75pt,font=\normalsize,baseline=(current bounding box.center)]
 \path[use as bounding box] (-1.05,-1.65) rectangle (4.25,1.1);
 \coordinate (L) at (0,0); \coordinate (R) at (2.4,0);
 \ifnum#1<3
  \ifnum#1=2 \def\gnyArrow{.75}\else\def\gnyArrow{.55}\fi
  \draw[gny fermion=\gnyArrow] (L) arc[start angle=180,end angle=0,x radius=1.2,y radius=.85];
  \draw[gny fermion=\gnyArrow] (R) arc[start angle=0,end angle=-180,x radius=1.2,y radius=.85];
  \ifnum#1=1
   \coordinate (u) at (.6,{.85*sqrt(3)/2});
   \coordinate (v) at (1.8,{.85*sqrt(3)/2});
   \draw[gny sigma] (u) to[bend right=65] (v);
   \foreach \w in {u,v} \fill (\w) circle (1.6pt);
  \fi
  \ifnum#1=2
   \draw[gny sigma] (1.2,.85)--(1.2,-.85);
   \fill (1.2,.85) circle (1.6pt); \fill (1.2,-.85) circle (1.6pt);
  \fi
 \else
  \coordinate (R) at (3,0);
  \coordinate (u) at (.8,.7); \coordinate (v) at (.8,-.7);
  \coordinate (s) at (2.2,.7); \coordinate (t) at (2.2,-.7);
  \draw[gny fermion] (L)--(u); \draw[gny fermion] (u)--(v); \draw[gny fermion] (v)--(L);
  \draw[gny fermion] (R)--(t); \draw[gny fermion] (t)--(s); \draw[gny fermion] (s)--(R);
  \draw[gny sigma] (u)--(s); \draw[gny sigma] (v)--(t);
  \foreach \w in {u,v,s,t} \fill (\w) circle (1.6pt);
 \fi
 \fill (L) circle (2pt); \node[left=3pt] at (L) {$#2$};
 \ifnum#4=1 \node[gny contact] at (R) {};
 \else \fill (R) circle (2pt); \fi
 \node[right=4pt] at (R) {$#3$};
 \ifnum#1=0
  \ifnum#4=1 \node at (1.2,-1.4) {$C_0$};
  \else \node at (1.2,-1.4) {$C_0$}; \fi
 \else \node at (1.2,-1.4) {$C_{#1}$}; \fi
\end{tikzpicture}%
}

\begin{document}
\notoc
\maketitle
\clearpage
\pagenumbering{roman}
\tableofcontents
\clearpage
\pagenumbering{arabic}
\raggedbottom

\section{Introduction and summary of results}\label{sec:intro}

Correlation functions of conserved currents provide a useful set of
observables in conformal field theory (CFT). Their correlation functions satisfy Ward identities and are generated
by functional derivatives with respect to background gauge fields
and the metric. While conformal symmetry
determines the two-point functions up to their central charges, three-point
functions contain further information about the theory. In three
dimensions, each of the parity-even correlators $\TJJ$ and $\TTT$ has two
independent conformal structures. A Ward identity fixes one linear
combination in terms of $C_J$ or $C_T$, leaving one dimensionless parameter
for each correlator \cite{OsbornPetkou}.

An important motivation to study those correlators comes from charge transport near quantum critical
points in condensed matter systems. The electrical, or spin, conductivity
is a linear response determined by the appropriate current two-point
function. At frequencies large compared with the temperature, the
current-current operator product expansion (OPE) expresses its thermal
corrections in terms of vacuum three-point coefficients and thermal
one-point functions. The stress-tensor contribution is fixed by $\TJJ$:
in $2+1$ dimensions it enters the conductivity at order $P/\omega^3$,
where $P$ is the pressure. This connection has been studied using
large-flavor expansions and holography \cite{ChowdhuryTransport},
thermal OPEs and quantum Monte Carlo \cite{KatzEtAl}, and conformal
perturbation theory near criticality
\cite{LucasGazitPodolskyWitczak,LucasSierensWitczak}. Scalar operators
with thermal expectation values can contribute earlier in this expansion.
Determining $\TJJ$ supplies the independent stress contribution beyond
the vacuum normalization $C_J$.

Quantum-critical transport has a long history. Damle and Sachdev \cite{DamleSachdevTransport} distinguished the collision-dominated and collisionless limits of the universal conductivity scaling function. Of particular relevance here, Chowdhury {\it et al.} \cite{ChowdhuryTransport} directly computed $\TJJ$ in a large-flavor expansion and matched it to holographic three-point functions, relating $\gamma$ to particle-like and vortex-like transport. Katz \textit{et al.} \cite{KatzEtAl} subsequently combined thermal OPEs, quantum Monte Carlo and holography to study the conductivity of the critical $O(N)$ vector model. The present calculation determines the interaction corrections to the vacuum three-point data entering this short-distance description.\par Chern--Simons-matter theories provide a further setting in which correlation functions, transport and duality can be studied together. Planar three-point calculations test bosonization and determine coupling-dependent parity-even and parity-odd structures \cite{AharonyBosonization,GurAriYacobyCS}. Gur-Ari \textit{et al.} \cite{GurAriHartnollMahajan} computed frequency-dependent longitudinal and Hall conductivities at nonzero temperature, exactly in the 't Hooft coupling for the large $N$ fermionic model. Ghosh and Mazumdar \cite{GhoshMazumdarThermal} obtained thermal two-point functions of scalar, current and stress-tensor operators to all orders in that coupling at leading large $N$. More recently, Kukolj \cite{KukoljHigherSpinContacts} analyzed local higher-spin Ward identities and contact terms, with perturbative checks including a collinear component of $\TJJ$. These results motivate extending our subgraph calculation beyond the parity-invariant models considered here.\par The stress-tensor three-point function has a parallel role in the shear
response. The frequency-dependent shear viscosity follows from a Kubo
formula for the retarded $T_{xy}$ two-point function. Its short-distance
expansion contains a stress contribution whose coefficient is determined
by $\TTT$, and hence by $C_T$ and the independent three-point shape.
These data constrain the high-frequency response and the shear spectral
sum rules \cite{ChowdhuryDavidPrakash,LucasSierensWitczak}. This gives
access to thermal response in strongly coupled systems without assuming
a quasiparticle description. The vacuum three-point data do not alone
fix the low-frequency viscosity, which also depends on thermal dynamics.

The same coefficients obey strong Lorentzian consistency conditions.
In a conformal collider experiment a local current or stress insertion
creates a state, and a distant detector measures its angular energy flux
\cite{HofmanMaldacena}. The expectation value is determined by $\TJJ$ or
$\TTT$. Positivity for every polarization constrains their shapes, rather
than only the positive two-point norms. For the parity-even conventions
used here these constraints read $|\gamma|\leq1/12$ and $|t_4|\leq4$.
They are field-theoretic consequences of unitarity and causality: the
lightcone bootstrap yields the collider inequalities
\cite{LiMeltzerPoland,HofmanEtAlProof}, while positivity of averaged
null energy provides their physical origin \cite{HartmanKunduTajdini}.
Early holographic studies connected these energy-flux constraints to bulk causality \cite{HofmanCausality2009,CamanhoEdelsteinCollider}. Buchel \textit{et al.} \cite{BuchelGBGravity} and Myers \textit{et al.} \cite{MyersQuasiTopological} studied stress-tensor three-point data and positivity constraints in Gauss--Bonnet and quasi-topological gravity, respectively. In the current sector, Myers \textit{et al.} \cite{MyersSachdevSinghTransport} related the allowed higher-derivative gauge coupling to energy-flux positivity and causality, and examined its consequences for quantum-critical conductivity. The free scalar and fermion structures lie at the endpoints. Our large
$N$ theories start at these endpoints, so the sign of the first departure
tests whether interactions move the correlator into the allowed region.
We discuss the energy-flux interpretation further in
sections~\ref{sec:tjj-general} and~\ref{sec:ttt-general}.

There are also applications to early-universe cosmology. De Sitter isometries act as conformal transformations on the future boundary, relating the momentum-space structures of $\TTT$ to graviton wavefunction coefficients and primordial tensor non-Gaussianity \cite{MaldacenaPimentel,McFaddenSkenderisCosmology}. The holographic dictionary relates cosmological bispectra to three-dimensional stress-tensor correlators, including the required contact contributions; explicit QFT calculations were developed in the closely related work \cite{BMSHolographicBispectra}. With spin-one fields present, the $\TJJ$ structures similarly describe a graviton coupled to two vector modes in the boundary wavefunction \cite{BaumannSpinning}. These connections underlie the cosmological bootstrap, which constrains inflationary correlators using symmetry, locality and factorization \cite{ArkaniHamedCosmologicalBootstrap,BaumannSpinning}; complementary momentum-space methods include the Mellin-space approach \cite{SleightTaronnaMellin}. Momentum-space Ward identities and their contact terms are therefore natural tools in this setting \cite{BMSMomentum,BMSRenormalised}. Related developments include the solution of scalar three-point conformal constraints and its relation to Feynman master integrals \cite{CorianoScalarMomentum}, and the construction of Lorentzian three-point functions using Ward identities and the momentum-space OPE \cite{GilliozLorentzianMomentum}. For broader accounts, see the review of momentum-space CFT correlators \cite{CorianoMaglioReview} and the cosmological-bootstrap review \cite{CosmologicalBootstrapReview}. These applications give a further reason to determine the three-point data directly in momentum space.

We will take the critical $\ON$ vector and Gross--Neveu--Yukawa (GNY) models as examples to compute these data. A comprehensive review
of the critical $\ON$ vector model, its methods, and its CFT data is given
in ref.~\cite{HenrikssonReview}. At large $N$, the
fundamental scalar and fermion propagators are free to leading order,
while an auxiliary singlet field mediates the first corrections. The
large $N$ expansion has long provided systematic access to the operator spectrum of these models \cite{MosheZinnJustinReview}. Critical-point methods yielded scaling dimensions and critical exponents beyond leading order in the $O(N)$ vector model \cite{VasilievCriticalExponents,HenrikssonReview} and the Gross--Neveu universality class \cite{GraceyEtaLargeN}; comparisons with the Yukawa description and dimensional expansions are discussed in ref.~\cite{FeiYukawaCFT}.\par For the two-point normalizations, Petkou \cite{PetkouCentralCharges} obtained the next-to-leading $C_J$ and $C_T$ of the $O(N)$ vector model, while Diab et al.~\cite{DiabEtAl} computed the corresponding corrections in the Gross--Neveu model and revisited the vector-model results. These quantities fix one Ward-identity combination of each three-point function, but leave its independent shape undetermined. To our knowledge, the present work gives the first explicit $1/N$ corrections to both $\TJJ$ and $\TTT$, including the independent collider parameters $\gamma$ and $t_4$, in the three-dimensional critical $O(N)$ vector and GNY models. Related techniques have been
developed for conformal gauge theories
\cite{HuhStrackSachdevCurrents,HuhStrackCentralCharges,GiombiTarnopolskyKlebanov,GuoLi}. We use the method of subgraphs,
applied to conformal three-point functions in ref.~\cite{LiSubgraphs},
to reduce the required multiscale loop integrals to their hard and soft
expansions. Conformal symmetry then allows the coefficients to be fixed
from a finite set of terms. Nonlinear background-source couplings require
contact vertices in addition to diagrams with separate operator
insertions. These vertices are essential when extracting the conformal
structures from a momentum expansion.

As an independent approach, we determine the three-point data from derivatives of the sphere free energy with respect to background sources \cite{BobevBuenoVreys,BuenoEtAl,BuenoEtAl2020,ParmentierBobev}. The physical motivation is that varying the metric and background gauge field generates stress-tensor and conserved-current insertions. Near the round sphere, the third derivative with respect to squashing probes an integrated projection of $\TTT$, while the mixed derivative taken once with respect to squashing and twice with respect to the gauge-source parameter probes $\TJJ$, including the required source contacts. Conformal symmetry and the two-point normalizations then allow these projections to determine $t_4$ and $\gamma$. In the large $N$ models studied here, the subleading free energy is obtained from singlet fluctuation determinants, providing an efficient alternative to the diagrammatic calculation. More broadly, this background-field approach offers a route to CFT data in other theories whenever suitable source deformations and tractable partition functions or fluctuation kernels are available; we develop its implementation in section~\ref{sec:sphere}.\par Our calculation concerns parity-even, separated-point conformal data in
$d=3$. We retain the source contacts needed to identify these data in the
generating functional. In the $O(N)$ vector model $N$ counts real fundamental
fields. In the GNY model $N$ counts two-component Majorana fermions.
A theory of $N_{\mathrm D}$ two-component Dirac fermions therefore
corresponds to $N=2N_{\mathrm D}$ in our formulas.

\subsection{Summary of results}\label{sec:summary}

In a parity-invariant three-dimensional CFT, the separated-point
correlators $\TJJ$ and $\TTT$ each have two independent parity-even
conformal structures. We define their coefficients by
\begin{align*}
 \TJJ&=\nsJ\TJJ_{\mathrm s}+\nfJ\TJJ_{\mathrm f},\\
 \TTT&=\nsT\TTT_{\mathrm s}+\nfT\TTT_{\mathrm f}.
\end{align*}
The subscripts $\mathrm s$ and $\mathrm f$ denote free scalar and free
fermion reference correlators.  For $\TJJ$, both references have the same flavor factor
$\tau^{ab}=-\Tr(t^at^b)$ and the same current normalization
$C_J^{\mathrm s}$ after this factor is removed. For $\TTT$, the references
are one real conformally coupled scalar and one two-component Majorana
fermion, with equal stress-tensor normalization $C_T^{\mathrm s}$.
These are the conventions of eqs.~\eqref{eq:tjj-decomposition}
and~\eqref{eq:ttt-decomposition}; the reference central charges are given
in eq.~\eqref{eq:free-c}. The momentum coefficients used for matching to
these references are given in eqs.~\eqref{eq:tjj-free-region-coefficients}
and~\eqref{eq:ttt-free-region-coefficients}.

Our results for the $O(N)$ vector model are
\begin{align}
 \nsJ\big|_{\ON}&=1+\frac1N\left(\frac{16}{9\pi^2}-1\right)+O(N^{-2}),&
 \nfJ\big|_{\ON}&=\frac1N\left(1-\frac{80}{9\pi^2}\right)+O(N^{-2}),
 \label{eq:on-j}\\
 \nsT\big|_{\ON}&=N+\frac{686}{3}-\frac{20360}{9\pi^2}+O(N^{-1}),&
 \nfT\big|_{\ON}&=\frac{20320}{9\pi^2}-\frac{686}{3}+O(N^{-1}).
 \label{eq:on-t}
\end{align}
For the GNY model we find
\begin{align}
 \nsJ\big|_{\GN}&=\frac1N\left(1-\frac{56}{9\pi^2}\right)+O(N^{-2}),&
 \nfJ\big|_{\GN}&=1-\frac1N\left(1+\frac8{9\pi^2}\right)+O(N^{-2}),
 \label{eq:gn-j}\\
 \nsT\big|_{\GN}&=\frac{20344}{9\pi^2}-\frac{686}{3}+O(N^{-1}),&
 \nfT\big|_{\GN}&=N+\frac{686}{3}-\frac{20336}{9\pi^2}+O(N^{-1}).
 \label{eq:gn-t}
\end{align}
The $\TJJ$ coefficients are of order one because we factor out
$\tau^{ab}=-\Tr(t^at^b)$. The $\TTT$ coefficients include the sum over
all fundamental fields and are of order $N$.

The sums obey the Ward identities
\begin{align}
 \nsJ+\nfJ&=\frac{C_J}{C_J^{\mathrm s}}
 =1-\frac{64}{9\pi^2N}+O(N^{-2}),\qquad (\ON\ \text{and GNY}),
 \label{eq:ward-j}\\
 \nsT+\nfT&=\frac{C_T}{C_T^{\mathrm s}}
 =\begin{cases}
 N-\dfrac{40}{9\pi^2}+O(N^{-1}),&\ON,\\[3pt]
 N+\dfrac8{9\pi^2}+O(N^{-1}),&\mathrm{GNY}.
 \end{cases}\label{eq:ward-t}
\end{align}
They agree with the known central charges \cite{DiabEtAl}, with the free
reference normalizations specified above.

The shape parameters used in the collider and bootstrap literature are
\begin{equation}
 \gamma=\frac{\nfJ-\nsJ}{12(\nsJ+\nfJ)},\qquad
 t_4=4\frac{\nsT-\nfT}{\nsT+\nfT}.
 \label{eq:shape-def}
\end{equation}
Expanding these ratios consistently through order $1/N$ gives
\begin{align}
 \gamma_{\ON}&=-\frac1{12}+\frac1{6N}\left(1-\frac{80}{9\pi^2}\right)+O(N^{-2}),\nonumber\\
 t_{4,\ON}&=4-\frac8N\left(\frac{20320}{9\pi^2}-\frac{686}{3}\right)+O(N^{-2}),
 \label{eq:shape-on}\\
 \gamma_{\GN}&=\frac1{12}-\frac1{6N}\left(1-\frac{56}{9\pi^2}\right)+O(N^{-2}),\nonumber\\
 t_{4,\GN}&=-4+\frac8N\left(\frac{20344}{9\pi^2}-\frac{686}{3}\right)+O(N^{-2}).
 \label{eq:shape-gn}
\end{align}
Numerically,
\begin{align}
 \gamma_{\ON}&=-\frac1{12}+\frac{0.01656}{N}+O(N^{-2}),&
 t_{4,\ON}&=4-\frac{0.75240}{N}+O(N^{-2}),\\
 \gamma_{\GN}&=\frac1{12}-\frac{0.06159}{N}+O(N^{-2}),&
 t_{4,\GN}&=-4+\frac{2.91392}{N}+O(N^{-2}).
\end{align}
Note the signs of these corrections are consistent with the conformal collider bounds $|\gamma|\leq1/12$ and $|t_4|\leq4$. In the $O(N)$ vector model, $\delta\gamma>0$ and $\delta t_4<0$, whereas in the GNY model $\delta\gamma<0$ and $\delta t_4>0$. Thus both models move inward from their free-field endpoints, as required by energy-flux positivity at sufficiently large $N$. 

\par \subsection{Comparison with the numerical bootstrap}\label{sec:bootstrap}

The mixed scalar--current correlator bootstrap for the $O(2)$ vector
model was studied in ref.~\cite{ReehorstMixed}, yielding bounds on
current--current--stress-tensor OPE coefficients and connecting these
data to the conductivity. Recent bootstrap studies determine
stress-tensor three-point data in the
Ising CFT \cite{ChangEtAl} and current and stress-tensor data in the
$O(2)$ model \cite{DommesEtAl}. Table~\ref{tab:bootstrap} compares these
with a direct small-$N$ extrapolation of eq.~\eqref{eq:shape-on}. The Ising
value follows from the reported $n_B,n_F$. In the conventions of
ref.~\cite{DommesEtAl}, $\gamma=\gamma_{JJT}$ and
$t_4=-48\gamma_{TTT}$.
\begin{table}[t]
\centering
\begin{tabular}{@{}lccc@{}}
\toprule
Theory and observable & $O(1/N)$ & Bootstrap & Difference\\
\midrule
Ising $t_4$, $N=1$ & $3.24760$ & $3.8893305\ldots$ \cite{ChangEtAl} & $16.50\%$\\
$O(2)$ $\gamma$, $N=2$ & $-0.07505$ & $-0.0776\pm0.0040$ \cite{DommesEtAl} & $3.28\%$\\
$O(2)$ $t_4$, $N=2$ & $3.62380$ & $3.7920\pm0.1392$ \cite{DommesEtAl} & $4.44\%$\\
\bottomrule
\end{tabular}
\caption{Direct extrapolations of the consistently expanded large $N$
expressions. 
The $O(2)$ intervals use eq.~(3.1) of ref.~\cite{DommesEtAl}, including its
parity-odd scalar and additional spin-two gap assumptions. The omitted
large $N$ terms have no assigned error bar. The rounded Ising central value
has a bootstrap uncertainty much smaller than the truncation discrepancy.}
\label{tab:bootstrap}
\end{table}

For the $O(2)$ vector model, the extrapolation gives
$\gamma=-0.07505$ and $t_4=3.62380$ at $N=2$.
These values differ from the bootstrap interval midpoints by $3.28\%$ and
$4.44\%$, respectively \cite{DommesEtAl}.
The pair lies within the joint allowed region obtained with the weaker
spectral assumptions at derivative cutoff $\Lambda=27$ in figure~5 of
ref.~\cite{DommesEtAl}. Under the stronger assumptions used in
table~\ref{tab:bootstrap}, our $\gamma$ estimate remains within the quoted
interval, whereas our $t_4$ estimate lies slightly below its lower bound.
This agreement is encouraging, although it does not provide a controlled
error estimate for the small-$N$ extrapolation.

For Ising, the extrapolation gives $t_4=3.24760$, compared with
$3.8893305\ldots$ from the bootstrap \cite{ChangEtAl}.
The relative difference in the full $t_4$ is $16.50\%$.
This distinction limits the quantitative interpretation at $N=1$.

The GNY model has been bootstrapped using fermion and mixed fermion--scalar correlators \cite{IliesiuGNYBootstrap,ErramilliGNYArchipelago}, with further scalar-spectrum and current-OPE bounds \cite{MitchellPolandGNY,BartlettTisdallAtlas}. To our knowledge, however, its collider parameters $\gamma$ and $t_4$ have not been determined numerically. Extracting them from current and stress-tensor bootstrap studies would provide a direct test of our large $N$ predictions.

This paper is organized as follows. Section~\ref{sec:background} explains sources, contacts, and the regulator.
Sections~\ref{sec:tjj} and~\ref{sec:ttt} present the $\TJJ$ and $\TTT$
calculations and their interpretation. Section~\ref{sec:sphere} provides
the independent squashed-sphere computation of our results. We conclude in
section~\ref{sec:discussion}. The appendices collect tensor bases, integral
formulas, and additional diagrammatic results.

\section{Background fields and contact terms}\label{sec:background} Our computations of the subleading corrections to $\TJJ$ and $\TTT$ use the subgraph expansion of Feynman diagrams, following its application to conformal three-point functions developed in ref.~\cite{LiSubgraphs}. These correlators have complicated tensor structures, and contact terms are essential for reproducing their precise form in the diagrammatic expansion. Keeping track of these contributions is one of the main technical challenges of this work. We introduce background metric and gauge fields to generate the contact vertices systematically and organize their contributions alongside diagrams with separate operator insertions. We therefore establish the background-field formalism before presenting the correlator calculations in sections~\ref{sec:tjj} and~\ref{sec:ttt}.\par 

The source formulation follows the treatment of current and stress-tensor
Ward identities in refs.~\cite{OsbornPetkou,ErdmengerOsborn} and the
momentum-space renormalization framework of
ref.~\cite{BMSRenormalised}. The correlators must be defined as
distributions, including coincident-point contributions.

\subsection{Generating functional and source variations}

We couple the stress-tensor and current to the background
metric $g_{\mu\nu}$ and gauge field $A_\mu^a$.
Define the generating functional as
\begin{equation}
Z[g,A]=\int\mathcal D\Phi\,e^{-S[\Phi;g,A]},\qquad W[g,A]=-\log Z[g,A],
\end{equation}
The insertions of $T$ and $J$ are given by
\begin{equation}
T^{\mu\nu}=-\frac2{\sqrt g}\frac{\delta S}{\delta g_{\mu\nu}},
 \qquad J^{a\mu}=\frac1{\sqrt g}\frac{\delta S}{\delta A_\mu^a}.
\label{eq:sources}
\end{equation}
Source differentiation gives the connected correlators
\begin{equation}
\begin{aligned}
\bigl\langle T^{\mu\nu}(x)J^{a\rho}(y)J^{b\sigma}(z)\bigr\rangle_{c,\mathrm{source}}
&=-2\left.
\frac{\delta^3 W[g,A]}
{\delta g_{\mu\nu}(x)\,\delta A_\rho^a(y)\,\delta A_\sigma^b(z)}
\right|_{g_{\alpha\beta}=\delta_{\alpha\beta},\,A=0},
\\[6pt]
\bigl\langle T^{\mu\nu}(x)T^{\rho\sigma}(y)T^{\alpha\beta}(z)\bigr\rangle_{c,\mathrm{source}}
&=-8\left.
\frac{\delta^3 W[g,A]}
{\delta g_{\mu\nu}(x)\,\delta g_{\rho\sigma}(y)\,\delta g_{\alpha\beta}(z)}
\right|_{g_{\gamma\delta}=\delta_{\gamma\delta},\,A=0}.
\end{aligned}
\label{eq:three-point-source-correlators}
\end{equation}
The subscript $c$ means that we take the connected part of the correlator, and the subscript $\mathrm{source}$ indicates that we include the contact terms generated by differentiating with respect to the sources.
These contact terms arise from the second and third variations of the action.
In momentum space, these contact terms must be included;
otherwise the Feynman diagrams generally do not lead to consistent results.


\subsection{Source couplings in the \texorpdfstring{$O(N)$}{O(N)} vector and GNY models}\label{sec:source-actions}

The contact terms follow from the nonlinear dependence of the action on the
background sources. We compute the required source variations in each model
and then set $g_{\mu\nu}=\delta_{\mu\nu}$ and $A=0$.
In this background, scaleless one-point functions vanish in analytic regularization.
The one-point contact terms from third source variations therefore vanish, so the contact terms only come from second source variations.

The critical $O(N)$ vector model couples to the sources through
\begin{equation}
S_{\ON}[g,A]=\frac12\int\dd^3x\sqrt g\left[
g^{\mu\nu}(D_\mu\phi)_i(D_\nu\phi)_i+
\frac18\mathcal R\phi_i\phi_i+\sigma\phi_i\phi_i\right],
\quad D_\mu\phi=\partial_\mu\phi-A_\mu^at^a\phi.
\label{eq:scalar-curved}
\end{equation}
Its mixed metric--gauge and second gauge variations give
\begin{equation}
\begin{aligned}
&\bigl\langle T^{\mu\nu}(x)J^{a\rho}(y)J^{b\sigma}(z)\bigr\rangle_{c,\mathrm{source}}
=\bigl\langle T^{\mu\nu}(x)J^{a\rho}(y)J^{b\sigma}(z)\bigr\rangle_c
\\
&\quad+\delta^{(3)}(x-y)\Bigl[
\delta^{\mu\nu}\bigl\langle J^{a\rho}(y)J^{b\sigma}(z)\bigr\rangle_c
-\delta^{\mu\rho}\bigl\langle J^{a\nu}(y)J^{b\sigma}(z)\bigr\rangle_c
-\delta^{\nu\rho}\bigl\langle J^{a\mu}(y)J^{b\sigma}(z)\bigr\rangle_c\Bigr]
\\
&\quad+\delta^{(3)}(x-z)\Bigl[
\delta^{\mu\nu}\bigl\langle J^{a\rho}(y)J^{b\sigma}(z)\bigr\rangle_c
-\delta^{\mu\sigma}\bigl\langle J^{a\rho}(y)J^{b\nu}(z)\bigr\rangle_c
-\delta^{\nu\sigma}\bigl\langle J^{a\rho}(y)J^{b\mu}(z)\bigr\rangle_c\Bigr]
\\
&\quad-\delta^{\rho\sigma}\delta^{(3)}(y-z)
\bigl\langle T^{\mu\nu}(x)(t^a\phi)_i(y)(t^b\phi)_i(y)\bigr\rangle_c.
\end{aligned}
\label{eq:scalar-tjj-source-expansion}
\end{equation}
The mixed variations act on $\sqrt g\,g^{\mu\nu}$ and produce the contacts
at $x=y$ and $x=z$. The second gauge variation produces the contact at
$y=z$, since the action is quadratic in $A$.

For the GNY model, the action is
\begin{equation}
S_{\GN}[g,A]=\frac12\int\dd^3x\sqrt g\left[
\bar\psi_i\gamma^Ae_A{}^\mu D_\mu\psi_i
+\sigma\bar\psi_i\psi_i+\frac{N}{g}\sigma^2\right],
\qquad i=1,\ldots,N,
\label{eq:fermion-curved}
\end{equation}
with
\begin{equation}
D_\mu\psi_i=\partial_\mu\psi_i
+\frac18\omega_{\mu AB}[\gamma^A,\gamma^B]\psi_i+A_\mu^a(t^a)_{ij}\psi_j.
\label{eq:fermion-covariant-derivative}
\end{equation}
Here $\psi_i$ are two-component Majorana fermions, $e_\mu{}^A$ is a vielbein satisfying $g_{\mu\nu}=e_\mu{}^Ae_\nu{}^B\delta_{AB}$, $e_A{}^\mu$ is its inverse, and $\omega_{\mu AB}$ is the associated torsion-free spin connection. The indices $A,B=1,2,3$ are frame indices, and $\gamma^A$ is a Euclidean gamma matrix satisfying $\{\gamma^A,\gamma^B\}=2\delta^{AB}\mathbf1$ and $\Tr\mathbf1=2$. The matrix $t^a$ is an $O(N)$ flavor generator, $\sigma$ is a flavor-singlet auxiliary scalar, and $g$ is the coupling tuned to the critical point.

The gauge field enters the action linearly, so
$\delta^2S_{\GN}/(\delta A_\rho^a(y)\delta A_\sigma^b(z))=0$.
The $\TJJ$ contact terms therefore come only from mixed metric--gauge
variations.
Using
\begin{equation}
\left.\frac{\delta(\sqrt{g(y)}\,e_A{}^\rho(y))}
{\delta g_{\mu\nu}(x)}\right|_{g=\delta}
=\delta^{(3)}(x-y)\left[
\frac12\delta^{\mu\nu}\delta_A{}^\rho
-\frac14\left(\delta^{\mu\rho}\delta_A{}^\nu
+\delta^{\nu\rho}\delta_A{}^\mu\right)\right],
\label{eq:fermion-vielbein-variation}
\end{equation}
we obtain the mixed metric--gauge variations
\begin{equation}
\begin{aligned}
\left.\frac{\delta^2S_{\GN}[g,A]}
{\delta g_{\mu\nu}(x)\,\delta A_\rho^a(y)}\right|_{g=\delta,A=0}
&=\left.\frac12\bar\psi_i(y)\gamma^A(t^a)_{ij}\psi_j(y)
\frac{\delta(\sqrt{g(y)}\,e_A{}^\rho(y))}{\delta g_{\mu\nu}(x)}
\right|_{g=\delta}
\\
&=\delta^{(3)}(x-y)\left[
\frac12\delta^{\mu\nu}J^{a\rho}(y)
-\frac14\delta^{\mu\rho}J^{a\nu}(y)
-\frac14\delta^{\nu\rho}J^{a\mu}(y)\right].
\end{aligned}
\label{eq:fermion-mixed-contact}
\end{equation}
The connected correlator with contact terms is thus
\begin{equation}
\begin{aligned}
&\bigl\langle T^{\mu\nu}(x)J^{a\rho}(y)J^{b\sigma}(z)\bigr\rangle_{c,\mathrm{source}}
=\bigl\langle T^{\mu\nu}(x)J^{a\rho}(y)J^{b\sigma}(z)\bigr\rangle_c
\\
&\quad+\delta^{(3)}(x-y)\Bigl[
\delta^{\mu\nu}\bigl\langle J^{a\rho}(y)J^{b\sigma}(z)\bigr\rangle_c
-\frac12\delta^{\mu\rho}\bigl\langle J^{a\nu}(y)J^{b\sigma}(z)\bigr\rangle_c
-\frac12\delta^{\nu\rho}\bigl\langle J^{a\mu}(y)J^{b\sigma}(z)\bigr\rangle_c\Bigr]
\\
&\quad+\delta^{(3)}(x-z)\Bigl[
\delta^{\mu\nu}\bigl\langle J^{a\rho}(y)J^{b\sigma}(z)\bigr\rangle_c
-\frac12\delta^{\mu\sigma}\bigl\langle J^{a\rho}(y)J^{b\nu}(z)\bigr\rangle_c
-\frac12\delta^{\nu\sigma}\bigl\langle J^{a\rho}(y)J^{b\mu}(z)\bigr\rangle_c\Bigr].
\end{aligned}
\label{eq:fermion-tjj-source-expansion}
\end{equation}

For $\TTT$, both models have contact terms from the second metric variation.
We express this through the vertex
\begin{equation}
\mathcal V_{\mu\nu,\rho\sigma}(x,y)=
\left.\frac{\delta^2S[g,A]}
{\delta g_{\mu\nu}(x)\,\delta g_{\rho\sigma}(y)}
\right|_{g=\delta,\,A=0},
\label{eq:metric-contact-vertex}
\end{equation}
where $S[g,A]$ is the action of the critical $O(N)$ vector model or the GNY model given above.
The full action must be varied before taking the flat-background limit.
This includes the scalar curvature coupling and the fermion spin connection.
The resulting vertex enters the source correlator as
\begin{equation}
\bigl\langle T^{\mu\nu}(x)T^{\rho\sigma}(y)T^{\alpha\beta}(z)\bigr\rangle_{c,\mathrm{source}}
=\bigl\langle T^{\mu\nu}(x)T^{\rho\sigma}(y)T^{\alpha\beta}(z)\bigr\rangle_c
-4\sum_{\mathrm{cyc}}
\bigl\langle\mathcal V_{\mu\nu,\rho\sigma}(x,y)T^{\alpha\beta}(z)\bigr\rangle_c.
\label{eq:scalar-ttt-source-expansion}
\end{equation}
The sum runs over the three cyclic permutations of $(x,\mu\nu)$,
$(y,\rho\sigma)$, and $(z,\alpha\beta)$.
Each term pairs one metric contact vertex with the remaining stress-tensor
insertion. 

\par For the $\langle JJJ\rangle$ correlator studied in ref.~\cite{LiSubgraphs}, the additional matter source-contact contributions vanish in both models. In the GNY model, the action is linear in the background gauge field, hence \begin{equation} \left.\frac{\delta^2S_{\GN}[g,A]}{\delta A_\mu^a(x)\,\delta A_\nu^b(y)}\right|_{g_{\alpha\beta}=\delta_{\alpha\beta},\,A=0}=0. \label{eq:jjj-gny-contact-vanishing} \end{equation} All higher variations with respect to the gauge source vanish as well. In the $O(N)$ vector model, the second variation instead gives \begin{align} \left.\frac{\delta^2S_{\ON}}{\delta A_\mu^a(x)\,\delta A_\nu^b(y)}\right|_{A=0} &=\delta^{\mu\nu}\delta^{(3)}(x-y)\,Q^{ab}(x),\nonumber\\ Q^{ab}&=(t^a\phi)_i(t^b\phi)_i=-\frac12\phi^T\{t^a,t^b\}\phi. \label{eq:jjj-scalar-contact-vertex} \end{align} The scalar bilinear is symmetric in flavor indices, whereas the current is antisymmetric, so \begin{equation} \bigl\langle Q^{ab}(x)J_\rho^c(z)\bigr\rangle_c=0,\qquad \operatorname{Tr}\!\left(\{t^a,t^b\}t^c\right)=0. \label{eq:jjj-scalar-contact-vanishing} \end{equation} The third gauge-source variation also vanishes because the $O(N)$ action is at most quadratic in $A$. The possible additional three-current contact contributions therefore vanish, and separate-insertion diagrams suffice to determine the $\langle JJJ\rangle$ data without explicitly introducing background fields. For $\TJJ$ and $\TTT$, mixed and metric source variations instead generate nonzero contact contributions, which the background-field formalism organizes systematically.\par \subsection{Contact terms in the momentum-space subgraph expansion}
\label{sec:contact-momentum}

Momentum-space correlators are Fourier transforms of the full position-space distributions. These distributions
include the contact terms of section~\ref{sec:source-actions}.
The contact terms must therefore be included in the subgraph
expansion, even though they vanish at distinct insertion points.

To determine their momentum dependence, assign incoming momenta
$(q,p,-p-q)$ to $(x,y,z)$, with
$|q|\ll|p|$. When two insertion points coincide, their momenta enter the
same local vertex. Momentum conservation then fixes the momentum carried
by the two-point function connecting this vertex to the remaining insertion.
There are following types of contact terms:
\begin{enumerate}
\item \emph{Contact between insertions carrying soft and hard momenta.}
The condition $x=y$ or $x=z$ leaves a two-point function evaluated at
$p+q$ or $p$, respectively. 
The two-point function can therefore be Taylor expanded
in the components of $q$. This contact contributes terms analytic in $q$
to the subgraph expansion.

\item \emph{Contact between the two insertions carrying hard momenta.}
When $y=z$, the momenta $p$ and $-p-q$ enter the same local vertex.
Their sum is $-q$, so the remaining two-point function is evaluated at
$q$. This function can be nonanalytic at $q=0$ and thus contribute
nonanalytic terms to the subgraph expansion.

\item \emph{Contact at three coincident insertion points.}
A contact supported at $x=y=z$ is a finite sum of derivatives of
$\delta^{(3)}(x-y)\delta^{(3)}(y-z)$.
After the overall momentum-conserving delta function is removed, its
Fourier transform is a polynomial in the external momenta, which are analytic in $q$.
\end{enumerate}

\subsubsection{Momentum-space correlators}
\label{sec:source-momentum-identities}

We now Fourier transform the correlators of
section~\ref{sec:source-actions} before expanding in the soft momentum.
Our convention is $\mathcal O(k)=\int\dd^3x\,e^{-ik\cdot x}\mathcal O(x)$.
In this subsection, a prime removes the overall factor
$(2\pi)^3\delta^{(3)}(q+p+r)$, and $r=-p-q$.
For a two-point function, the prime removes the analogous delta function
of its two momenta.
For example,
\begin{equation}
\begin{aligned}
&\int\dd^3x\,\dd^3y\,\dd^3z\,
 e^{-i(q\cdot x+p\cdot y+r\cdot z)}\delta^{(3)}(x-y)
 \langle J^{a\rho}(y)J^{b\sigma}(z)\rangle_c
\\
&\quad=\int\dd^3y\,\dd^3z\,
 e^{-i((q+p)\cdot y+r\cdot z)}
 \langle J^{a\rho}(y)J^{b\sigma}(z)\rangle_c
\\
&\quad=(2\pi)^3\delta^{(3)}(q+p+r)
 \langle J^{a\rho}(-r)J^{b\sigma}(r)\rangle_c' .
\end{aligned}
\label{eq:fourier-contact-collapse}
\end{equation}
The contacts at $x=z$ and $y=z$ similarly leave two-point functions
with momenta $(p,-p)$ and $(q,-q)$, respectively.

\paragraph{The $O(N)$ vector model: $\TJJ$.}
Fourier transforming eq.~\eqref{eq:scalar-tjj-source-expansion} gives
\begin{equation}
\begin{aligned}
&\langle T^{\mu\nu}(q)J^{a\rho}(p)J^{b\sigma}(r)\rangle_{c,\mathrm{source}}'
=\langle T^{\mu\nu}(q)J^{a\rho}(p)J^{b\sigma}(r)\rangle_c'
\\
&\quad+\delta^{\mu\nu}\langle J^{a\rho}(-r)J^{b\sigma}(r)\rangle_c'
-\delta^{\mu\rho}\langle J^{a\nu}(-r)J^{b\sigma}(r)\rangle_c'
-\delta^{\nu\rho}\langle J^{a\mu}(-r)J^{b\sigma}(r)\rangle_c'
\\
&\quad+\delta^{\mu\nu}\langle J^{a\rho}(p)J^{b\sigma}(-p)\rangle_c'
-\delta^{\mu\sigma}\langle J^{a\rho}(p)J^{b\nu}(-p)\rangle_c'
-\delta^{\nu\sigma}\langle J^{a\rho}(p)J^{b\mu}(-p)\rangle_c'
\\
&\quad-\delta^{\rho\sigma}\langle T^{\mu\nu}(q)Q^{ab}(-q)\rangle_c'.
\end{aligned}
\label{eq:on-tjj-momentum-source}
\end{equation}
Here $Q^{ab}$ is the scalar bilinear in
eq.~\eqref{eq:jjj-scalar-contact-vertex}; $Q^{ab}(-q)$ is its Fourier
transform as a local composite operator.
Since the stress tensor is a flavor singlet, its two-point function with $Q^{ab}$ receives a contribution only from the component proportional to $\phi_i\phi_i$:
\begin{equation}
\langle T^{\mu\nu}(q)Q^{ab}(-q)\rangle_c'
=\frac{\tau^{ab}}{N}
 \langle T^{\mu\nu}(q)(\phi_i\phi_i)(-q)\rangle_c'.
\label{eq:on-tjj-singlet-contact}
\end{equation}
\paragraph{The GNY model: $\TJJ$.}
Fourier transforming eq.~\eqref{eq:fermion-tjj-source-expansion} gives
\begin{equation}
\begin{aligned}
&\langle T^{\mu\nu}(q)J^{a\rho}(p)J^{b\sigma}(r)\rangle_{c,\mathrm{source}}'
=\langle T^{\mu\nu}(q)J^{a\rho}(p)J^{b\sigma}(r)\rangle_c'
\\
&\quad+\delta^{\mu\nu}\langle J^{a\rho}(-r)J^{b\sigma}(r)\rangle_c'
-\tfrac12\delta^{\mu\rho}\langle J^{a\nu}(-r)J^{b\sigma}(r)\rangle_c'
-\tfrac12\delta^{\nu\rho}\langle J^{a\mu}(-r)J^{b\sigma}(r)\rangle_c'
\\
&\quad+\delta^{\mu\nu}\langle J^{a\rho}(p)J^{b\sigma}(-p)\rangle_c'
-\tfrac12\delta^{\mu\sigma}\langle J^{a\rho}(p)J^{b\nu}(-p)\rangle_c'
-\tfrac12\delta^{\nu\sigma}\langle J^{a\rho}(p)J^{b\mu}(-p)\rangle_c'.
\end{aligned}
\label{eq:gny-tjj-momentum-source}
\end{equation}
The action is linear in the gauge source, so there is no counterpart
of the last line of eq.~\eqref{eq:on-tjj-momentum-source}.

\paragraph{The $O(N)$ vector model: $\TTT$.}
The metric contacts require the Fourier transform of the full vertex
in eq.~\eqref{eq:metric-contact-vertex}.
For either model $M\in\{\ON,\GN\}$, its momentum space representation is
\begin{equation}
\mathcal V_{M;\mu\nu,\rho\sigma}(k_1,k_2)
=\int\dd^3x\,\dd^3y\,
 e^{-i(k_1\cdot x+k_2\cdot y)}
 \mathcal V_{M;\mu\nu,\rho\sigma}(x,y).
\label{eq:metric-contact-fourier}
\end{equation}
For the scalar model this gives
\begin{equation}
\begin{aligned}
&\langle T^{\mu\nu}(q)T^{\rho\sigma}(p)T^{\alpha\beta}(r)\rangle_{c,\mathrm{source}}'
=\langle T^{\mu\nu}(q)T^{\rho\sigma}(p)T^{\alpha\beta}(r)\rangle_c'
\\
&\quad-4\langle T^{\mu\nu}(q)
 \mathcal V_{\ON;\rho\sigma,\alpha\beta}(p,r)\rangle_c'
\\
&\quad-4\langle T^{\rho\sigma}(p)
 \mathcal V_{\ON;\mu\nu,\alpha\beta}(q,r)\rangle_c'
\\
&\quad-4\langle T^{\alpha\beta}(r)
 \mathcal V_{\ON;\mu\nu,\rho\sigma}(q,p)\rangle_c'.
\end{aligned}
\label{eq:on-ttt-momentum-source}
\end{equation}
\paragraph{The GNY model: $\TTT$.}
For the fermion model, the equation takes the same form
\begin{equation}
\begin{aligned}
&\langle T^{\mu\nu}(q)T^{\rho\sigma}(p)T^{\alpha\beta}(r)\rangle_{c,\mathrm{source}}'
=\langle T^{\mu\nu}(q)T^{\rho\sigma}(p)T^{\alpha\beta}(r)\rangle_c'
\\
&\quad-4\langle T^{\mu\nu}(q)
 \mathcal V_{\GN;\rho\sigma,\alpha\beta}(p,r)\rangle_c'
\\
&\quad-4\langle T^{\rho\sigma}(p)
 \mathcal V_{\GN;\mu\nu,\alpha\beta}(q,r)\rangle_c'
\\
&\quad-4\langle T^{\alpha\beta}(r)
 \mathcal V_{\GN;\mu\nu,\rho\sigma}(q,p)\rangle_c'.
\end{aligned}
\label{eq:gny-ttt-momentum-source}
\end{equation}
Explicit momentum-space expressions for its contraction with arbitrary
symmetric traceless source polarizations are given in
eqs.~\eqref{eq:on-explicit-metric-vertex} and~\eqref{eq:gny-explicit-metric-vertex}.
Their diagrammatic contributions are explained alongside
figures~\ref{fig:tjj}--\ref{fig:gny-ttt-contacts} in
sections~\ref{sec:tjj} and~\ref{sec:ttt}.

\subsection{Analytic regularization and source renormalization}\label{sec:regulator}

We use the measure $\intk$ and remove the overall momentum delta function.
The auxiliary propagators are regulated as
\begin{equation}
 G^{\ON}_{\sigma,\Delta}(k)=-\frac{16\mu^{2\Delta}}{N(k^2)^{-1/2+\Delta}},
 \qquad G^{\GN}_{\sigma,\Delta}(k)=\frac{16\mu^{2\Delta}}{N(k^2)^{1/2+\Delta}}.
 \label{eq:regulated-propagators}
\end{equation}
We set $\mu=1$ in displayed Laurent coefficients, so $\log p^2$ means
$\log(p^2/\mu^2)$.
The Ward identities fix the renormalization factors \cite{DiabEtAl}
\begin{align}
 Z_T^{\ON}&=1+\frac8{15\pi^2N}\left(\frac1\Delta-4\right)+O(N^{-2}),\\
 Z_T^{\GN}&=1+\frac8{15\pi^2N}\left(\frac1\Delta+4\right)+O(N^{-2}),
 \qquad Z_J=1+O(N^{-2}).\label{eq:z-factors}
\end{align}
Vertices generated by second variations with respect to the metric require additional local renormalization in appendix~\ref{app:counterterms}.

\subsection{Hard and soft momentum regions}\label{sec:regions}

We take momenta $(q,p,-p-q)$ with $|q|\ll|p|$. Hard regions have loop
momenta of order $|p|$; soft regions contain momenta of order $|q|$.
Their sum gives the regulated asymptotic expansion. The subgraph
construction and its relation to conformal matching are described in
section~\ref{sec:subgraphs}.
We write $D_i^{\mathrm{soft},(j)}$ for degree $j$ under
$q\mapsto\lambda q$ at fixed $p$, with logarithms counted with their
accompanying power. External placements must be expanded before taking
$q\to0$. In particular, exchanging hard legs also shifts $p\mapsto-p-q$,
which can feed lower-degree terms into the cubic coefficient.
For the correlators, the superscripts $\mathrm{hard},(0)$ and
$\mathrm{soft},(3)$ select the degree-zero hard term and the degree-three
nonanalytic soft term, respectively.
These degrees refer to the momentum expansion, independently of the
order in $1/N$.

\section{\texorpdfstring{$\langle TJJ\rangle$}{TJJ} in the \texorpdfstring{$O(N)$}{O(N)} vector and GNY models}\label{sec:tjj}

\input{subgraph-method}

\subsection{Conformal structures, collider bounds, and conductivity}\label{sec:tjj-general}

Our position-space normalizations are
\begin{align}
 \langle J_\mu^a(x)J_\nu^b(0)\rangle
 &=\tau^{ab}C_J\frac{I_{\mu\nu}(x)}{|x|^4},
 \qquad I_{\mu\nu}(x)=\delta_{\mu\nu}-2\frac{x_\mu x_\nu}{x^2},\\
 \langle T_{\mu\nu}(x)T_{\rho\sigma}(0)\rangle
 &=C_T\frac{\mathcal I_{\mu\nu,\rho\sigma}(x)}{|x|^6},\\
 \mathcal I_{\mu\nu,\rho\sigma}
 &=\frac12(I_{\mu\rho}I_{\nu\sigma}+I_{\mu\sigma}I_{\nu\rho})
 -\frac13\delta_{\mu\nu}\delta_{\rho\sigma}.
 \label{eq:two-points}
\end{align}
The free reference values are
\begin{equation}
 C_J^{\mathrm s}=\frac1{16\pi^2},\qquad C_T^{\mathrm s}=\frac3{32\pi^2}.
 \label{eq:free-c}
\end{equation}
We normalize the free-scalar and free-fermion $\TJJ$ reference correlators to have the same flavor factor $\tau^{ab}$ and the same current two-point coefficient $C_J=C_J^{\mathrm s}$.
At separated points,
\begin{equation}
 \TJJ=\nsJ\TJJ_{\mathrm s}+\nfJ\TJJ_{\mathrm f},
 \qquad \nsJ+\nfJ=\frac{C_J}{C_J^{\mathrm s}}.
 \label{eq:tjj-decomposition}
\end{equation}
The shape $\gamma$ is related to $\nsJ$ and $\nfJ$ by eq.~\eqref{eq:shape-def}. Our
convention gives $\gamma_{\mathrm s}=-1/12$ and
$\gamma_{\mathrm f}=+1/12$.

For a current-created state of energy $E$ and spatial polarization
$\boldsymbol\epsilon$, the energy flux along $\boldsymbol n\in S^1$ is
\begin{equation}
 \frac{\langle\mathcal E(\boldsymbol n)\rangle_{\epsilon\cdot J}}{E/(2\pi)}
 =1+a_2\left(\frac{|\boldsymbol\epsilon\cdot\boldsymbol n|^2}
 {|\boldsymbol\epsilon|^2}-\frac12\right),\qquad a_2=-24\gamma.
 \label{eq:j-flux}
\end{equation}
Positivity for all polarizations gives the parity-even collider bounds
\cite{HofmanMaldacena,LiMeltzerPoland}
\begin{equation}
 -2\leq a_2\leq2,\qquad -\frac1{12}\leq\gamma\leq\frac1{12},
 \qquad \nsJ,\nfJ\geq0.\label{eq:j-bounds}
\end{equation}

For a fixed detector direction the two orthogonal current polarizations
diagonalize the energy-flux matrix. After removing its positive overall
normalization, the eigenvalues are $1\pm12\gamma$. A negative eigenvalue
would permit a polarized state with negative measured null energy.
Equation~\eqref{eq:j-bounds} is therefore a constraint on a physical
measurement. The independent
bootstrap and causality arguments in
refs.~\cite{HofmanEtAlProof,HartmanKunduTajdini} show why this constraint
applies beyond free and holographic examples. In a parity-violating
theory the extra structure supplies off-diagonal matrix elements, and
positivity constrains it jointly with $\gamma$
\cite{ChowdhuryParityCollider}.

The same data determine the stress contribution to the current OPE.
For a fixed Abelian generator, suppressing its flavor factor, the
high-frequency conductivity at $\Omega\gg T_{\rm th}$ has the form
\begin{equation}
\begin{aligned}
 \sigma(\ii\Omega)&=\sigma_\infty+
 \sum_{\mathcal O}b_{\mathcal O}
 \frac{\langle\mathcal O\rangle_{T_{\rm th}}}{\Omega^{\Delta_{\mathcal O}}}
 +\mathcal B_J\frac{P}{\Omega^3}+\cdots,\\
 \sigma_\infty&=\frac{\pi^2C_J}{2},\qquad
 \mathcal B_J=72\,\gamma\,\frac{C_J}{C_T}.
\end{aligned}
\label{eq:conductivity}
\end{equation}
Here $P$ denotes the equilibrium pressure of the full CFT at temperature
$T_{\rm th}$. In a homogeneous, isotropic thermal state,
$\langle T_{xx}\rangle_{T_{\rm th}}=\langle T_{yy}\rangle_{T_{\rm th}}=P$,
and the energy density satisfies $\varepsilon_{\mathrm{th}}=2P$ in three spacetime
dimensions. 
The expression for $\mathcal B_J$ follows from the stress-tensor term in
the current OPE \cite{KatzEtAl}, translated to the two-point
normalizations stated above.
The sum runs over scalar operators with nonzero thermal one-point
functions; their contributions are separate from the displayed
stress-tensor term.
The parameter $\gamma$
controls this definite part of the high-frequency response.  The relation between $\gamma$, momentum-space $\TJJ$, and quantum
critical transport was made explicit in ref.~\cite{ChowdhuryTransport}.

\subsection{The critical \texorpdfstring{$O(N)$}{O(N)} vector model}\label{sec:tjj-on}

The flat-space auxiliary-field action for the $O(N)$ model is
\begin{equation}
 S_{\ON}=\int\dd^3x\left[\frac12(\partial_\mu\phi_i)^2
 +\frac12\sigma\phi_i\phi_i+\frac{N}{4g}\sigma^2\right].
\end{equation}
At large $N$, $\Delta_\sigma=2+O(1/N)$, so $\sigma^2$ is irrelevant
at criticality. The leading large-$N$ propagators and interaction vertex are
\begin{equation}
\begin{array}{c@{\quad=\quad}l}
\begin{tikzpicture}[baseline=-0.5ex,line width=0.8pt]
  \draw (0,0) -- (2.2,0);
  \node[left] at (0,0) {$\phi_i$};
  \node[right] at (2.2,0) {$\phi_j$};
  \node[above] at (1.1,0) {$p$};
\end{tikzpicture}
& \displaystyle\frac{\delta_{ij}}{p^2}\,,
\\[16pt]
\begin{tikzpicture}[baseline=-0.5ex,line width=0.8pt]
  \draw[dashed] (0,0) -- (2.2,0);
  \node[left] at (0,0) {$\sigma$};
  \node[right] at (2.2,0) {$\sigma$};
  \node[above] at (1.1,0) {$p$};
\end{tikzpicture}
& \displaystyle-\frac{16|p|}{N}\,,
\\[20pt]
\begin{tikzpicture}[baseline=-0.5ex,line width=0.8pt]
  \draw[dashed] (0,0) -- (1.1,0);
  \draw (2.2,0.65) -- (1.1,0) -- (2.2,-0.65);
  \fill (1.1,0) circle (1.8pt);
  \node[left] at (0,0) {$\sigma$};
  \node[right] at (2.2,0.65) {$\phi_i$};
  \node[right] at (2.2,-0.65) {$\phi_j$};
\end{tikzpicture}
& -\delta_{ij}\,.
\end{array}
\label{eq:on-rules}
\end{equation}
Solid lines denote scalar fields, and dashed lines denote the auxiliary field.
The current and traceless stress insertion are
\begin{align}
 J_\mu^a&=\phi_i(t^a)_{ij}\partial_\mu\phi_j,\\
 T_{\mu\nu}&=\partial_\mu\phi_i\partial_\nu\phi_i
 -\frac13\delta_{\mu\nu}(\partial_\rho\phi_i)^2
 -\frac18\left(\partial_\mu\partial_\nu
 -\frac13\delta_{\mu\nu}\partial^2\right)(\phi_i\phi_i).
 \label{eq:scalar-operators}
\end{align}

Fourier transforming these operators gives
\begin{equation}
\begin{aligned}
J_\mu^a(p)
&=\frac{\mathrm{i}}{2}\int\frac{\dd^3k}{(2\pi)^3}
(2k+p)_\mu\phi_i(-k)(t^a)_{ij}\phi_j(k+p)\,,
\\
T_{\mu\nu}(p)
&=\frac12\int\frac{\dd^3k}{(2\pi)^3}
\Bigl[k_\mu(k+p)_\nu+k_\nu(k+p)_\mu
\\
&\qquad-\frac23\delta_{\mu\nu}\,k\cdot(k+p)
+\frac14\bigl(p_\mu p_\nu-\frac13\delta_{\mu\nu}p^2\bigr)\Bigr]
\phi_i(-k)\phi_i(k+p)\,.
\end{aligned}
\label{eq:scalar-fourier-insertions}
\end{equation}
The corresponding two-leg insertion rules are
\begin{equation}
\begin{array}{@{}c@{\enspace=\enspace}l@{}}
\begin{tikzpicture}[baseline=-0.5ex,line width=0.8pt]
  \tikzset{momentum/.style={
    postaction={decorate},
    decoration={markings,mark=at position 0.55 with {
      \arrow{Stealth[length=2.4mm,width=1.7mm]}
    }}
  }}
  \draw[momentum] (1.1,0.65) -- node[midway,above] {$k$} (0,0);
  \draw[momentum] (0,0) -- node[midway,below=4pt] {$k+p$} (1.1,-0.65);
  \fill (0,0) circle (2pt);
  \node[left=5pt] at (0,0) {$J_\mu^a(p)$};
  \node[right] at (1.1,0.65) {$\phi_i$};
  \node[right] at (1.1,-0.65) {$\phi_j$};
\end{tikzpicture}
&\mathrm{i}\,(2k+p)_\mu\,(t^a)_{ij}\,,
\\[16pt]
\begin{tikzpicture}[baseline=-0.5ex,line width=0.8pt]
  \tikzset{momentum/.style={
    postaction={decorate},
    decoration={markings,mark=at position 0.55 with {
      \arrow{Stealth[length=2.4mm,width=1.7mm]}
    }}
  }}
  \draw[momentum] (1.1,0.65) -- node[midway,above] {$k$} (0,0);
  \draw[momentum] (0,0) -- node[midway,below=4pt] {$k+p$} (1.1,-0.65);
  \fill (0,0) circle (2pt);
  \node[left=5pt] at (0,0) {$T_{\mu\nu}(p)$};
  \node[right] at (1.1,0.65) {$\phi_i$};
  \node[right] at (1.1,-0.65) {$\phi_j$};
\end{tikzpicture}
&\begin{aligned}
&\Bigl[k_\mu(k+p)_\nu+k_\nu(k+p)_\mu\\
&\quad-\frac23\delta_{\mu\nu}\,k\cdot(k+p)\\
&\quad+\frac14\bigl(p_\mu p_\nu-\frac13\delta_{\mu\nu}p^2\bigr)\Bigr]\delta_{ij}\,.
\end{aligned}
\end{array}
\label{eq:scalar-diagrammatic-insertions}
\end{equation}
The arrows label momentum flow through the insertion.

\begin{figure}[t]
\centering\includegraphics[width=\textwidth]{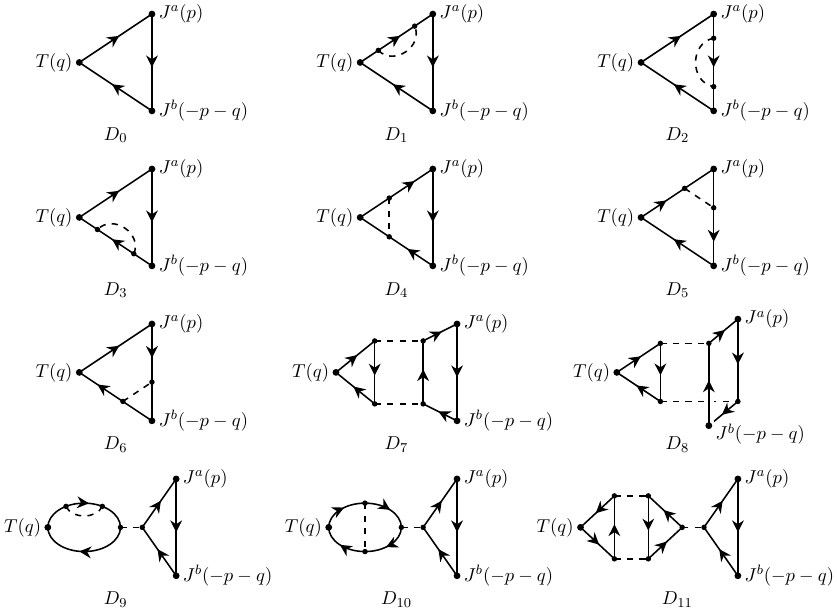}
\caption{Separate-insertion diagrams for $\TJJ$ in the critical $O(N)$ vector model through order $1/N$. Solid and dashed lines denote fundamental and auxiliary propagators.
Crossed external placements are included in the full result.
These diagrams supply the separate-insertion term in
eq.~\eqref{eq:on-tjj-momentum-source}.}\label{fig:tjj}
\end{figure}

\begin{figure}[t]
\centering
\includegraphics[width=\textwidth]{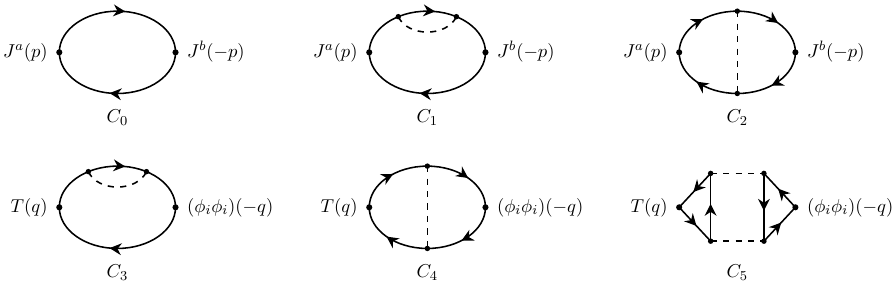}\par\vspace{4pt}
\resizebox{\textwidth}{!}{%
\begin{tikzpicture}[line width=.75pt,font=\normalsize,
 scalar flow/.style={postaction={decorate},decoration={markings,
 mark=at position .5 with {\arrow{Stealth[length=1.8mm,width=1.8mm]}}}},
 scalar double flow/.style={postaction={decorate},decoration={markings,
 mark=at position .25 with {\arrow{Stealth[length=1.8mm,width=1.8mm]}},
 mark=at position .75 with {\arrow{Stealth[length=1.8mm,width=1.8mm]}}}},
 auxiliary/.style={dashed,line width=.75pt}]
\begin{scope}
 \coordinate (T) at (0,0); \coordinate (s) at (1.5,0);
 \draw[scalar flow] (T) .. controls (0,.87) and (1.5,.87) .. (s);
 \draw[scalar flow] (s) .. controls (1.5,-.87) and (0,-.87) .. (T);
 \draw[auxiliary] (.234375,.489375) .. controls (.38,.12) and (1.12,.12) .. (1.265625,.489375);
 \fill (.234375,.489375) circle (1.6pt); \fill (1.265625,.489375) circle (1.6pt);
 \coordinate (u) at (2.35,0); \coordinate (O) at (3.65,0);
 \draw[auxiliary] (s) -- (u);
 \draw[scalar flow] (u) .. controls (2.35,.67) and (3.65,.67) .. (O);
 \draw[scalar flow] (O) .. controls (3.65,-.67) and (2.35,-.67) .. (u);
 \foreach \v in {T,s,u,O}{\fill (\v) circle (1.8pt);}
 \node[left] at (T) {$T(q)$};
 \node[right=3pt] at (O) {$(\phi_i\phi_i)(-q)$};
 \node at (1.83,-1.05) {$C_6$};
\end{scope}
\begin{scope}[shift={(7.4,0)}]
 \coordinate (T) at (0,0); \coordinate (s) at (1.5,0);
 \draw[scalar double flow] (T) .. controls (0,.87) and (1.5,.87) .. (s);
 \draw[scalar double flow] (s) .. controls (1.5,-.87) and (0,-.87) .. (T);
 \draw[auxiliary] (.75,.6525) -- (.75,-.6525);
 \fill (.75,.6525) circle (1.6pt); \fill (.75,-.6525) circle (1.6pt);
 \coordinate (u) at (2.35,0); \coordinate (O) at (3.65,0);
 \draw[auxiliary] (s) -- (u);
 \draw[scalar flow] (u) .. controls (2.35,.67) and (3.65,.67) .. (O);
 \draw[scalar flow] (O) .. controls (3.65,-.67) and (2.35,-.67) .. (u);
 \foreach \v in {T,s,u,O}{\fill (\v) circle (1.8pt);}
 \node[left] at (T) {$T(q)$};
 \node[right=3pt] at (O) {$(\phi_i\phi_i)(-q)$};
 \node at (1.83,-1.05) {$C_7$};
\end{scope}
\begin{scope}[shift={(14.8,0)}]
 \coordinate (T) at (0,0);
 \coordinate (a) at (.55,.55); \coordinate (b) at (.55,-.55);
 \coordinate (c) at (1.4,.55); \coordinate (d) at (1.4,-.55);
 \coordinate (s) at (1.95,0);
 \draw[scalar flow] (a) -- (T); \draw[scalar flow] (T) -- (b);
 \draw[scalar flow] (b) -- (a);
 \draw[auxiliary] (a) -- (c); \draw[auxiliary] (b) -- (d);
 \draw[scalar flow] (s) -- (c); \draw[scalar flow] (c) -- (d);
 \draw[scalar flow] (d) -- (s);
 \coordinate (u) at (2.7,0); \coordinate (O) at (3.9,0);
 \draw[auxiliary] (s) -- (u);
 \draw[scalar flow] (u) .. controls (2.7,.67) and (3.9,.67) .. (O);
 \draw[scalar flow] (O) .. controls (3.9,-.67) and (2.7,-.67) .. (u);
 \foreach \v in {T,a,b,c,d,s,u,O}{\fill (\v) circle (1.8pt);}
 \node[left] at (T) {$T(q)$};
 \node[right=3pt] at (O) {$(\phi_i\phi_i)(-q)$};
 \node at (1.95,-1.05) {$C_8$};
\end{scope}
\end{tikzpicture}%
}
\caption{Two-point kernels entering the $\TJJ$ contact contributions in the critical $O(N)$ vector model.
The upper row contains the current kernels $C_0,C_1,C_2$, which supply
the second and third lines of eq.~\eqref{eq:on-tjj-momentum-source}.
The middle and lower rows contain mixed kernels of the stress tensor
and the bilinear $\phi_i\phi_i$, entering its last line through
eq.~\eqref{eq:on-tjj-singlet-contact}.
The added kernels $C_6,C_7,C_8$ connect the corresponding matter block
in $C_3,C_4,C_5$ through one additional auxiliary-field propagator to a
free bilinear bubble. In particular, $C_8$ contains three matter loops.
The kernels $C_1,C_3,C_6$ include both self-energy placements; one is drawn.
The mixed kernels obey $C_{i+3}=-C_i$ for $i=3,4,5$, so their nonlocal terms cancel;
see eq.~\eqref{eq:on-tjj-mixed-contact-pairs}.}
\label{fig:tjj-contacts}
\end{figure}
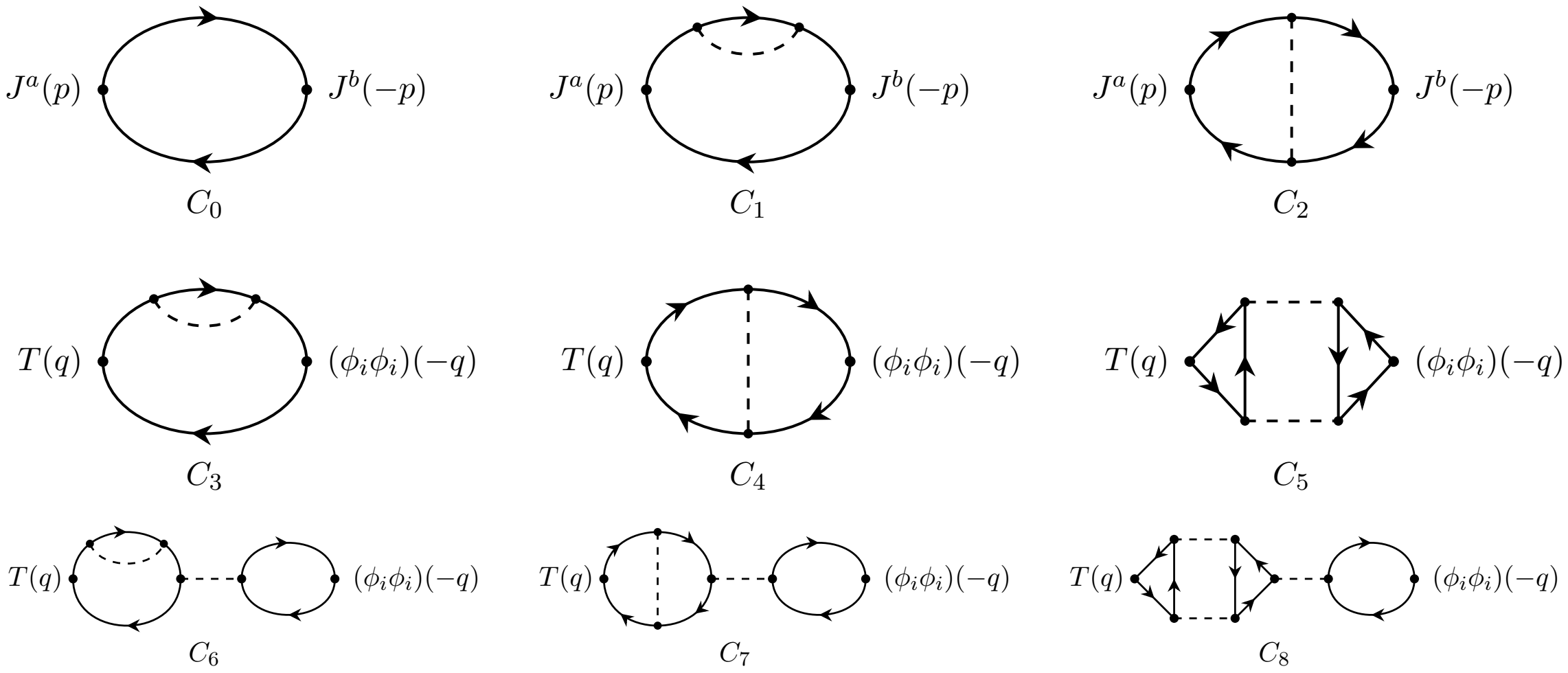

The separate-insertion term on the right of
eq.~\eqref{eq:on-tjj-momentum-source} gives $D_0,\ldots,D_{11}$ in
figure~\ref{fig:tjj}, with their external assignments.
Their individual contributions are listed in appendix~\ref{app:tjj-results}.
The remaining terms are contact contributions, whose two-point kernels
are shown in figure~\ref{fig:tjj-contacts}.

The second and third lines of eq.~\eqref{eq:on-tjj-momentum-source}
attach the current kernels $C_0,C_1,C_2$ in the upper row of
figure~\ref{fig:tjj-contacts} at momenta $p+q$ and $p$.
Their Kronecker tensors are the two mixed metric--gauge vertices.
The last line instead attaches the mixed stress-tensor--scalar kernels
$C_3,\ldots,C_8$ in the middle and lower rows to the second gauge variation.
In $C_6,C_7,C_8$, the subdiagrams containing the
stress-tensor insertion is the same as in $C_3,C_4,C_5$,
respectively.
Each such subdiagram connects to a free bilinear
bubble through one auxiliary-field propagator.
At the order displayed, the diagrams obey
$C_6=-C_3$, $C_7=-C_4$, and $C_8=-C_5$.

The reference tensors $\HJ$, $\SJ$, and $\FJ$ are defined in
appendix~\ref{app:tjj-basis}, eqs.~\eqref{eq:tjj-hard-structure}--\eqref{eq:tjj-structures}.
The selected terms of the free reference correlators are
\begin{equation}
\begin{aligned}
 \TJJ_{\mathrm s}^{\mathrm{hard},(0)}
 &=\TJJ_{\mathrm f}^{\mathrm{hard},(0)}
 =\frac{\tau^{ab}|p|}{32}\HJ,\\
 \bigl(\TJJ_{\mathrm s}^{\mathrm{soft},(3)},
       \TJJ_{\mathrm f}^{\mathrm{soft},(3)}\bigr)
 &=\frac{\tau^{ab}|q|^3}{128|p|^2}\bigl(\SJ,\FJ\bigr).
\end{aligned}
\label{eq:tjj-free-region-coefficients}
\end{equation}
Thus $\HJ$ is shared by the two normalized references, while $\SJ$ and
$\FJ$ distinguish their cubic nonanalytic soft terms.

The leading hard-region contribution sums $D_0,\ldots,D_8$ in
figure~\ref{fig:tjj}; $D_9,D_{10},D_{11}$ vanish at order $q^0$.
Its mixed contacts attach the current kernels $C_0,C_1,C_2$ in
figure~\ref{fig:tjj-contacts} to both metric--gauge vertices.
These attachments use the symmetric identity tensor $I$ in
eq.~\eqref{eq:symmetric-identity} and the traceless projection
$\mathcal P_J$ in appendix~\ref{app:tjj-basis}.
The kernels below are evaluated at $p$, since $p+q=p$ at this order.
The last line before the equality is the mixed vertex from the
stress-tensor counterterm.
With the regulator removed after summation, the leading hard-region contribution is
\begin{equation}
\begin{aligned}
 &\sum_{i=0}^{8}D_i^{(0)}+(Z_T^{\ON}-1)D_0^{(0)}
 \\
 &\quad-2\mathcal P_J\!\left[
 \sum_{i=0}^{2}\left(I_{\rho\sigma,\mu\lambda}C_{i,\lambda\nu}
 +I_{\rho\sigma,\nu\lambda}C_{i,\mu\lambda}\right)\right]
 \\
 &\quad-2(Z_T^{\ON}-1)\mathcal P_J\!\left[
 I_{\rho\sigma,\mu\lambda}C_{0,\lambda\nu}
 +I_{\rho\sigma,\nu\lambda}C_{0,\mu\lambda}\right]
 \\
 &=\frac{\tau^{ab}|p|}{32}\left(1-\frac{64}{9\pi^2N}\right)\HJ
 +O(N^{-2}).
\end{aligned}
\label{eq:on-tjj-hard}
\end{equation}
The cubic nonanalytic soft-region contribution instead includes $D_9,D_{10},D_{11}$.
Its current contacts are analytic near $q=0$.
The mixed contacts $C_3,\ldots,C_8$ depend only on $q$;
their terms proportional to $|q|$ cancel in pairs and do not contribute
to the cubic coefficient.
After the lower nonanalytic powers cancel, we obtain
\begin{equation}
\begin{aligned}
 &\sum_{i=0}^{11}D_i^{\mathrm{soft},(3)}
 +(Z_T^{\ON}-1)D_0^{\mathrm{soft},(3)}
 \\
 &=\frac{\tau^{ab}|q|^3}{128|p|^2}
 \left[\left(1+\frac1N\left(\frac{16}{9\pi^2}-1\right)\right)\SJ
 +\frac1N\left(1-\frac{80}{9\pi^2}\right)\FJ\right]
 +O(N^{-2}).
\end{aligned}
\label{eq:on-tjj-soft}
\end{equation}
These equations give the $q^0$ hard-region coefficient and the cubic
nonanalytic soft-region coefficient of the renormalized connected
$\TJJ$ correlator, including the contact terms generated by source
differentiation.

\subsection{The GNY model}\label{sec:tjj-gn}

The action for the GNY model is given by
\begin{equation}
 S_{\GN}=\int\dd^3x\left[\frac12\bar\psi_i\gamma^\mu\partial_\mu\psi_i
 +\frac12\sigma\bar\psi_i\psi_i+\frac{N}{2g}\sigma^2\right],
\end{equation}
where the $\sigma^2$ coefficient is tuned to criticality.
At large $N$, the scaling dimension of $\sigma$ is $\Delta_\sigma=1+O(1/N)$.
The Feynman rules are
\begin{equation}
\begin{array}{c@{\quad=\quad}l}
\begin{tikzpicture}[baseline=-0.5ex,line width=0.8pt]
  \draw (0,0) -- (2.2,0);
  \node[left] at (0,0) {$\psi_i$};
  \node[right] at (2.2,0) {$\bar\psi_j$};
  \node[above] at (1.1,0) {$p$};
\end{tikzpicture}
& \displaystyle-\frac{\mathrm{i}\,\gamma^\mu p_\mu}{p^2}\,\delta_{ij}\,,
\\[16pt]
\begin{tikzpicture}[baseline=-0.5ex,line width=0.8pt]
  \draw[dashed] (0,0) -- (2.2,0);
  \node[left] at (0,0) {$\sigma$};
  \node[right] at (2.2,0) {$\sigma$};
  \node[above] at (1.1,0) {$p$};
\end{tikzpicture}
& \displaystyle\frac{16}{N|p|}\,,
\\[20pt]
\begin{tikzpicture}[baseline=-0.5ex,line width=0.8pt]
  \draw[dashed] (0,0) -- (1.1,0);
  \draw (2.2,0.65) -- (1.1,0) -- (2.2,-0.65);
  \fill (1.1,0) circle (1.8pt);
  \node[left] at (0,0) {$\sigma$};
  \node[right] at (2.2,0.65) {$\bar\psi_i$};
  \node[right] at (2.2,-0.65) {$\psi_j$};
\end{tikzpicture}
& -\delta_{ij}\,\mathbf1\,.
\end{array}
\label{eq:gn-rules}
\end{equation}
Solid lines denote fermion fields, and dashed lines denote the auxiliary field.
We take $\{\gamma_\mu,\gamma_\nu\}=2\delta_{\mu\nu}\mathbf1$ and
$\Tr\mathbf1=2$. A closed Majorana loop carries $-1/2$. The operators are
\begin{align}
 J_\mu^a&=\frac12\bar\psi_i\gamma_\mu(t^a)_{ij}\psi_j,\\
 T_{\mu\nu}&=\frac18\bar\psi_i
 \left(\gamma_\mu\overleftrightarrow\partial_\nu
 +\gamma_\nu\overleftrightarrow\partial_\mu
 -\frac23\delta_{\mu\nu}\gamma^\rho\overleftrightarrow\partial_\rho\right)\psi_i,
\end{align}
where $\bar\psi\overleftrightarrow\partial_\mu\psi
=\bar\psi\partial_\mu\psi-(\partial_\mu\bar\psi)\psi$.

Fourier transforming these operators gives
\begin{equation}
\begin{aligned}
J_\mu^a(p)
&=\frac12\int\frac{\dd^3k}{(2\pi)^3}
\bar\psi_i(-k)\gamma_\mu(t^a)_{ij}\psi_j(k+p)\,,
\\
T_{\mu\nu}(p)
&=\frac{\mathrm{i}}8\int\frac{\dd^3k}{(2\pi)^3}
\bar\psi_i(-k)\Bigl[\gamma_\mu(2k+p)_\nu+\gamma_\nu(2k+p)_\mu
\\
&\qquad-\frac23\delta_{\mu\nu}\gamma^\rho(2k+p)_\rho\Bigr]\psi_i(k+p)\,.
\end{aligned}
\label{eq:fermion-fourier-insertions}
\end{equation}
The corresponding two-leg insertion rules are
\begin{equation}
\begin{array}{@{}c@{\enspace=\enspace}l@{}}
\begin{tikzpicture}[baseline=-0.5ex,line width=0.8pt]
  \tikzset{momentum/.style={
    postaction={decorate},
    decoration={markings,mark=at position 0.55 with {
      \arrow{Stealth[length=2.4mm,width=1.7mm]}
    }}
  }}
  \draw[momentum] (1.1,0.65) -- node[midway,above] {$k$} (0,0);
  \draw[momentum] (0,0) -- node[midway,below=4pt] {$k+p$} (1.1,-0.65);
  \fill (0,0) circle (2pt);
  \node[left=5pt] at (0,0) {$J_\mu^a(p)$};
  \node[right] at (1.1,0.65) {$\bar\psi_i$};
  \node[right] at (1.1,-0.65) {$\psi_j$};
\end{tikzpicture}
&\gamma_\mu\,(t^a)_{ij}\,,
\\[16pt]
\begin{tikzpicture}[baseline=-0.5ex,line width=0.8pt]
  \tikzset{momentum/.style={
    postaction={decorate},
    decoration={markings,mark=at position 0.55 with {
      \arrow{Stealth[length=2.4mm,width=1.7mm]}
    }}
  }}
  \draw[momentum] (1.1,0.65) -- node[midway,above] {$k$} (0,0);
  \draw[momentum] (0,0) -- node[midway,below=4pt] {$k+p$} (1.1,-0.65);
  \fill (0,0) circle (2pt);
  \node[left=5pt] at (0,0) {$T_{\mu\nu}(p)$};
  \node[right] at (1.1,0.65) {$\bar\psi_i$};
  \node[right] at (1.1,-0.65) {$\psi_j$};
\end{tikzpicture}
&\begin{aligned}
&\frac{\mathrm{i}}4\Bigl[\gamma_\mu(2k+p)_\nu+\gamma_\nu(2k+p)_\mu\\
&\qquad-\frac23\delta_{\mu\nu}\gamma^\rho(2k+p)_\rho\Bigr]\delta_{ij}\,.
\end{aligned}
\end{array}
\label{eq:fermion-diagrammatic-insertions}
\end{equation}
The arrows label momentum flow through the insertion.

\begin{figure}[tbp]
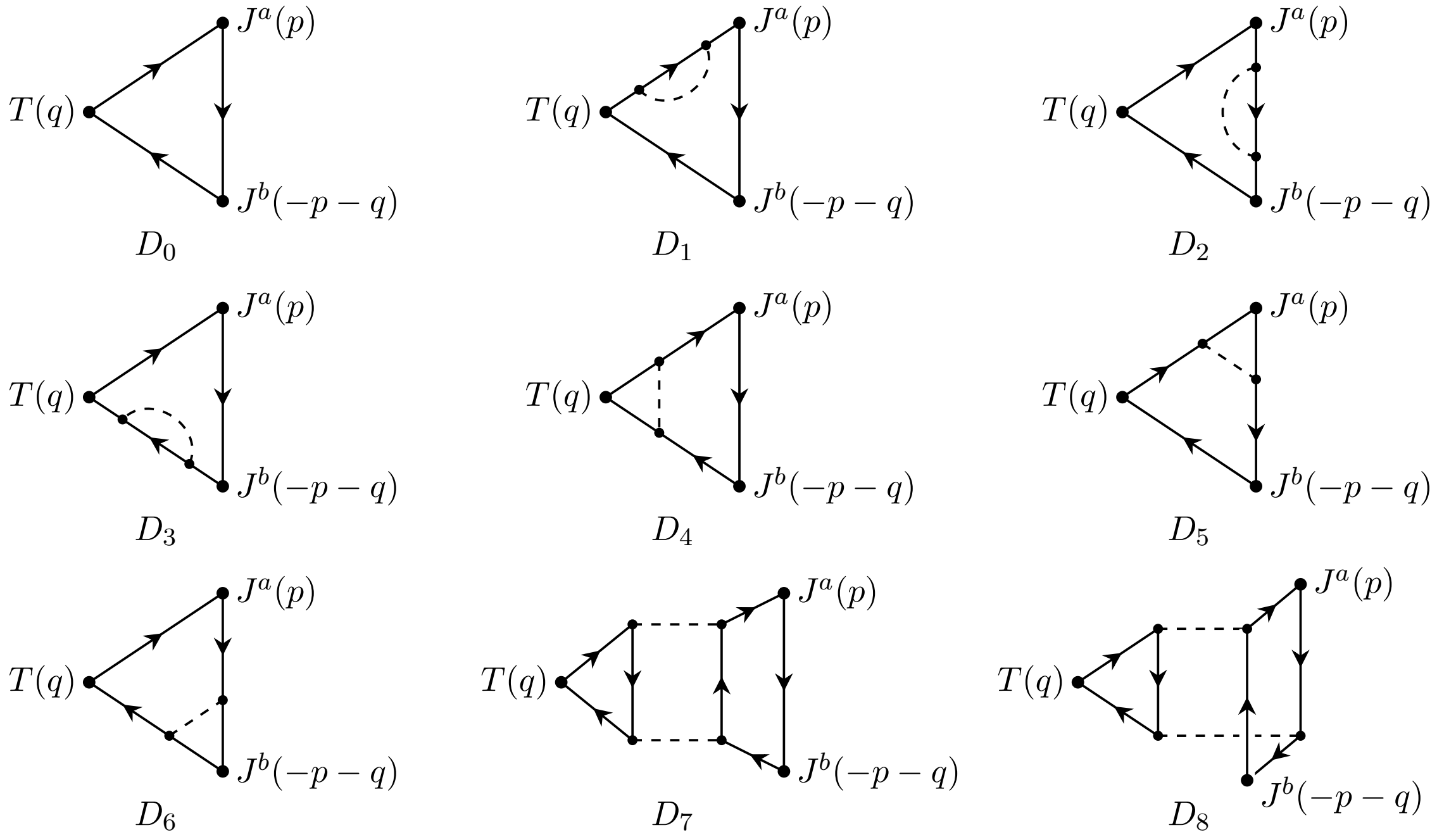

\centering
\GNYThreePointFigure{J^a(p)}{J^b(-p-q)}{0}
\caption{Separate-insertion diagrams for $\TJJ$ in the GNY model through order $1/N$.
Solid lines with arrows denote fermion propagators; dashed lines denote the
auxiliary-field propagator. Arrows specify the orientation used for fermion
contractions. The labels match appendix~\ref{app:gny-tjj-results}, whose
coefficients include the contraction and symmetry factors.
The diagrams $D_9,D_{10},D_{11}$ vanish and are omitted.
The displayed diagrams supply the separate-insertion term in
eq.~\eqref{eq:gny-tjj-momentum-source}.}
\label{fig:gny-tjj}
\end{figure}

\begin{figure}[tbp]
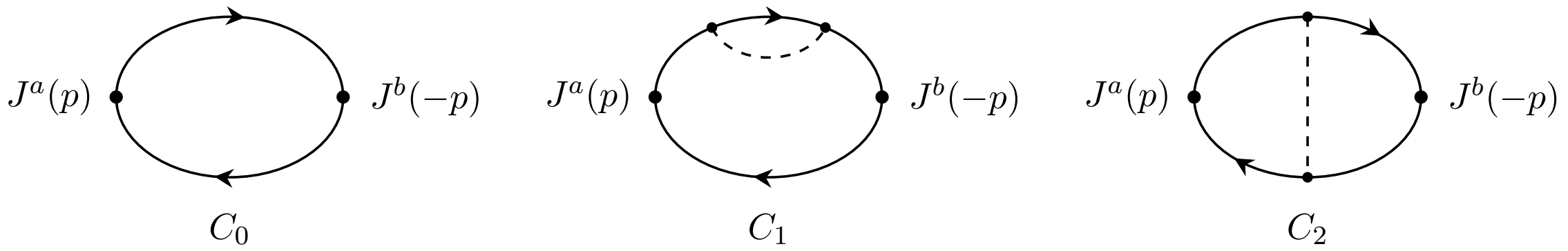

\centering
\resizebox{\textwidth}{!}{%
\begin{tabular}{@{}c@{\hspace{.4cm}}c@{\hspace{.4cm}}c@{}}
\GNYContactPanel{0}{J^a(p)}{J^b(-p)}{0}&
\GNYContactPanel{1}{J^a(p)}{J^b(-p)}{0}&
\GNYContactPanel{2}{J^a(p)}{J^b(-p)}{0}
\end{tabular}%
}
\caption{Current two-point kernels entering the $\TJJ$ contact contributions in the GNY model.
The kernel $C_1$ includes self-energy insertions on either fermion line;
one placement is drawn. These kernels attach to the mixed metric--gauge
vertex in eq.~\eqref{eq:fermion-mixed-contact}.
The two attachments use momenta $p$ and $p+q$, respectively.
These attachments supply the second and third lines of
eq.~\eqref{eq:gny-tjj-momentum-source}.
Their values are given in eq.~\eqref{eq:gny-contact-current-diagrams}.}
\label{fig:gny-tjj-contacts}
\end{figure}

The separate-insertion term in eq.~\eqref{eq:gny-tjj-momentum-source}
gives $D_0,\ldots,D_8$ in figure~\ref{fig:gny-tjj}.
The other two lines attach the current kernels $C_0,C_1,C_2$ in
figure~\ref{fig:gny-tjj-contacts} at momenta $p+q$ and $p$.
Their factors of $1/2$ arise from the inverse vielbein variation in
eq.~\eqref{eq:fermion-vielbein-variation}.
The action is linear in the gauge source, so there is no contact between
the two current insertions.

The diagrams in appendix~\ref{app:gny-tjj-results} determine both tensor
coefficients. Their hard sum includes the current kernels $C_0,C_1,C_2$
in eq.~\eqref{eq:gny-contact-current-diagrams}.
The mixed metric--gauge vertex gives half the corresponding attachment factor in the critical $O(N)$ vector model.
Its counterterm gives the last line before the equality below, as in
eq.~\eqref{eq:gny-tjj-mixed-counterterm}.
With all kernels evaluated at $p$, the leading hard-region contribution is
\begin{equation}
\begin{aligned}
 &\sum_{i=0}^{8}D_i^{(0)}+(Z_T^{\GN}-1)D_0^{(0)}
 \\
 &\quad-\mathcal P_J\!\left[
 \sum_{i=0}^{2}\left(I_{\rho\sigma,\mu\lambda}C_{i,\lambda\nu}
 +I_{\rho\sigma,\nu\lambda}C_{i,\mu\lambda}\right)\right]
 \\
 &\quad-(Z_T^{\GN}-1)\mathcal P_J\!\left[
 I_{\rho\sigma,\mu\lambda}C_{0,\lambda\nu}
 +I_{\rho\sigma,\nu\lambda}C_{0,\mu\lambda}\right]
 \\
 &=\frac{\tau^{ab}|p|}{32}\left(1-\frac{64}{9\pi^2N}\right)\HJ
 +O(N^{-2}).
\end{aligned}
\label{eq:gn-tjj-hard}
\end{equation}
Here $D_9,D_{10},D_{11}$ vanish.
The remaining diagrams give the cubic nonanalytic soft-region contribution
\begin{equation}
\begin{aligned}
 &\sum_{i=0}^{8}D_i^{\mathrm{soft},(3)}
 +(Z_T^{\GN}-1)D_0^{\mathrm{soft},(3)}
 \\
 &=\frac{\tau^{ab}|q|^3}{128|p|^2}
 \left[\frac1N\left(1-\frac{56}{9\pi^2}\right)\SJ
 +\left(1-\frac1N\left(1+\frac8{9\pi^2}\right)\right)\FJ\right]
 +O(N^{-2}),
\end{aligned}
\label{eq:gn-tjj-soft}
\end{equation}
where the $1/\Delta$ pole and logarithms cancel.
The contact diagrams are analytic near $q=0$, so they do not enter the
cubic nonanalytic coefficient.
We thus obtain eq.~\eqref{eq:gn-j}, and eq.~\eqref{eq:shape-def} gives
$\gamma_{\GN}$ in eq.~\eqref{eq:shape-gn}.

\section{\texorpdfstring{$\langle TTT\rangle$}{TTT} in the \texorpdfstring{$O(N)$}{O(N)} vector and GNY models}\label{sec:ttt}

We use the subgraph expansion of section~\ref{sec:subgraphs} with three
metric insertions.
Replacing the two current insertions with stress-tensor insertions only changes the tensor structures. The contributions from the hard and
soft momentum regions are evaluated by the same procedure and
matched to the corresponding conformal structures.

\subsection{Conformal structures, collider bounds, and shear response}\label{sec:ttt-general}

For a parity-invariant three-dimensional CFT,
\begin{equation}
 \TTT=\nsT\TTT_{\mathrm s}+\nfT\TTT_{\mathrm f},\qquad
 \nsT+\nfT=\frac{C_T}{C_T^{\mathrm s}}.
 \label{eq:ttt-decomposition}
\end{equation}
The references are one real conformally coupled scalar and one
two-component Majorana fermion, whose $C_T$ values coincide in $d=3$.
The shape $t_4$ takes values $+4$ and $-4$ in these two theories.
For a state created by a spatial symmetric traceless polarization
$\epsilon_{ij}T_{ij}$,
\begin{equation}
 \frac{\langle\mathcal E(\boldsymbol n)\rangle_{\epsilon\cdot T}}{E/(2\pi)}
 =1+t_4\left(\frac{|\epsilon_{ij}n_in_j|^2}
 {\epsilon_{ij}^*\epsilon_{ij}}-\frac14\right).
\end{equation}
Energy positivity gives
\begin{equation}
 -4\leq t_4\leq4,\qquad \nsT,\nfT\geq0.
 \label{eq:t-bounds}
\end{equation}
The two stress polarizations have normalized flux eigenvalues
$1\pm t_4/4$. The endpoints have a polarization with zero energy in the
chosen detector direction, as realized by the free scalar and fermion
theories. The positive opposite-structure weights generated at the first
subleading order lift this zero and move both models inward at large
$N$. The collider condition supplies a nontrivial sign check on the
calculation, without determining the magnitude of the correction.
It constrains the coefficient of the stress contribution to the shear
OPE and hence its sum-rule weight.
The analogous positivity problem with parity-odd stress structures is
discussed in ref.~\cite{ChowdhuryParityCollider}.
The bootstrap convention is $\gamma_{TTT}=-t_4/48$
\cite{DommesEtAl}; this is distinct from the $\TJJ$ parameter $\gamma$.

The shear response probes the stress OPE. At zero spatial momentum,
\begin{equation}
 \eta(\omega)=-\frac1\omega\operatorname{Im}G^R_{xy,xy}(\omega,0),
 \qquad G^R_{xy,xy}(t,\boldsymbol x)
 =-\ii\theta(t)\langle[T_{xy}(t,\boldsymbol x),T_{xy}(0)]\rangle.
 \label{eq:kubo}
\end{equation}
For this $-\ii$ retarded convention, the positive-frequency spectral density is
\begin{equation}
 \rho_\eta(\omega)=-\operatorname{Im}G^R_{xy,xy}(\omega,0),\qquad
 \omega>0.
\end{equation}
It gives $\eta(\omega)=\rho_\eta(\omega)/\omega$.
For the thermal shear response, we adopt the pressure-contact convention of ref.~\cite{ChowdhuryDavidPrakash}, with $G^E_{xy,xy}(0,0)=P$.\footnote{Relative to the raw covariant-metric second derivative of $W$ in section~\ref{sec:background}, this prescription adds the real local pressure contact $2P$ to the Euclidean shear kernel. The spectral density and dissipative viscosity are unchanged.} The high-frequency expansion then reads \begin{equation}\begin{aligned} G^E_{xy,xy}(\Omega,0)&=\mathcal C_T|\Omega|^3+\sum_{\mathcal O}\widetilde b_{\mathcal O}\langle\mathcal O\rangle_{T_{\rm th}}|\Omega|^{3-\Delta_{\mathcal O}}+\mathcal B_\eta(t_4)P+\cdots,\\ \mathcal C_T&=\frac{\pi^2C_T}{48},\qquad \mathcal B_\eta(t_4)=\frac14-\frac{t_4}{16}. \end{aligned}\label{eq:shear-ope}\end{equation} Here $P$ is the pressure defined below eq.~\eqref{eq:conductivity}. The sum in eq.~\eqref{eq:shear-ope} contains scalar operators only; the pressure term arises from the stress tensor in the $TT$ OPE and is fixed by $\TTT$.\par The same three-point data determine the stress-tensor contribution to the shear spectral sum rule. In the pressure-contact convention specified above, its weight is $P-\mathcal B_\eta(t_4)P=(3/4+t_4/16)P$ \cite{ChowdhuryDavidPrakash}. Scalar operators with nonzero thermal expectation values and dimensions $\Delta_{\mathcal O}\leq3$ require the corresponding OPE contributions and ultraviolet subtractions. Thus our results determine the stress contribution to high-frequency shear response and integrated spectral weight; the full thermal response also requires thermal one-point functions and dynamical information \cite{ChowdhuryDavidPrakash,LucasSierensWitczak}.

\subsection{The critical \texorpdfstring{$O(N)$}{O(N)} vector model}\label{sec:ttt-on}

\begin{figure}[tbp]
\centering\includegraphics[width=\textwidth]{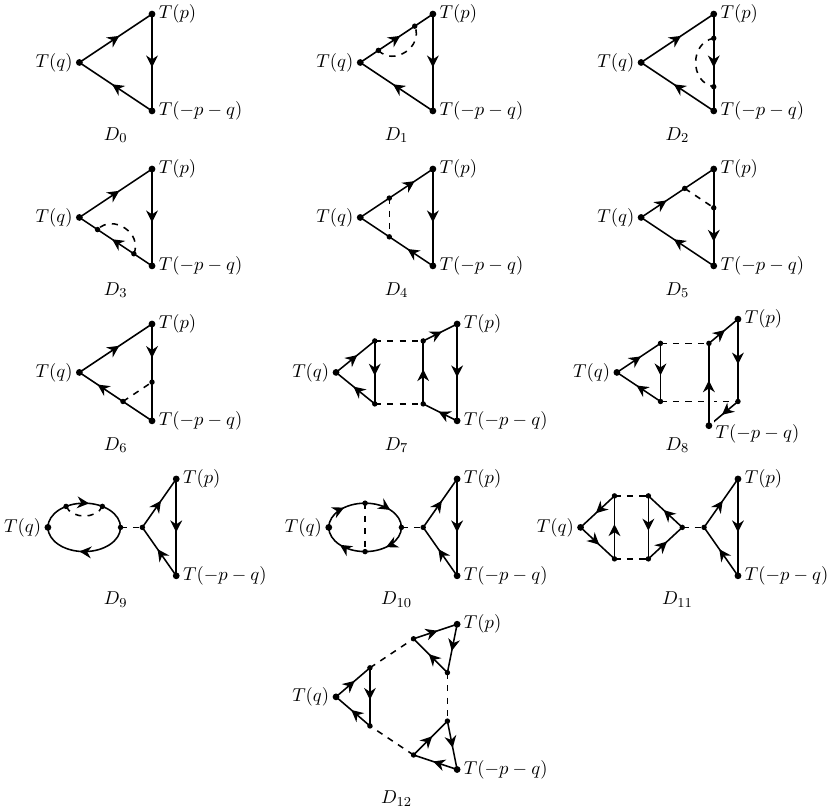}
\caption{Separate-insertion diagrams for $\TTT$ in the critical $O(N)$ vector model through order $N^0$, organized with the soft stress tensor carrying $q$. The full result includes cyclic external
assignments. These diagrams supply the separate-insertion term in
eq.~\eqref{eq:on-ttt-momentum-source}.}\label{fig:ttt}
\end{figure}

The separate-point term in eq.~\eqref{eq:on-ttt-momentum-source}
gives $D_0,\ldots,D_{12}$ in figure~\ref{fig:ttt}.
Each stress tensor is a flavor singlet, so the free triangle is of order
$N$ and the first correction of order $N^0$.
The contact terms are represented in figure~\ref{fig:ttt-contacts}.
The regulated calculation also requires the local source correction in
appendix~\ref{app:counterterms}.

The tensors $L_i$ and combinations $\HT,\ST,\FT$ are defined in
appendix~\ref{app:ttt-basis}. The selected terms of the free reference
correlators are
\begin{equation}
\begin{aligned}
 \TTT_{\mathrm s}^{\mathrm{hard},(0)}
 &=\TTT_{\mathrm f}^{\mathrm{hard},(0)}
 =\frac{|p|^3}{512}\HT,\\
 \bigl(\TTT_{\mathrm s}^{\mathrm{soft},(3)},
       \TTT_{\mathrm f}^{\mathrm{soft},(3)}\bigr)
 &=\frac{|q|^3}{2048}\bigl(\ST,\FT\bigr).
\end{aligned}
\label{eq:ttt-free-region-coefficients}
\end{equation}
The common hard tensor $\HT$ fixes the sum of structure weights, while
$\ST$ and $\FT$ distinguish the scalar and fermion cubic soft terms.
The interacting hard-region contribution requires all separate-point diagrams
and all three metric-source pairs.
For each pair, $C_0$ and $C_1,\ldots,C_8$ are the contacts in
figure~\ref{fig:ttt-contacts}, with the normalization of
appendix~\ref{app:on-ttt-contacts}.
The sum $\sum_{\mathrm{pairs}}$ includes the pairs $(p,-p-q)$,
$(q,-p-q)$, and $(q,p)$, with their stress-tensor indices permuted as well.
These permutations act on the full kernels before the $q^0$ coefficient
is extracted.
The symbol $\delta\TTT_{\mathrm{source}}^{(0)}$ denotes that coefficient
in the sum of the three corresponding versions of the source counterterm
in eq.~\eqref{eq:scalar-source-correction}.
The additional assignments of $D_7,D_8$ are written separately as
$(D_7+D_8)_{\mathrm{cyc}}$
Adding these assignments, the contact terms, and the counterterms gives
\begin{equation}
\begin{aligned}
 &\sum_{i=0}^{12}D_i^{(0)}
 +(D_7+D_8)_{\mathrm{cyc}}^{(0)}
 +3(Z_T^{\ON}-1)D_0^{(0)}
 \\
 &\quad-4\sum_{\mathrm{pairs}}\left[
 C_0+\sum_{i=1}^{8}C_i+2(Z_T^{\ON}-1)C_0\right]^{(0)}
 +\delta\TTT_{\mathrm{source}}^{(0)}
 \\
 &=\frac{N|p|^3}{512}\left(1-\frac{40}{9\pi^2N}\right)\HT+O(N^{-1}).
\end{aligned}
\label{eq:on-ttt-hard}
\end{equation}
The cubic nonanalytic soft sum uses only the contact pair $(p,-p-q)$.
The other two pairs are analytic near $q=0$ and therefore do not enter
this coefficient.
Using eq.~\eqref{eq:scalar-source-correction} for the source counterterm gives
\begin{equation}
\begin{aligned}
 &\sum_{i=0}^{12}D_i^{\mathrm{soft},(3)}
 +(D_7+D_8)_{\mathrm{cyc}}^{\mathrm{soft},(3)}
 +3(Z_T^{\ON}-1)D_0^{\mathrm{soft},(3)}
 \\
 &\quad-4\left[C_0+\sum_{i=1}^{8}C_i
 +2(Z_T^{\ON}-1)C_0\right]^{\mathrm{soft},(3)}
 +\delta\TTT_{\mathrm{source}}^{\mathrm{soft},(3)}
 \\
 &=\frac{|q|^3}{2048}\left[\left(N+\frac{686}{3}-\frac{20360}{9\pi^2}\right)\ST+\left(\frac{20320}{9\pi^2}-\frac{686}{3}\right)\FT\right]+O(N^{-1}),
\end{aligned}
\label{eq:on-ttt-soft}
\end{equation}
where the lower nonanalytic powers cancel.
Matching the cubic coefficients gives eq.~\eqref{eq:on-t}.
The individual contributions are listed in appendix~\ref{app:ttt-results}.

The diagrams in figure~\ref{fig:ttt-contacts} give the first contact
term in eq.~\eqref{eq:on-ttt-momentum-source}, with the second metric
variation given explicitly in eq.~\eqref{eq:on-explicit-metric-vertex}.
The other two contact terms follow by permuting the momenta and
stress-tensor indices together. 

\begin{figure}[t]
\centering
\begingroup
\tikzset{
  ttt contact/.style={line width=0.75pt,scale=0.78,transform shape},
  tc scalar/.style={postaction={decorate},
    decoration={markings,mark=at position #1 with {
      \arrow{Stealth[length=2mm,width=2mm]}
    }}},
  tc scalar/.default=0.55,
  tc sigma/.style={dashed,line width=0.75pt},
  tc source/.style={rectangle,fill=black,inner sep=2pt}
}
\newcommand{\TTContactLabels}{%
  \fill (L) circle (1.8pt);
  \node[left=3pt] at (L) {$T(q)$};
  \node[tc source] at (R) {};
  \node[right=4pt] at (R) {$\mathcal V_{\ON}$};
}
\newcommand{\TTContactBubble}[1][0.55]{%
  \coordinate (L) at (0,0);
  \coordinate (R) at (2.4,0);
  \draw[tc scalar=#1] (L) arc[start angle=180,end angle=0,x radius=1.2,y radius=0.85];
  \draw[tc scalar=#1] (R) arc[start angle=0,end angle=-180,x radius=1.2,y radius=0.85];
  \TTContactLabels
}
\newcommand{\TTContactBridge}{%
  \coordinate (L) at (0,0);
  \coordinate (a) at (1.2,0);
  \coordinate (b) at (1.8,0);
  \coordinate (R) at (3,0);
  \draw[tc sigma] (a) -- (b);
  \draw[tc scalar] (b) arc[start angle=180,end angle=0,x radius=0.6,y radius=0.55];
  \draw[tc scalar] (R) arc[start angle=0,end angle=-180,x radius=0.6,y radius=0.55];
  \foreach \w in {a,b} {\fill (\w) circle (1.6pt);}
  \TTContactLabels
}
\resizebox{\textwidth}{!}{%
\begin{tikzpicture}[ttt contact]
\begin{scope}[shift={(0,0)}]
  \TTContactBubble
  \node at (1.2,-1.4) {$C_0$};
\end{scope}
\begin{scope}[shift={(6.2,0)}]
  \TTContactBubble
  \coordinate (u) at (0.6,{0.85*sqrt(3)/2});
  \coordinate (v) at (1.8,{0.85*sqrt(3)/2});
  \draw[tc sigma] (u) to[bend right=65] (v);
  \foreach \w in {u,v} {\fill (\w) circle (1.6pt);}
  \node at (1.2,-1.4) {$C_1$};
\end{scope}
\begin{scope}[shift={(12.4,0)}]
  \TTContactBubble[0.75]
  \coordinate (u) at (1.2,0.85);
  \coordinate (v) at (1.2,-0.85);
  \draw[tc sigma] (u) -- (v);
  \foreach \w in {u,v} {\fill (\w) circle (1.6pt);}
  \node at (1.2,-1.4) {$C_2$};
\end{scope}
\begin{scope}[shift={(0,-3.5)}]
  \coordinate (L) at (0,0);
  \coordinate (R) at (3,0);
  \coordinate (u) at (0.8,0.7);
  \coordinate (v) at (0.8,-0.7);
  \coordinate (s) at (2.2,0.7);
  \coordinate (t) at (2.2,-0.7);
  \draw[tc scalar] (L) -- (u);
  \draw[tc scalar] (u) -- (v);
  \draw[tc scalar] (v) -- (L);
  \draw[tc scalar] (R) -- (t);
  \draw[tc scalar] (t) -- (s);
  \draw[tc scalar] (s) -- (R);
  \draw[tc sigma] (u) -- (s);
  \draw[tc sigma] (v) -- (t);
  \foreach \w in {u,v,s,t} {\fill (\w) circle (1.6pt);}
  \TTContactLabels
  \node at (1.5,-1.4) {$C_3$};
\end{scope}
\begin{scope}[shift={(6.2,-3.5)}]
  \TTContactBridge
  \draw[tc scalar] (L) arc[start angle=180,end angle=0,x radius=0.6,y radius=0.55];
  \draw[tc scalar] (a) arc[start angle=0,end angle=-180,x radius=0.6,y radius=0.55];
  \coordinate (u) at (0.3,{0.55*sqrt(3)/2});
  \coordinate (v) at (0.9,{0.55*sqrt(3)/2});
  \draw[tc sigma] (u) to[bend right=65] (v);
  \foreach \w in {u,v} {\fill (\w) circle (1.6pt);}
  \node at (1.5,-1.4) {$C_4$};
\end{scope}
\begin{scope}[shift={(12.4,-3.5)}]
  \TTContactBridge
  \draw[tc scalar=0.75] (L) arc[start angle=180,end angle=0,x radius=0.6,y radius=0.55];
  \draw[tc scalar=0.75] (a) arc[start angle=0,end angle=-180,x radius=0.6,y radius=0.55];
  \coordinate (u) at (0.6,0.55);
  \coordinate (v) at (0.6,-0.55);
  \draw[tc sigma] (u) -- (v);
  \foreach \w in {u,v} {\fill (\w) circle (1.6pt);}
  \node at (1.5,-1.4) {$C_5$};
\end{scope}
\begin{scope}[shift={(0,-7)}]
  \coordinate (L) at (-0.3,0);
  \coordinate (u) at (0.3,0.65);
  \coordinate (v) at (0.3,-0.65);
  \coordinate (s) at (1.1,0.65);
  \coordinate (t) at (1.1,-0.65);
  \coordinate (a) at (1.7,0);
  \coordinate (b) at (2.2,0);
  \coordinate (R) at (3.4,0);
  \draw[tc scalar] (L) -- (u);
  \draw[tc scalar] (u) -- (v);
  \draw[tc scalar] (v) -- (L);
  \draw[tc scalar] (a) -- (t);
  \draw[tc scalar] (t) -- (s);
  \draw[tc scalar] (s) -- (a);
  \draw[tc sigma] (u) -- (s);
  \draw[tc sigma] (v) -- (t);
  \draw[tc sigma] (a) -- (b);
  \draw[tc scalar] (b) arc[start angle=180,end angle=0,x radius=0.6,y radius=0.55];
  \draw[tc scalar] (R) arc[start angle=0,end angle=-180,x radius=0.6,y radius=0.55];
  \foreach \w in {u,v,s,t,a,b} {\fill (\w) circle (1.6pt);}
  \TTContactLabels
  \node at (1.5,-1.4) {$C_6$};
\end{scope}
\begin{scope}[shift={(6.2,-7)}]
  \TTContactBubble[0.75]
  \coordinate (u) at (1.2,0.85);
  \draw[tc sigma] (u) to[bend right=45] (R);
  \fill (u) circle (1.6pt);
  \node at (1.2,-1.4) {$C_7$};
\end{scope}
\begin{scope}[shift={(12.4,-7)}]
  \coordinate (L) at (0,0);
  \coordinate (u) at (0.8,0.65);
  \coordinate (v) at (0.8,-0.65);
  \coordinate (R) at (2.5,0.65);
  \coordinate (t) at (2.5,-0.65);
  \draw[tc scalar] (L) -- (u);
  \draw[tc scalar] (u) -- (v);
  \draw[tc scalar] (v) -- (L);
  \draw[tc sigma] (u) -- (R);
  \draw[tc sigma] (v) -- (t);
  \draw[tc scalar] (R) arc[start angle=90,end angle=-90,x radius=0.5,y radius=0.65];
  \draw[tc scalar] (t) arc[start angle=-90,end angle=-270,x radius=0.5,y radius=0.65];
  \foreach \w in {u,v,t} {\fill (\w) circle (1.6pt);}
  \fill (L) circle (1.8pt);
  \node[left=3pt] at (L) {$T(q)$};
  \node[tc source] at (R) {};
  \node[right=8pt,yshift=2pt] at (R) {$\mathcal V_{\ON}$};
  \node at (1.5,-1.4) {$C_8$};
\end{scope}
\end{tikzpicture}%
}
\endgroup
\caption{Metric-contact diagrams contributing to $\TTT$ in the critical $O(N)$ vector model. The vertex $\mathcal V_{\ON}$ is the
second metric derivative of the action; each source pair has a factor
$-4$. The leading contribution is of order $N$ and the interacting
correction is of order $N^0$.
The displayed assignment supplies the first contact line of
eq.~\eqref{eq:on-ttt-momentum-source}; the other two lines follow by
permuting the source momenta and indices.}\label{fig:ttt-contacts}
\end{figure}
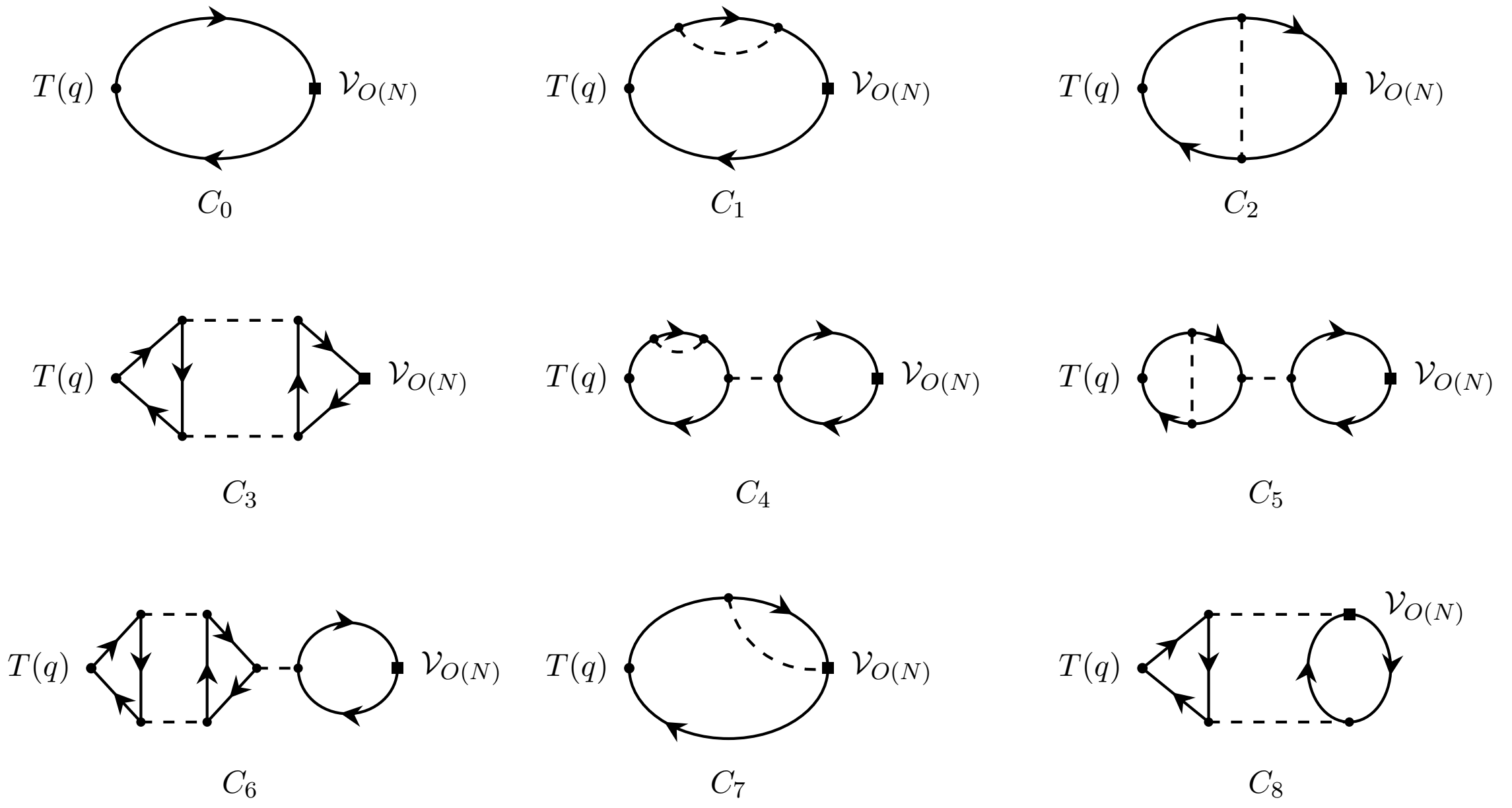

\subsection{The GNY model}\label{sec:ttt-gn}

\begin{figure}[tbp]
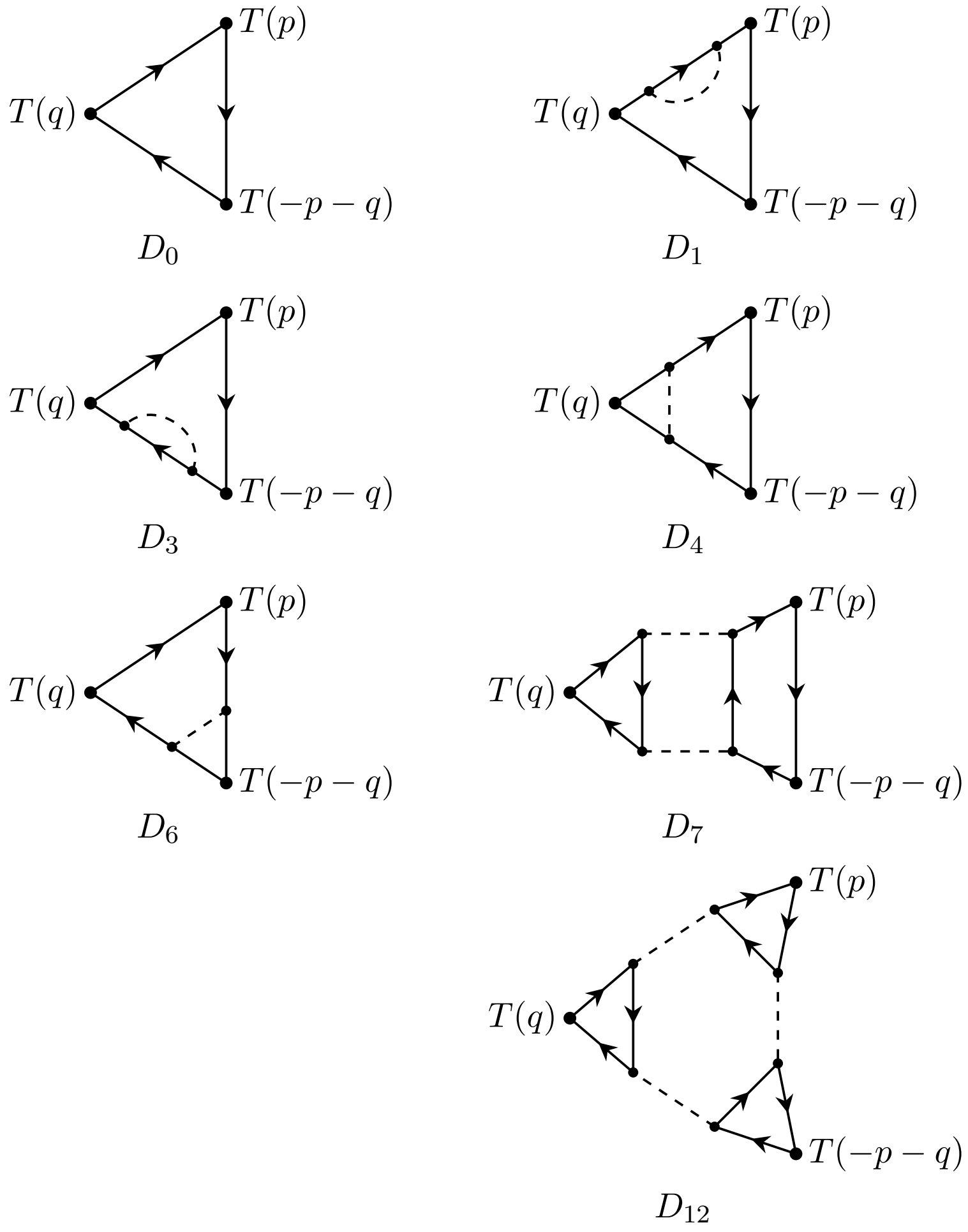

\centering
\GNYThreePointFigure{T(p)}{T(-p-q)}{1}
\caption{Separate-point diagrams for $\TTT$ in the GNY model through order $N^0$.
Solid lines with arrows denote fermion propagators; dashed lines denote the
auxiliary-field propagator. The labels match appendix~\ref{app:gny-ttt-results}.
The full sum includes the two additional cyclic assignments of $D_7,D_8$.
The diagram $D_{12}$ has three fermion triangles connected by three
auxiliary-field propagators.
These diagrams supply the separate-insertion term in
eq.~\eqref{eq:gny-ttt-momentum-source}.}
\label{fig:gny-ttt}
\end{figure}

\begin{figure}[tbp]
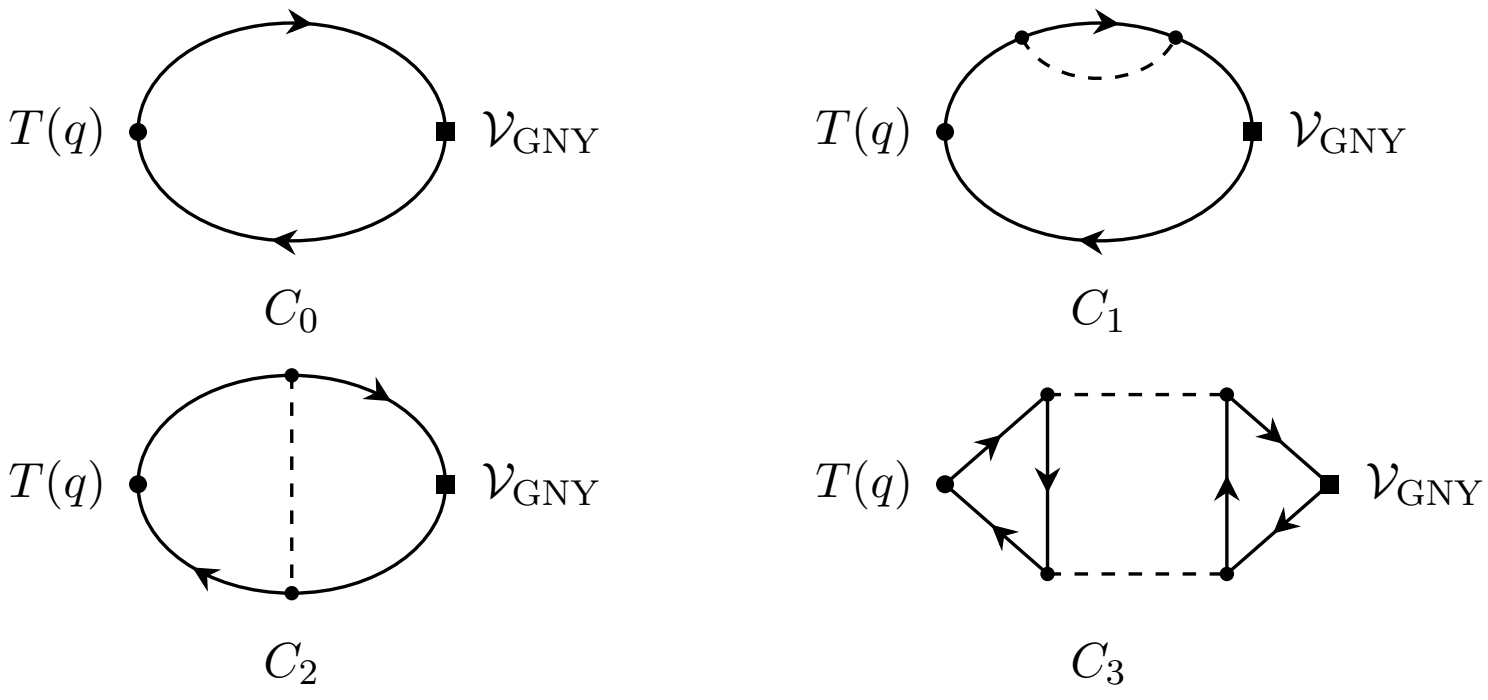

\centering
\resizebox{.65\textwidth}{!}{%
\begin{tabular}{@{}c@{\hspace{1cm}}c@{}}
\GNYContactPanel{0}{T(q)}{\mathcal V_{\GN}}{1}&
\GNYContactPanel{1}{T(q)}{\mathcal V_{\GN}}{1}\\[3pt]
\GNYContactPanel{2}{T(q)}{\mathcal V_{\GN}}{1}&
\GNYContactPanel{3}{T(q)}{\mathcal V_{\GN}}{1}
\end{tabular}%
}
\caption{Metric-contact diagrams for $\TTT$ in the GNY model through order $N^0$.
Other source pairs follow by cyclic permutation, with a factor $-4$
per pair; counterterms are added separately.}
\label{fig:gny-ttt-contacts}
\end{figure}

The fermion theory uses the propagators and insertions of
section~\ref{sec:tjj-gn}.
The separate-point term in eq.~\eqref{eq:gny-ttt-momentum-source}
gives $D_0,\ldots,D_8,D_{12}$ in figure~\ref{fig:gny-ttt}.
Their values are listed in appendix~\ref{app:gny-ttt-results}.
They include the additional cyclic assignments of $D_7,D_8$.
The metric contact contributions are given by the kernels $C_0,C_1,C_2,C_3$
shown in figure~\ref{fig:gny-ttt-contacts} and evaluated in
appendix~\ref{app:gny-full-contacts}.
When the second metric-variation vertex carries momenta $(p,-p-q)$, these kernels give
the first contact term in eq.~\eqref{eq:gny-ttt-momentum-source}.
The other two contact terms follow by permuting the momenta and
stress-tensor indices together.



The hard-region contribution includes all three metric-source pairs, with the same
permutation and expansion convention as in section~\ref{sec:ttt-on}.
Its source counterterm is obtained in eq.~\eqref{eq:fermion-source-correction}.
The full hard sum is
\begin{equation}
\begin{aligned}
 &\sum_{i=0}^{8}D_i^{(0)}+D_{12}^{(0)}
 +(D_7+D_8)_{\mathrm{cyc}}^{(0)}
 +3(Z_T^{\GN}-1)D_0^{(0)}
 \\
 &\quad-4\sum_{\mathrm{pairs}}\left[
 C_0+C_1+C_2+C_3+2(Z_T^{\GN}-1)C_0\right]^{(0)}
 +\delta\TTT_{\mathrm{source}}^{(0)}
 \\
 &=\frac{N|p|^3}{512}\left(1+\frac8{9\pi^2N}\right)\HT+O(N^{-1}).
\end{aligned}
\label{eq:gn-ttt-hard}
\end{equation}
The cubic nonanalytic soft coefficient is
\begin{equation}
\begin{aligned}
 &\sum_{i=0}^{8}D_i^{\mathrm{soft},(3)}+D_{12}^{\mathrm{soft},(3)}
 +(D_7+D_8)_{\mathrm{cyc}}^{\mathrm{soft},(3)}
 +3(Z_T^{\GN}-1)D_0^{\mathrm{soft},(3)}
 \\
 &\quad-4\left[C_0+C_1+C_2+C_3
 +2(Z_T^{\GN}-1)C_0\right]^{\mathrm{soft},(3)}
 +\delta\TTT_{\mathrm{source}}^{\mathrm{soft},(3)}
 \\
 &=\frac{|q|^3}{2048}\left[\left(\frac{20344}{9\pi^2}-\frac{686}{3}\right)\ST+\left(N+\frac{686}{3}-\frac{20336}{9\pi^2}\right)\FT\right]+O(N^{-1}).
\end{aligned}
\label{eq:gn-ttt-soft}
\end{equation}
All regulator poles and logarithms cancel.
The lower nonanalytic powers also vanish.
The resulting $t_4$ is given in eq.~\eqref{eq:shape-gn}.

\section{Squashed-sphere computation of the three-point data}\label{sec:sphere}
\input{squashed-sphere}

\section{Discussion}\label{sec:discussion}

We have determined the first subleading large $N$ contributions to the
two parity-even structures of $\TJJ$ and $\TTT$ in the critical $O(N)$
vector and GNY models. The subgraph expansion isolates the momentum
coefficients needed for conformal matching, including the source
contacts. The resulting sums reproduce $C_J$ and $C_T$, and the shape
corrections move $\gamma$ and $t_4$ inward from their free-field collider
endpoints. The squashed-sphere method provides an efficient independent determination of both the normalizations and shapes, in agreement with the diagrammatic results. The small-$N$ extrapolations give
encouraging agreement with the Ising and $O(2)$ bootstrap data. These
results supply three-point information relevant to quantum critical
conductivity and shear response beyond the central charges alone.

A natural next step is to compute the corresponding $1/N$ corrections
in conformal gauge theories, in particular the QED$_3$, QED$_3$-GNY, and scalar QED$_3$. The two-point analyses in
refs.~\cite{HuhStrackSachdevCurrents,HuhStrackCentralCharges,GiombiTarnopolskyKlebanov,GuoLi} provide a starting point.
These gauge theories are important candidate descriptions of deconfined
quantum critical points and Dirac spin liquids
\cite{WangDQCP,HermeleSpinLiquid}. Their global flavor and topological
currents give several distinct $\TJJ$ observables. Together with $\TTT$
and thermal one-point functions, these would determine stress-mediated
high-frequency charge or spin response and shear response in strongly
coupled condensed matter systems. Matching the stress tensor and the
appropriate currents across dual descriptions would also test
three-dimensional dualities at the level of three-point data
\cite{SeibergDualityWeb,WangDQCP}.

Recent developments have made the role of gauge fluctuations in thermal transport particularly concrete. Guo \textit{et al.} \cite{GuoThermalHall} computed the leading matter and next-to-leading gauge contributions to the thermal Hall response, including a Dirac Chern--Simons description of a transition involving semion topological order. Maity \textit{et al.} \cite{MaityThermalHall} extended this analysis to the finite-temperature thermal Hall response of an abelian chiral spin liquid. These calculations concern finite-temperature energy-current response, rather than a complete determination of the vacuum $\TTT$ structures. The two types of data are complementary: vacuum OPE coefficients constrain short-distance thermal response, whereas the full transport coefficients also require thermal dynamics and, for thermal Hall transport, magnetization contributions.\par The approach can also be extended to gauge theories with Chern--Simons
couplings. The planar correlators of these theories already give
nontrivial tests of three-dimensional bosonization
\cite{AharonyBosonization}. Subleading corrections would probe the
duality beyond the planar limit. In that setting one must include the
parity-odd structures and their mixing in the energy-flux matrix:
the allowed parity-even and parity-odd coefficients satisfy joint
collider constraints \cite{ChowdhuryParityCollider}. Keeping the
background gauge and gravitational contact terms under control is
also necessary when comparing duality maps \cite{ClossetContactTerms}.

The bootstrap comparison suggests that useful three-point estimates
can survive extrapolation to small $N$. For the $O(2)$ vector model, the
predicted pair of collider parameters lies within the joint allowed region
obtained with the weaker spectral assumptions of ref.~\cite{DommesEtAl}.
Under the stronger assumptions used in table~\ref{tab:bootstrap},
$\gamma$ remains within its quoted interval, whereas $t_4$ lies slightly
below its lower bound; both predictions differ from the interval midpoints
by only a few percent. The Ising comparison is less precise.
This motivates analogous comparisons for candidate DQCP theories and
QED$_3$ with $N_f=4$ two-component Dirac fermions. Gauge-theory
bootstrap studies have made progress in identifying compatible regions
of conformal data \cite{LiDQCPBootstrap,LiPolandGauge,AlbayrakQEDBootstrap}, and the
$SO(5)$ bootstrap has provided evidence for a deconfined tricritical
scenario \cite{ChesterSuDQCP}, with further evidence for multicriticality
from the bootstrap cone analysis of ref.~\cite{LiShenDQCP}. Large $N$ predictions for the current
and stress-tensor OPE coefficients could provide valuable input and
consistency tests for bootstrap systems with spinning external
operators. The truncation errors should remain explicit when using
such predictions as assumptions: small-$N$ agreement for the present
models is encouraging evidence, rather than a general error estimate.

Finally, the relative economy of the squashed-sphere calculation has a
simple structural explanation. Once conformal symmetry restricts each
correlator to two parity-even coefficients, one needs only a
normalization and one independent projection of the tensor structure.
The sphere derivatives furnish precisely these projections, following
the background-field framework of Parmentier and Bobev
\cite{ParmentierBobev}.
The shorter computation reorganizes the perturbation theory and could provide an efficient route to $\TJJ$ and $\TTT$ in other large $N$
CFTs whenever the corresponding source-dependent fluctuation kernels
can be constructed. For gauge theories the determinant must include
coupled fluctuations, gauge fixing, ghosts, and zero modes, while the
chosen background profiles must still distinguish the independent
conformal structures. The diagrammatic method remains useful for
arbitrary tensor projections and for relating the sphere integrals to
momentum-space correlators.

\acknowledgments   
This research was supported by the Startup Funding
4007022314 of the Southeast University, the Southeast University Global Engagement of
Excellence Fund No. 3360682401E, and the National Natural Science Foundation of China
funding No. 12375061.
We acknowledge the assistance of ChatGPT Codex in the computations and the preparation of the manuscript draft.

\appendix
\input{appendices}
\bibliographystyle{JHEP} \bibliography{references}
\end{document}

%% file: subgraph-method.tex
\subsection{Method of subgraphs and conformal matching}\label{sec:subgraphs}

The diagrammatic calculation follows the application of the subgraph
method to conformal three-point functions developed in
ref.~\cite{LiSubgraphs}. The useful limit is $|q|\ll|p|$ for external
momenta $(q,p,-p-q)$. Instead of determining a multiloop function at
arbitrary momentum ratios, we compute a few coefficients of its
asymptotic expansion. The finite number of conformal structures then
turns these coefficients into complete separated-point three-point data.
We apply this construction to both $\TJJ$ and $\TTT$. Related developments
in the graphical characterization of expansion regions, including their
relation to Landau equations and Newton polytopes, are presented in
refs.~\cite{GardiRegions,MaFacetRegions}.

For Euclidean integrals, the expansion can be organized graphically
\cite{SmirnovAsymptotics,SmirnovBook}. In schematic notation,
\begin{equation}
 \mathcal I_G(p,q)\underset{|q|/|p|\to0}{\sim}
 \sum_{h\in\mathfrak H(G)}
 \mathcal I_{G/h}(q)\circ
 \mathcal T_{q,k_{\rm s}}\mathcal I_h(p,q,k_{\rm s}).
 \label{eq:subgraph-master}
\end{equation}
An allowed hard subgraph $h$ contains the hard external vertices and is
one-particle irreducible after those vertices are identified. The Taylor
operator expands its integrand in the soft external momentum and the
soft momenta $k_{\rm s}$ entering from the contracted graph $G/h$.
The symbol $\circ$ inserts the resulting momentum polynomial as a local
vertex in that graph. Each expanded integral is evaluated over its full
integration domain with a common regulator. Scaleless terms vanish;
intermediate ultraviolet and infrared poles are combined before the
regulator is removed. This graph expansion accounts for soft regions
that a uniform Taylor expansion of the original integrand would miss.

The distinction is already visible in a massless triangle:
\begin{equation}
 I(p,q)=\intk\frac1{k^2(k-q)^2(k+p)^2}.
 \label{eq:subgraph-triangle}
\end{equation}
For $k$ hard, Taylor expansion of $(k-q)^{-2}$ gives terms analytic in
$q$. For $k$ soft, the hard propagator instead expands as
\begin{equation}
 \frac1{(p+k)^2}=\frac1{p^2}-\frac{2p\cdot k}{p^4}+\cdots.
\end{equation}
The first soft term is
\begin{equation}
 I_{\rm soft}^{\rm leading}(p,q)
 =\frac1{p^2}\intk\frac1{k^2(k-q)^2}
 =\frac1{8p^2|q|}.
 \label{eq:subgraph-triangle-soft}
\end{equation}
It is nonanalytic at $q=0$ and cannot arise from the hard Taylor series.
For our interacting diagrams the same separation reduces multiscale
integrals to single-scale tensor integrals and their insertions. The
auxiliary propagator powers remain analytically regulated throughout.

The conformal interpretation is a momentum-space OPE. The two hard
operators fuse into a local operator, whose correlator with the soft
insertion carries the nonanalytic soft dependence
\cite{LiSubgraphs,BMSMomentum}. The stress contribution involves
$\langle T(q)T(-q)\rangle\propto C_T|q|^3$. Accordingly the cubic
soft terms have overall scales $|q|^3/|p|^2$ in $\TJJ$ and $|q|^3$ in
$\TTT$. Their tensor coefficients distinguish the two parity-even
structures, while the leading hard term checks their Ward-identity sum.
This is why the required three-point constants can be extracted without
summing the momentum expansion to all orders.

%% file: squashed-sphere.tex
\subsection{Free energy and conformal collider parameters}\label{sec:sphere-sources}

We now compute the three-point coefficients independently using the squashed sphere framework of Parmentier and Bobev
\cite{ParmentierBobev}.
We will extract the required CFT data from free-energy derivatives.
This provides an efficient alternative to the flat-space diagrammatic calculation. 
Consider the unit-radius Hopf metric and background gauge one-form
\begin{align}
 \dd s^2&=\frac14\left[\dd\theta^2+\sin^2\theta\,\dd\varphi^2
 +(1+\varepsilon)(\dd\psi+\cos\theta\,\dd\varphi)^2\right],
 \label{eq:hopf-metric}\\
 A&=\frac\lambda2(\dd\psi+\cos\theta\,\dd\varphi).
 \label{eq:hopf-gauge}
\end{align}
The ranges are $0\leq\theta\leq\pi$, $0\leq\varphi<2\pi$, and
$0\leq\psi<4\pi$, with the usual sphere identifications.
We take the parity-even free energy
\begin{equation}
 F(\varepsilon,\lambda)=-\log|Z[g(\varepsilon),A(\lambda)]|
 =N F_0(\varepsilon,\lambda)+F_1(\varepsilon,\lambda)+O(N^{-1}),
 \label{eq:sphere-expansion}
\end{equation}
The metric responses \cite{BobevBuenoVreys,BuenoEtAl,BuenoEtAl2020,ParmentierBobev}
and the mixed response are\footnote{The theory-independent factors in these relations can be fixed by considering $N$ free real scalars or $N$ free Majorana fermions, where the conformal data are known.}
\begin{equation}
 \begin{gathered}
 \left.\partial_\varepsilon^2 F\right|_0=-\frac{\pi^4 C_T}{3},
 \qquad
 \left.\partial_\varepsilon^3 F\right|_0=\frac{\pi^4 C_Tt_4}{630},\\[4pt]
 \left.\partial_\lambda^2F\right|_0=2\pi^4 N C_J,
 \qquad
 \left.\partial_\varepsilon\partial_\lambda^2F\right|_0
 =\frac{16\pi^4 N C_J\gamma}{5},
 \end{gathered}
 \label{eq:sphere-stress-response}
\end{equation}
where all derivatives are evaluated at $(\varepsilon,\lambda)=(0,0)$.
Therefore, the conformal collider parameters are given by
\begin{equation}
 t_4=-210\left.\frac{\partial_\varepsilon^3F}
 {\partial_\varepsilon^2F}\right|_0,
 \qquad
 \gamma=\frac58\left.\frac{\partial_\varepsilon\partial_\lambda^2F}
 {\partial_\lambda^2F}\right|_0.
 \label{eq:sphere-extraction}
\end{equation}
At $N=\infty$, the free energy per real scalar or Majorana fermion reads
\begin{align}
 F_0^{\ON}-F_0^{\ON}(0,0)
 &=-\frac{\pi^2}{64}\varepsilon^2+\frac{\pi^2}{10080}\varepsilon^3
 +\frac{\pi^2}{16}\lambda^2-\frac{\pi^2}{120}\varepsilon\lambda^2+\cdots,
 \label{eq:sphere-leading-on}\\
 F_0^{\GN}-F_0^{\GN}(0,0)
 &=-\frac{\pi^2}{64}\varepsilon^2-\frac{\pi^2}{10080}\varepsilon^3
 +\frac{\pi^2}{16}\lambda^2+\frac{\pi^2}{120}\varepsilon\lambda^2+\cdots.
 \label{eq:sphere-leading-gn}
\end{align}
This leads to
\begin{equation}
\begin{aligned}
 t_{4,\ON}&=4, &\qquad \gamma_{\ON}&=-\frac1{12},\\
 t_{4,\GN}&=-4, &\qquad \gamma_{\GN}&=\frac1{12},
\end{aligned}
\end{equation}
as expected.
For the $1/N$ correction, define
\begin{equation}
 F_1-F_1(0,0)=f_2\varepsilon^2+f_3\varepsilon^3
 +a_{02}\lambda^2+a_{12}\varepsilon\lambda^2
 +O(\varepsilon^4,\varepsilon^2\lambda^2,\lambda^4).
 \label{eq:sphere-f1-coefficients}
\end{equation}
In what follows, we determine these four coefficients from the
matter spectrum and the singlet fluctuation kernel.

\subsection{The critical \texorpdfstring{$O(N)$}{O(N)} saddle and fluctuation kernel}\label{sec:sphere-on}

For scalar harmonics $Y_{nma}$ on the round sphere, take
\begin{equation}
 n=1,2,\ldots,\qquad m=-\frac{n-1}{2},\ldots,\frac{n-1}{2},
 \qquad a=1,\ldots,n.
 \label{eq:sphere-harmonics}
\end{equation}
They satisfy $-\nabla_0^2Y=(n^2-1)Y$ and $-\ii\partial_\psi Y=mY$.
The degeneracy at each $m$ is $n$. The charged scalar operator at the
homogeneous saddle and its eigenvalues are
\begin{align}
 L_A&=-D^2+\frac{\mathcal{R}}{8}+\sigma_*(\varepsilon,\lambda),
 \qquad \mathcal{R}=6-2\varepsilon,\\
 \Lambda_{nm}&=n^2-\frac14-\frac{4m^2\varepsilon}{1+\varepsilon}
 -\frac\varepsilon4+\frac{4m\lambda+\lambda^2}{1+\varepsilon}
 +\sigma_*(\varepsilon,\lambda).
 \label{eq:sphere-scalar-spectrum}
\end{align}
The leading free energy per real scalar takes the form
\begin{equation}
 F_0^{\ON}(\varepsilon,\lambda)
 =\frac12\log\det L_A
 =\frac12\sum_{n,m} n\log\Lambda_{nm}.
 \label{eq:sphere-scalar-free-energy}
\end{equation}
Stationarity with respect to the homogeneous auxiliary field gives the gap equation
$\sum_{n,m}n/\Lambda_{nm}=0$, which should be understood as a regularized equation.
The solution is
\begin{equation}
 \sigma_*(\varepsilon,\lambda)
 =\frac{\varepsilon^2}{10}
 +\frac{202-21\pi^2}{84}\varepsilon^3
 -\frac23\lambda^2
 +\frac{-418+45\pi^2}{45}\varepsilon\lambda^2
 +O(\varepsilon^4,\varepsilon^2\lambda^2,\lambda^4).
 \label{eq:sphere-scalar-saddle}
\end{equation}
Substituting this saddle into eq.~\eqref{eq:sphere-scalar-free-energy}
yields the leading free-energy expansion in eq.~\eqref{eq:sphere-leading-on}.

We expand $\sigma$ around the homogeneous saddle as
\begin{equation}
 \sigma(x)=\sigma_*(\varepsilon,\lambda)
 +\ii\sqrt{\frac{2}{N}}\,\eta(x).
 \label{eq:sphere-scalar-fluctuation}
\end{equation}
To compute the leading $1/N$ correction, we keep only the quadratic effective action for $\eta$:
\begin{equation}
 S_{\mathrm{eff}}^{(2)}[\eta]
 =\frac12\int\dd V_g(x)\dd V_g(y)\,
 \eta(x)G(x,y)G(y,x)\eta(y).
\label{eq:sphere-scalar-quadratic-action}
\end{equation}
The squashed-sphere volume element is $\dd V_g=\rho\,\dd V_0$,
where $\rho=\sqrt{1+\varepsilon}$ and $\dd V_0$ is the round-sphere
volume element.
Using harmonics normalized with respect to $\dd V_0$,
we write the scalar Green function as
\begin{equation}
 G(x,y)=\rho^{-1}\overline G(x,y),\qquad
 \overline G(x,y)=\sum_{n,m,a}
 \frac{Y_{nma}(x)Y_{nma}^*(y)}{\Lambda_{nm}}.
\end{equation}
The kernel in \eqref{eq:sphere-scalar-quadratic-action} is a pointwise product,
\begin{equation}
 B_{\ON}(x,y)=G(x,y)G(y,x).
 \label{eq:sphere-scalar-bubble}
\end{equation}

To expand $\overline G$ in $\varepsilon$ and $\lambda$, introduce
the commuting operators
$R_r=(-\nabla_0^2+3/4)^{-r}$, $Q=-4\partial_\psi^2+1/4$, and $U=-4\ii\partial_\psi$.
The coefficients $g_{ij}(x,y)$ of $\varepsilon^i\lambda^j$ in $\overline G(x,y)$ are
\begin{align}
 g_{10}&=QR_2,\qquad g_{01}=-UR_2,\\
 g_{02}&=U^2R_3-\frac13R_2,\qquad g_{11}=UR_2-2UQR_3,\\
 g_{12}&=\left(\frac{463}{45}-\pi^2\right)R_2-\left(\frac23Q+2U^2\right)R_3
 +3U^2QR_4,\\
 g_{20}&=Q^2R_3-\left(Q-\frac3{20}\right)R_2,\notag\\
 g_{30}&=Q^3R_4-2Q\left(Q-\frac3{20}\right)R_3
 +\left(Q+\frac{21\pi^2-223}{84}\right)R_2.
 \label{eq:sphere-scalar-inverse}
\end{align}
Exchanging $x$ and $y$ in $\overline G(x,y)$ changes the sign of
terms with odd powers of $\lambda$.
Multiplying $\overline G(x,y)$ by $\overline G(y,x)$ therefore gives
the following expansion, whose right-hand side contains pointwise products
of kernels evaluated at $(x,y)$:
\begin{equation}
 \begin{aligned}
 \overline B_{\ON}
 &=G_0^2+2\varepsilon G_0g_{10}\\
 &\quad+\varepsilon^2\bigl(2G_0g_{20}+g_{10}^2\bigr)
 +\varepsilon^3\bigl(2G_0g_{30}+2g_{10}g_{20}\bigr)\\
 &\quad+\lambda^2\bigl(2G_0g_{02}-g_{01}^2\bigr)
 +\varepsilon\lambda^2\bigl(2G_0g_{12}+2g_{10}g_{02}-2g_{01}g_{11}\bigr)\\
 &\quad+O(\varepsilon^4,\varepsilon^2\lambda^2,\lambda^4),
 \end{aligned}
 \label{eq:sphere-scalar-mixed-bubble}
\end{equation}
and the correction to the free energy is given by $F_1^{\ON}=\frac12\log\det (\rho^{-1}\overline B_{\ON})$.
Here $G_0(x,y)=R_1(x,y)$ denotes the Green function on the
round sphere at zero gauge source.


\subsection{The GNY saddle and spinor kernel}\label{sec:sphere-gn}

The parity-even fermion effective action is even in the singlet mass.
We use its massless symmetric saddle, $\sigma_*=0$, near the round
background. At this saddle, the Hermitian Dirac operator in a
left-invariant orthonormal frame is
\begin{equation}
 \begin{aligned}
 H_A={}&-2\ii(\tau_2\cos\psi-\tau_1\sin\psi)\partial_\theta
 -\frac{2\ii(\tau_1\cos\psi+\tau_2\sin\psi)}{\sin\theta}
 (\partial_\varphi-\cos\theta\,\partial_\psi)\\
 &-\frac{2\ii}{\rho}\tau_3\partial_\psi
 +\left(\frac\rho2+\frac1\rho\right)\mathbf1
 +\frac\lambda\rho\tau_3.
 \end{aligned}
 \label{eq:sphere-dirac-operator}
\end{equation}
Here $\tau_a$ are Pauli matrices.
The coordinate derivatives act on the spinor components in this frame.
The leading free energy per Majorana fermion is
$F_0^{\GN}=-\tfrac12\log\det|H_A|$.

The inverse operator $P_A=H_A^{-1}$ has a kernel $P_A(x,y)$ with
respect to $\dd V_g$.
Using harmonics normalized with respect to $\dd V_0$, we write this kernel as
\begin{equation}
 P_A(x,y)=\rho^{-1}\overline P_A(x,y).
\end{equation}
The quadratic fluctuation kernel is
\begin{equation}
 B_{\GN}(x,y)=-\operatorname{tr}[P_A(x,y)P_A(y,x)]
 =\rho^{-2}\overline B_{\GN}(x,y).
 \label{eq:sphere-fermion-bubble}
\end{equation}
The trace acts on the two spinor components.
The minus sign follows from the relation between the fermion propagator
and the Hermitian inverse, $(\gamma^\mu D_\mu)^{-1}=-\ii P_A$.

To expand $\overline P_A$ in $\varepsilon$ and $\lambda$, write
$\overline P_A=\sum_{i,j\geq0}\varepsilon^i\lambda^jP_{ij}$.
At zero source, the corresponding operators are
\begin{equation}
 \begin{aligned}
 H_0={}&-2\ii(\tau_2\cos\psi-\tau_1\sin\psi)\partial_\theta
 -\frac{2\ii(\tau_1\cos\psi+\tau_2\sin\psi)}{\sin\theta}
 (\partial_\varphi-\cos\theta\,\partial_\psi)
 -2\ii\tau_3\partial_\psi+\frac32,\\
 P_{00}={}&H_0^{-1}=\frac{H_0-1}{-\nabla_0^2+3/4}.
 \end{aligned}
\end{equation}
For pure metric variations,
$H_A|_{\lambda=0}=H_0+\varepsilon V_1+\varepsilon^2V_2+\varepsilon^3V_3
+O(\varepsilon^4)$, where
\begin{equation}
 V_1=\ii\tau_3\partial_\psi-\frac14,\quad
 V_2=-\frac{3\ii}{4}\tau_3\partial_\psi+\frac5{16},\quad
 V_3=\frac{5\ii}{8}\tau_3\partial_\psi-\frac9{32}.
\end{equation}
Expanding the inverse gives the operator recursion
\begin{equation}
 P_{k0}=-P_{00}\sum_{a=1}^kV_aP_{k-a,0},\qquad k=1,2,3.
 \label{eq:sphere-fermion-metric-inverse}
\end{equation}

To obtain the gauge and mixed coefficients, let $K_A=H_A-\rho/2$.
On the spin-$j$ representation with $j=(n-1)/2$, this operator satisfies
\begin{equation}
 K_A^2=n^2-\frac{\varepsilon}{1+\varepsilon}(-2\ii\partial_\psi+\tau_3)^2
 +\frac{2\lambda(-2\ii\partial_\psi+\tau_3)+\lambda^2}{1+\varepsilon}.
 \label{eq:sphere-spin-identity}
\end{equation}
This identity gives
$H_A^{-1}=(K_A-\rho/2)(K_A^2-\rho^2/4)^{-1}$.
For the component with $\tau_3$ eigenvalue $v=\pm1$, set
$Q_v=-2\ii\partial_\psi+v$ and $W_v=Q_v^2+1/4$.
The coefficients $\mathcal{C}_{ij}^{(v)}$ of $\varepsilon^i\lambda^j$
in $(K_A^2-\rho^2/4)^{-1}$ on this component are
\begin{align}
 \mathcal{C}_{00}^{(v)}&=R_1,\qquad
 \mathcal{C}_{10}^{(v)}=W_vR_2,\\
 \mathcal{C}_{01}^{(v)}&=-2Q_vR_2,\qquad
 \mathcal{C}_{02}^{(v)}=4Q_v^2R_3-R_2,\\
 \mathcal{C}_{11}^{(v)}&=2Q_vR_2-4Q_vW_vR_3,\\
 \mathcal{C}_{12}^{(v)}&=R_2-(2W_v+8Q_v^2)R_3+12W_vQ_v^2R_4.
 \label{eq:sphere-fermion-resolvents}
\end{align}
Here $R_r$ are the same operators used in section~\ref{sec:sphere-on}.
Multiplying this expansion on the left by $K_A-\rho/2$ and collecting
powers of $\varepsilon$ and $\lambda$ gives $P_{01},P_{02},P_{11},P_{12}$.

Substituting these coefficients into the product of the two inverse kernels
gives the bubble expansion below.
In each product $P_{ij}P_{kl}$, the first kernel is evaluated at $(x,y)$
and the second at $(y,x)$.
Charge conjugation cancels the terms with odd powers of $\lambda$, so
\begin{equation}
 \begin{aligned}
 \overline B_{\GN}(x,y)
 =-\operatorname{tr}\Bigl[&P_{00}P_{00}
 +\varepsilon\bigl(P_{10}P_{00}+P_{00}P_{10}\bigr)\\
 &+\varepsilon^2\bigl(P_{20}P_{00}+P_{10}P_{10}+P_{00}P_{20}\bigr)\\
 &+\varepsilon^3\bigl(P_{30}P_{00}+P_{20}P_{10}+P_{10}P_{20}+P_{00}P_{30}\bigr)\\
 &+\lambda^2\bigl(P_{02}P_{00}+P_{01}P_{01}+P_{00}P_{02}\bigr)\\
 &+\varepsilon\lambda^2\bigl(P_{12}P_{00}+P_{11}P_{01}+P_{10}P_{02}\\
 &\hspace{54pt}+P_{02}P_{10}+P_{01}P_{11}+P_{00}P_{12}\bigr)\Bigr]\\
 &+O(\varepsilon^4,\varepsilon^2\lambda^2,\lambda^4).
 \end{aligned}
 \label{eq:sphere-fermion-mixed-bubble}
\end{equation}
The corresponding free-energy correction is
$F_1^{\GN}=\frac12\log\det\bigl(\rho^{-1}\overline B_{\GN}\bigr)$,
where the determinant is evaluated using the round-sphere measure.

\subsection{Harmonic projection and regularization}\label{sec:sphere-sums}

Project either round-measure bubble onto the scalar harmonics:
\begin{equation}
 \overline b_{nm}=\int\dd V_0(x)\dd V_0(y)\,
 Y_{nma}^*(x)\overline B(x,y)Y_{nma}(y).
 \label{eq:sphere-bubble-projection}
\end{equation}
Homogeneity removes dependence on $a$ and leaves degeneracy $n$.
The physical eigenvalues are $b_{nm}=\rho^{-1}\overline b_{nm}$, with
\begin{equation}
 b_{nm}^{\ON}(0,0)=\frac1{8n},\qquad b_{nm}^{\GN}(0,0)=\frac n8.
\end{equation}
Write the relative expansion as
\begin{equation}
 \frac{\overline b_{nm}}{b_{nm}(0,0)}
 =1+r_1\varepsilon+r_2\varepsilon^2+r_3\varepsilon^3
 +r_{02}\lambda^2+r_{12}\varepsilon\lambda^2+\cdots.
 \label{eq:sphere-relative-eigenvalues}
\end{equation}
The coefficients $r_1,r_2,r_3,r_{02},r_{12}$ are obtained by applying
the projection in eq.~\eqref{eq:sphere-bubble-projection} to the corresponding Taylor coefficients of the bubble kernel and dividing by $b_{nm}(0,0)$.

To regulate the free energy, use
\begin{align}
 \mathcal Z_{\ON}(s)&=\sum_{n,m}n[b_{nm}^{\ON}]^s,
 &F_1^{\ON}&=\frac12\partial_s\mathcal Z_{\ON}(0),\\
 \mathcal Z_{\GN}(s)&=\sum_{n,m}n[b_{nm}^{\GN}]^{-s},
 &F_1^{\GN}&=-\frac12\partial_s\mathcal Z_{\GN}(0).
 \label{eq:sphere-zeta}
\end{align}
Both series initially converge for $\operatorname{Re}s>3$ and are
continued to zero.
The direct summands are
\begin{align}
 q_{2,n}&=\frac n2\sum_m\left(r_2-\frac12r_1^2\right),&
 q_{3,n}&=\frac n2\sum_m\left(r_3-r_1r_2+\frac13r_1^3\right),\\
 q_{02,n}&=\frac n2\sum_m r_{02},&
 q_{12,n}&=\frac n2\sum_m(r_{12}-r_1r_{02}).
 \label{eq:sphere-q-definitions}
\end{align}
With zeta regularization, the convergent expressions are
\begin{align}
 f_2^{\ON}&=\sum_{n=1}^{\infty}
 \left(q_{2,n}^{\ON}+\frac{11n^2}{60}-\frac1{20}\right)
 -\frac1{40}+\frac1{15},\\
 f_3^{\ON}&=\sum_{n=1}^{\infty}
 \left(q_{3,n}^{\ON}-\frac{9n^2}{70}+\frac1{70}\right)
 +\frac1{140}+\frac{59}{1575},\\
 f_2^{\GN}&=\sum_{n=1}^{\infty}
 \left(q_{2,n}^{\GN}+\frac{n^2}{15}+\frac1{20}\right)
 +\frac1{40}-\frac1{15},\\
 f_3^{\GN}&=\sum_{n=1}^{\infty}
 \left(q_{3,n}^{\GN}-\frac{4n^2}{105}-\frac1{70}\right)
 -\frac1{140}-\frac{61}{1575},
 \label{eq:sphere-metric-sums}\\
 a_{02}^{\ON}&=\sum_{n=1}^{\infty}q_{02,n}^{\ON},\qquad
 a_{12}^{\ON}=\sum_{n=1}^{\infty}q_{12,n}^{\ON}+\frac4{225},\\
 a_{02}^{\GN}&=\sum_{n=1}^{\infty}q_{02,n}^{\GN},\qquad
 a_{12}^{\GN}=\sum_{n=1}^{\infty}q_{12,n}^{\GN}+\frac4{225}.
 \label{eq:sphere-mixed-sums}
\end{align}

\subsection{Numerical results and comparison with the diagrams}\label{sec:sphere-matching}

The numerical coefficients of the singlet determinant in
 eq.~\eqref{eq:sphere-f1-coefficients} are
\begin{equation}
 \begin{aligned}
 f_2^{\ON}&\simeq 0.06944444444444444444,
 &f_2^{\GN}&\simeq -0.01388888888888888890,\\
 f_3^{\ON}&\simeq -0.00062509132623677169,
 &f_3^{\GN}&\simeq 0.00062509132623677163,\\
 a_{02}^{\ON}&\simeq -0.444444444444444441,
 &a_{02}^{\GN}&\simeq -0.44444444444444444445,\\
 a_{12}^{\ON}&\simeq 0.075604517795933740,
 &a_{12}^{\GN}&\simeq -0.12004896224037819917.
 \end{aligned}
 \label{eq:sphere-numerical-coefficients}
\end{equation}
The parameter $t_4$ in eq.~\eqref{eq:sphere-extraction} is given by
\begin{align}
 t_{4,\ON}&=4+\frac{40320f_3^{\ON}+256f_2^{\ON}}{\pi^2N}
 +O(N^{-2}),
 \label{eq:sphere-t4-matching-on}\\
 t_{4,\GN}&=-4+\frac{40320f_3^{\GN}-256f_2^{\GN}}{\pi^2N}
 +O(N^{-2}).
 \label{eq:sphere-t4-matching}
\end{align}
Likewise the parameter $\gamma$ is
\begin{align}
 \gamma_{\ON}&=-\frac1{12}+\frac{10}{\pi^2N}
 \left(a_{12}^{\ON}+\frac2{15}a_{02}^{\ON}\right)+O(N^{-2}),
 \label{eq:sphere-gamma-matching-on}\\
 \gamma_{\GN}&=\frac1{12}+\frac{10}{\pi^2N}
 \left(a_{12}^{\GN}-\frac2{15}a_{02}^{\GN}\right)+O(N^{-2}).
 \label{eq:sphere-gamma-matching}
\end{align}
%
Inserting the values in eq.~\eqref{eq:sphere-numerical-coefficients} into
eqs. \eqref{eq:sphere-t4-matching-on}, \eqref{eq:sphere-t4-matching}, \eqref{eq:sphere-gamma-matching-on}, and \eqref{eq:sphere-gamma-matching} gives
\begin{align}
 t_{4,\ON}&=4-\frac{0.75240143315868\ldots}{N}+O(N^{-2}),\\
 t_{4,\GN}&=-4+\frac{2.91392001752855\ldots}{N}+O(N^{-2}),\\
 \gamma_{\ON}&=-\frac1{12}+\frac{0.0165612094188\ldots}{N}+O(N^{-2}),\\
 \gamma_{\GN}&=\frac1{12}-\frac{0.0615928465931\ldots}{N}+O(N^{-2}).
 \label{eq:sphere-final-comparison}
\end{align}
They agree with the exact expressions in
eqs.~\eqref{eq:shape-on} and~\eqref{eq:shape-gn}.
The squashed-sphere calculation determines the physical three-point data without separately evaluating flat-space Feynman diagrams.
Its agreement with the diagrammatic calculation provides a consistency check.

%% file: appendices.tex
\section{Momentum-space tensor basis for \texorpdfstring{$\TJJ$}{TJJ}}\label{app:tjj-basis}

Set $U_\mu=q_\mu/|q|$, $V_\mu=p_\mu/|p|$, and $W=U\cdot V$.
The soft stress pair is projected with
\begin{equation}
 P_{\alpha\beta}^{\rho\sigma}
 =\frac12(\delta_\alpha^\rho\delta_\beta^\sigma
 +\delta_\alpha^\sigma\delta_\beta^\rho)
 -\frac13\delta_{\alpha\beta}\delta^{\rho\sigma}.
 \label{eq:traceless-projector}
\end{equation}
To keep the basis compact, let
$\mathcal P_J[X_{\rho\sigma\mu\nu}]
=P_{\alpha\beta}^{\rho\sigma}X_{\rho\sigma\mu\nu}$. Then
\begin{align}
 K_1&=\mathcal P_J[U_\rho U_\sigma\delta_{\mu\nu}],&
 K_2&=\mathcal P_J[U_\rho U_\sigma U_\mu U_\nu],\\
 K_3&=\mathcal P_J[U_\rho U_\sigma U_\mu V_\nu],&
 K_4&=\mathcal P_J[U_\rho U_\sigma V_\mu U_\nu],\\
 K_5&=\mathcal P_J[U_\rho U_\sigma V_\mu V_\nu],&
 K_6&=\mathcal P_J[(U_\rho V_\sigma+V_\rho U_\sigma)\delta_{\mu\nu}],\\
 K_7&=\mathcal P_J[(U_\rho V_\sigma+V_\rho U_\sigma)U_\mu V_\nu],&
 K_8&=\mathcal P_J[(U_\rho V_\sigma+V_\rho U_\sigma)V_\mu U_\nu],\\
 K_9&=\mathcal P_J[(U_\rho V_\sigma+V_\rho U_\sigma)V_\mu V_\nu],&
 K_{10}&=\mathcal P_J[V_\rho V_\sigma\delta_{\mu\nu}],\\
 K_{11}&=\mathcal P_J[V_\rho V_\sigma U_\mu V_\nu],&
 K_{12}&=\mathcal P_J[V_\rho V_\sigma V_\mu U_\nu],\\
 K_{13}&=\mathcal P_J[V_\rho V_\sigma V_\mu V_\nu],&
 K_{14}&=\mathcal P_J[(\delta_{\rho\mu}U_\sigma+\delta_{\sigma\mu}U_\rho)U_\nu],\\
 K_{15}&=\mathcal P_J[(\delta_{\rho\mu}U_\sigma+\delta_{\sigma\mu}U_\rho)V_\nu],&
 K_{16}&=\mathcal P_J[(\delta_{\rho\mu}V_\sigma+\delta_{\sigma\mu}V_\rho)U_\nu],\\
 K_{17}&=\mathcal P_J[(\delta_{\rho\mu}V_\sigma+\delta_{\sigma\mu}V_\rho)V_\nu],&
 K_{18}&=\mathcal P_J[(\delta_{\rho\nu}U_\sigma+\delta_{\sigma\nu}U_\rho)U_\mu],\\
 K_{19}&=\mathcal P_J[(\delta_{\rho\nu}U_\sigma+\delta_{\sigma\nu}U_\rho)V_\mu],&
 K_{20}&=\mathcal P_J[(\delta_{\rho\nu}V_\sigma+\delta_{\sigma\nu}V_\rho)U_\mu],\\
 K_{21}&=\mathcal P_J[(\delta_{\rho\nu}V_\sigma+\delta_{\sigma\nu}V_\rho)V_\mu],&
 K_{22}&=\mathcal P_J[\delta_{\rho\mu}\delta_{\sigma\nu}
 +\delta_{\rho\nu}\delta_{\sigma\mu}].
 \label{eq:k-basis}
\end{align}
Under $\mu\leftrightarrow\nu$, the pairs
$(K_3,K_4)$, $(K_7,K_8)$, $(K_{11},K_{12})$, $(K_{14},K_{18})$,
$(K_{15},K_{19})$, $(K_{16},K_{20})$, and $(K_{17},K_{21})$ exchange.
The remaining tensors are invariant.

The structures in section~\ref{sec:tjj} are
\begin{align}
 \HJ={}&K_{10}+K_{13}-K_{17}-K_{21}+K_{22},\label{eq:tjj-hard-structure}\\
 \SJ={}&K_1+K_2+K_7+K_8-K_{14}-K_{17}-K_{18}-K_{21}+K_{22}
 \nonumber\\
 &-W(K_3+K_4-K_{15}-K_{19})+(W^2-1)K_5-2WK_9+2K_{13},\\
 \FJ={}&\SJ+(1+W^2)(K_1-K_5)-2W(K_6-K_9)+2(K_{10}-K_{13}).
 \label{eq:tjj-structures}
\end{align}

\section{Momentum-space tensor basis for \texorpdfstring{$\TTT$}{TTT}}\label{app:ttt-basis}

For three stress pairs let
\begin{equation}
 \mathcal P[X_{abcdef}]
 =P_{\alpha\beta}^{ab}P_{\mu\nu}^{cd}P_{\rho\sigma}^{ef}X_{abcdef}.
\end{equation}
Every entry below includes this common projection:
\begin{align}
 L_1&=\mathcal P[V_cV_dV_eV_fV_aV_b],&L_2&=\mathcal P[V_dV_fV_aV_b\delta_{ce}],\\
 L_3&=\mathcal P[V_dV_eV_fV_b\delta_{ca}],&L_4&=\mathcal P[V_aV_b\delta_{cf}\delta_{de}],\\
 L_5&=\mathcal P[V_fV_b\delta_{ca}\delta_{de}],&L_6&=\mathcal P[V_eV_f\delta_{cb}\delta_{da}],\\
 L_7&=\mathcal P[V_cV_dV_fV_b\delta_{ea}],&L_8&=\mathcal P[V_dV_b\delta_{cf}\delta_{ea}],\\
 L_9&=\mathcal P[V_dV_f\delta_{cb}\delta_{ea}],&L_{10}&=\mathcal P[\delta_{cb}\delta_{df}\delta_{ea}],\\
 L_{11}&=\mathcal P[V_cV_d\delta_{eb}\delta_{fa}],&L_{12}&=\mathcal P[V_cV_dV_eV_fV_bU_a],\\
 L_{13}&=\mathcal P[V_dV_eV_fV_bU_cU_a],&L_{14}&=\mathcal P[V_cV_dV_fV_bU_eU_a],\\
 L_{15}&=\mathcal P[V_cV_dV_eV_fU_aU_b],&L_{16}&=\mathcal P[V_dV_eV_fU_cU_aU_b],\\
 L_{17}&=\mathcal P[V_eV_fU_cU_dU_aU_b],&L_{18}&=\mathcal P[V_cV_dV_fU_eU_aU_b],\\
 L_{19}&=\mathcal P[V_dV_fU_cU_eU_aU_b],&L_{20}&=\mathcal P[V_cV_dU_eU_fU_aU_b],\\
 L_{21}&=\mathcal P[V_dV_fV_bU_a\delta_{ce}],&L_{22}&=\mathcal P[V_fV_bU_dU_a\delta_{ce}],\\
 L_{23}&=\mathcal P[V_dV_bU_fU_a\delta_{ce}],&L_{24}&=\mathcal P[V_dV_fU_aU_b\delta_{ce}],\\
 L_{25}&=\mathcal P[V_fU_dU_aU_b\delta_{ce}],&L_{26}&=\mathcal P[V_dU_fU_aU_b\delta_{ce}],\\
 L_{27}&=\mathcal P[U_dU_fU_aU_b\delta_{ce}],&L_{28}&=\mathcal P[V_dV_eV_fU_b\delta_{ca}],\\
 L_{29}&=\mathcal P[V_eV_fU_dU_b\delta_{ca}],&L_{30}&=\mathcal P[V_dV_fU_eU_b\delta_{ca}],\\
 L_{31}&=\mathcal P[V_bU_a\delta_{cf}\delta_{de}],&L_{32}&=\mathcal P[U_aU_b\delta_{cf}\delta_{de}],\\
 L_{33}&=\mathcal P[V_fU_b\delta_{ca}\delta_{de}],&L_{34}&=\mathcal P[U_fU_b\delta_{ca}\delta_{de}],\\
 L_{35}&=\mathcal P[V_cV_dV_fU_b\delta_{ea}],&L_{36}&=\mathcal P[V_dV_fU_cU_b\delta_{ea}],\\
 L_{37}&=\mathcal P[V_cV_dU_fU_b\delta_{ea}],&L_{38}&=\mathcal P[V_dU_b\delta_{cf}\delta_{ea}],\\
 L_{39}&=\mathcal P[U_dU_b\delta_{cf}\delta_{ea}].&&
 \label{eq:l-basis}
\end{align}
The structures in section~\ref{sec:ttt} are
\begin{align}
 \HT={}&L_1+4L_2-4L_3-6L_4+8L_5-2L_6-4L_7+8L_8+8L_9-8L_{10}-2L_{11},\label{eq:ttt-hard-structure}\\
 \ST={}&2L_1-8L_3+4L_6-8L_7+32L_9-32L_{10}+4L_{11}
 \nonumber\\
 &-4WL_{12}+8L_{13}+8L_{14}+(W^2-7)L_{15}-4WL_{16}+2L_{17}
 \nonumber\\
 &-4WL_{18}+16L_{19}+2L_{20}+16L_{24}-16L_{27}+8WL_{28}-8L_{29}
 \nonumber\\
 &-32L_{30}-16L_{32}+32L_{34}+8WL_{35}-32L_{36}-8L_{37}+32L_{39},\\
 \FT={}&4\big[2L_2-2L_3-2L_5+2L_6-2L_7-2L_8+6L_9-6L_{10}
 +2L_{11}+2L_{13}+2L_{14}
 \nonumber\\
 &-2L_{15}-WL_{16}+L_{17}-WL_{18}+3L_{19}+L_{20}
 -4WL_{21}+2L_{22}+2L_{23}
 \nonumber\\
 &+(W^2+2)L_{24}-WL_{25}-WL_{26}-3L_{27}+2WL_{28}-4L_{29}
 -6L_{30}-3L_{32}
 \nonumber\\
 &+2WL_{33}+6L_{34}+2WL_{35}-6L_{36}-4L_{37}+2WL_{38}+6L_{39}\big].
 \label{eq:ttt-structures}
\end{align}
The projected tensors form a spanning set. Matching must respect any
dimension-specific identities among their products.

\section{Integral formulas and individual diagram results}\label{app:diagrams}

\subsection{Tensor integrals}

The scalar convolution integral is
\begin{equation}
 \intk\frac1{(k^2)^\alpha((p-k)^2)^\beta}
 =\frac{(p^2)^{3/2-\alpha-\beta}}{(4\pi)^{3/2}}
 \frac{\Gamma(\alpha+\beta-3/2)\Gamma(3/2-\alpha)\Gamma(3/2-\beta)}
 {\Gamma(\alpha)\Gamma(\beta)\Gamma(3-\alpha-\beta)}.
 \label{eq:scalar-master}
\end{equation}
After Feynman parameterization, tensor numerators reduce to rotational
averages. A useful moment, including $n$ powers of the shifted numerator
weight, is
\begin{equation}
 J_{m,n}(\alpha,\beta)=
 \frac{(p^2)^{3/2+m-\alpha-\beta}
 \Gamma(\alpha+\beta-3/2-m)\Gamma(3/2+m-\beta)
 \Gamma(3/2+m-\alpha+n)}
 {(4\pi)^{3/2}2^m\Gamma(\alpha)\Gamma(\beta)
 \Gamma(3+2m-\alpha-\beta+n)}.
 \label{eq:tensor-moment}
\end{equation}
Gamma functions specify analytic continuation outside the initial
convergence domain. In the source calculation below, $m=0,1$ and
$n=0,1,2$ suffice.

The following subsections give the individual diagrams for both models.
Their labels follow figures~\ref{fig:tjj} and~\ref{fig:ttt} for the critical $O(N)$ vector model,
and figures~\ref{fig:gny-tjj} and~\ref{fig:gny-ttt} for the GNY model.
The solid lines represent propagators of the fundamental fields: $\phi_i$ in the critical $O(N)$ vector model and $\psi_i$ in the GNY model.
The tensors $K_i$ and $L_i$ are listed in appendices~\ref{app:tjj-basis} and~\ref{app:ttt-basis}.

All integrals use the measure $\dd^3k/(2\pi)^3$ and the propagators in
eq.~\eqref{eq:regulated-propagators}, with regulator scale $\mu=1$.
The results include the pole and finite terms as $\Delta\to0$.
They include graph symmetry factors but exclude operator counterterms and source contacts.
The contact diagrams are listed separately.
The $TJJ$ results include the flavor factor $\tau^{ab}$ used in section~\ref{sec:tjj}.
The $TTT$ results include the flavor trace, so $D_0$ is of order $N$ and the first correction is of order $N^0$.

We denote the $q$-independent contribution from the hard momentum region by $D_i^{(0)}$.
The contribution from all soft momentum regions at order $\lambda^j$ under
$q\mapsto\lambda q$ is denoted by $D_i^{\mathrm{soft},(j)}$.
Logarithms are grouped with the power of $\lambda$ that multiplies them.
For the GNY model, the listed contributions from soft momentum regions also include
polynomials in the components of $q$.
These polynomials are omitted when extracting the nonanalytic terms in the summed correlator.

\subsection{\texorpdfstring{$O(N)$ model: $\TJJ$}{O(N) model: TJJ}}
\label{app:tjj-results}

\subsubsection{Terms independent of the soft momentum}

At fixed nonzero $p$, the terms independent of $q$ use only
$K_{10},K_{13},K_{17},K_{21},K_{22}$.
These tensors are independent of the direction of the soft momentum $U$.

The results for Fig.~\ref{fig:tjj} are
\begingroup
\allowdisplaybreaks
\begin{adjustwidth}{-5mm}{-5mm}
\small
\begin{align}
&D_0^{(0)} =\frac{\tau^{ab}|p|}{32}(K_{10}+K_{13}-K_{22})\,,\label{eq:tjj-leading-triangle}\\
&D_1^{(0)}=D_3^{(0)} =-\frac{4\left(\frac1\Delta-\log p^2\right)}{3\pi^2N}D_0^{(0)}+\frac{\tau^{ab}|p|}{144\pi^2N} \Bigl[7K_{10}-3K_{13}+9K_{22}+2(K_{17}+K_{21})\Bigr]\,,\\
&D_2^{(0)} =-\frac{4\left(\frac1\Delta-\log p^2\right)}{3\pi^2N}D_0^{(0)}+\frac{\tau^{ab}|p|}{144\pi^2N} \Bigl[-3K_{10}-29K_{13}+11K_{22}-2(K_{17}+K_{21})\Bigr]\,,\\
&D_4^{(0)} =\frac{4\left(\frac1\Delta-\log p^2\right)}{15\pi^2N}D_0^{(0)}+\frac{\tau^{ab}|p|}{3600\pi^2N} \Bigl[-53K_{10}-3K_{13}-27K_{22}-10(K_{17}+K_{21})\Bigr]\,,\\
&D_5^{(0)} =\frac{4\left(\frac1\Delta-\log p^2\right)}{3\pi^2N}D_0^{(0)}+\frac{\tau^{ab}|p|}{\pi^2N} \Bigl[-\frac{77}{720}K_{10}+\frac{41}{720}K_{13}+\frac{197}{3600}K_{22}-\frac{161}{1800}K_{21}-\frac1{360}K_{17}\notag\\
&\quad{}+\frac{\left(\frac1\Delta-\log p^2\right)}{30}(K_{21}-K_{22})\Bigr]\,,\\
&D_6^{(0)} =\left.D_5^{(0)}\right|_{K_{21}\leftrightarrow K_{17}}\,,\\
&D_7^{(0)} =\frac8{15\pi^2N} \left(\frac1\Delta-2\log p^2\right)D_0^{(0)}+\frac{\tau^{ab}|p|}{\pi^2N} \Bigl[-\frac2{225}K_{10}+\frac7{150}K_{13}-\frac2{25}K_{22}-\frac1{90}(K_{21}+K_{17})\Bigr]\,,\\
&D_8^{(0)} =\frac{\tau^{ab}|p|}{\pi^2N} \Bigl[\frac1{180}K_{10}-\frac{13}{180}K_{13}+\frac{59}{900}K_{22}+\frac{23}{900}(K_{21}+K_{17})\notag\\
&\quad{}+\frac1{60}\left(\frac1\Delta-2\log p^2\right)(2K_{22}-K_{21}-K_{17})\Bigr]\,,\label{eq:tjj-hard-diagrams}\\
&D_i^{(0)}=0\,,\qquad i=9,10,11\,.
\end{align}
\end{adjustwidth}
\endgroup
The equality between $D_1^{(0)}$ and $D_3^{(0)}$ holds at $q=0$.
At nonzero $q$, the crossed placements instead exchange
$(a,\mu,p)$ with $(b,\nu,-p-q)$.
This exchange also relates $D_5$ and $D_6$.

\subsubsection{Contributions from soft momentum regions}

The linear terms are
\begingroup\allowdisplaybreaks
\begin{adjustwidth}{-5mm}{-5mm}
\small
\begin{align}
&D_1^{\mathrm{soft},(1)}=D_3^{\mathrm{soft},(1)} =-\frac{\tau^{ab}|q|}{48\pi^2N}K_{5}\,,\\
&D_4^{\mathrm{soft},(1)} =\frac{\tau^{ab}|q|}{120\pi^2N}K_{5}\,,\\
&D_7^{\mathrm{soft},(1)} =\frac{\tau^{ab}|q|}{60\pi^2N}K_{5}\,,\\
&D_9^{\mathrm{soft},(1)} =\frac{\tau^{ab}|q|}{24\pi^2N}K_{5}\,,\\
&D_{10}^{\mathrm{soft},(1)} =-\frac{\tau^{ab}|q|}{120\pi^2N}K_{5}\,,\\
&D_{11}^{\mathrm{soft},(1)} =-\frac{\tau^{ab}|q|}{60\pi^2N}K_{5}\,,\\
&D_i^{\mathrm{soft},(1)}=0\,,\qquad i=0,2,5,6,8\,.
\end{align}
\end{adjustwidth}
\endgroup
The quadratic terms distinguish the two self-energy placements adjacent to $T$:
\begingroup\allowdisplaybreaks
\begin{adjustwidth}{-5mm}{-5mm}
\small
\begin{align}
&D_1^{\mathrm{soft},(2)} =\frac{\tau^{ab}|q|^2}{144\pi^2N|p|} \left(K_{3}+4K_{4}-WK_{5}+2K_{9}-2K_{15}-2K_{19}\right)\,,\\
&D_3^{\mathrm{soft},(2)} =\frac{\tau^{ab}|q|^2}{144\pi^2N|p|} \left(-7K_{3}-4K_{4}+7WK_{5}-2K_{9}+2K_{15}+2K_{19}\right)\,,\\
&D_4^{\mathrm{soft},(2)} =\frac{\tau^{ab}|q|^2}{120\pi^2N|p|}(K_{3}-WK_{5})\,,\\
&D_7^{\mathrm{soft},(2)} =\frac{\tau^{ab}|q|^2}{60\pi^2N|p|}(K_{3}-WK_{5})\,,\\
&D_9^{\mathrm{soft},(2)} =\frac{\tau^{ab}|q|^2}{24\pi^2N|p|}(K_{3}-WK_{5})\,,\\
&D_{10}^{\mathrm{soft},(2)} =-\frac{\tau^{ab}|q|^2}{120\pi^2N|p|}(K_{3}-WK_{5})\,,\\
&D_{11}^{\mathrm{soft},(2)} =-\frac{\tau^{ab}|q|^2}{60\pi^2N|p|}(K_{3}-WK_{5})\,,\\
&D_i^{\mathrm{soft},(2)}=0\,,\qquad i=0,2,5,6,8\,.
\end{align}
\end{adjustwidth}
\endgroup
The difference between $D_1$ and $D_3$ follows from exchanging the complete
current momenta, including the shift $p\mapsto-p-q$.
This shift also enters their cubic terms.

The results for Fig.~\ref{fig:tjj} are
\begingroup\allowdisplaybreaks
\begin{adjustwidth}{-5mm}{-5mm}
\small
\begin{align}
&D_0^{\mathrm{soft},(3)} =\frac{\tau^{ab}|q|^3}{128|p|^2}\Bigl[\bigl(K_{1}+K_{2}+K_{7}+K_{8}-K_{14}\bigr)\notag\\
&\quad{}-\bigl(K_{17}+K_{18}+K_{21}-K_{22}\bigr)-W\,\bigl(K_{3}+K_{4}-K_{15}-K_{19}\bigr)\notag\\
&\quad{}+\bigl(W^{2} - 1\bigr)\,K_{5}-2 W\,K_{9}+2\,K_{13}\Bigr]\,,\\[4pt]
&D_1^{\mathrm{soft},(3)} =\frac{\tau^{ab}|q|^3}{1152\pi^2N|p|^2}\Bigl[\bigl(-\frac{12}{\Delta} +12\log q^2 -23\bigr)\,K_{1}\notag\\
&\quad{}+3\bigl(-\frac{4}{\Delta} +4\log q^2 +3\bigr)\,K_{2}+W\bigl(\frac{12}{\Delta} -12\log q^2 -17\bigr)\,K_{3}\notag\\
&\quad{}+3 W\bigl(\frac{4}{\Delta} -4\log q^2 -3\bigr)\,K_{4}+\bigl(\frac{12 \left(1 - W^{2}\right)}{\Delta} +12 \left(W^{2} - 1\right)\log q^2 +17 W^{2} + 23\bigr)\,K_{5}\notag\\
&\quad{}+\bigl(-\frac{12}{\Delta} +12\log q^2 -13\bigr)\,\bigl(K_{7}-K_{18}\bigr)\notag\\
&\quad{}+\bigl(-\frac{12}{\Delta} +12\log q^2 -29\bigr)\,\bigl(K_{8}-K_{14}\bigr)+2 W\bigl(\frac{12}{\Delta} -12\log q^2 +13\bigr)\,K_{9}\notag\\
&\quad{}+6\bigl(-\frac{4}{\Delta} +4\log q^2 -11\bigr)\,K_{13}+W\bigl(-\frac{12}{\Delta} +12\log q^2 -13\bigr)\,\bigl(K_{15}+K_{19}\bigr)\notag\\
&\quad{}+3\bigl(\frac{4}{\Delta} -4\log q^2 +11\bigr)\,\bigl(K_{17}+K_{21}-K_{22}\bigr)\Bigr]\,,\\[4pt]
&D_2^{\mathrm{soft},(3)} =\frac{\tau^{ab}|q|^3}{576\pi^2N|p|^2}\Bigl[2\bigl(-\frac{3}{\Delta} +3\log p^2 -2\bigr)\,\bigl(K_{1}+K_{2}-K_{14}-K_{18}+K_{22}\bigr)\notag\\
&\quad{}+2 W\bigl(\frac{3}{\Delta} -3\log p^2 +5\bigr)\,\bigl(K_{3}+K_{4}-K_{15}-K_{19}\bigr)+\bigl(\frac{6 \left(1 - W^{2}\right)}{\Delta} +6 \left(W^{2} - 1\right)\log p^2 +7 - 13 W^{2}\bigr)\,K_{5}\notag\\
&\quad{}+2\bigl(-\frac{3}{\Delta} +3\log p^2 -5\bigr)\,\bigl(K_{7}+K_{8}-K_{17}-K_{21}\bigr)+2 W\bigl(\frac{6}{\Delta} -6\log p^2 +13\bigr)\,K_{9}\notag\\
&\quad{}+2\bigl(-\frac{6}{\Delta} +6\log p^2 -13\bigr)\,K_{13}\Bigr]\,,\\[4pt]
&D_3^{\mathrm{soft},(3)} =\frac{\tau^{ab}|q|^3}{1152\pi^2N|p|^2}\Bigl[\bigl(-\frac{12}{\Delta} +12\log q^2 -23\bigr)\,K_{1}\notag\\
&\quad{}+\bigl(-\frac{12}{\Delta} +12\log q^2 -55\bigr)\,K_{2}+3 W\bigl(\frac{4}{\Delta} -4\log q^2 +37\bigr)\,K_{3}\notag\\
&\quad{}+W\bigl(\frac{12}{\Delta} -12\log q^2 +55\bigr)\,K_{4}+\bigl(\frac{12 \left(1 - W^{2}\right)}{\Delta} +12 \left(W^{2} - 1\right)\log q^2 +23 - 111 W^{2}\bigr)\,K_{5}\notag\\
&\quad{}+3\bigl(-\frac{4}{\Delta} +4\log q^2 -15\bigr)\,\bigl(K_{7}-K_{18}\bigr)\notag\\
&\quad{}+\bigl(-\frac{12}{\Delta} +12\log q^2 -29\bigr)\,\bigl(K_{8}-K_{14}\bigr)+6 W\bigl(\frac{4}{\Delta} -4\log q^2 +15\bigr)\,K_{9}\notag\\
&\quad{}+6\bigl(-\frac{4}{\Delta} +4\log q^2 -11\bigr)\,K_{13}+3 W\bigl(-\frac{4}{\Delta} +4\log q^2 -15\bigr)\,\bigl(K_{15}+K_{19}\bigr)\notag\\
&\quad{}+3\bigl(\frac{4}{\Delta} -4\log q^2 +11\bigr)\,\bigl(K_{17}+K_{21}-K_{22}\bigr)\Bigr]\,,\\[4pt]
&D_4^{\mathrm{soft},(3)} =\frac{\tau^{ab}|q|^3}{14400\pi^2N|p|^2}\Bigl[\bigl(\frac{30}{\Delta} +30\log p^2 -60\log q^2 +20 W^{2} + 107\bigr)\,K_{1}\notag\\
&\quad{}+3\bigl(\frac{10}{\Delta} +10\log p^2 -20\log q^2 +29\bigr)\,\bigl(K_{2}-K_{17}-K_{21}\bigr)+W\bigl(-\frac{30}{\Delta} -30\log p^2 +60\log q^2 -157\bigr)\,K_{3}\notag\\
&\quad{}+W\bigl(-\frac{30}{\Delta} -30\log p^2 +60\log q^2 -37\bigr)\,K_{4}\notag\\
&\quad{}+2\bigl(\frac{15 \left(W^{2} - 1\right)}{\Delta} +15 \left(W^{2} - 1\right)\log p^2 +30 \left(1 - W^{2}\right)\log q^2 +81 W^{2} - 16\bigr)\,K_{5}-40 W\,K_{6}\notag\\
&\quad{}+\bigl(\frac{30}{\Delta} +30\log p^2 -60\log q^2 +67\bigr)\,\bigl(K_{7}+K_{8}\bigr)+12 W\bigl(-\frac{5}{\Delta} -5\log p^2 +10\log q^2 -12\bigr)\,K_{9}+40\,K_{10}\notag\\
&\quad{}+4\bigl(\frac{15}{\Delta} +15\log p^2 -30\log q^2 +46\bigr)\,K_{13}+3\bigl(-\frac{10}{\Delta} -10\log p^2 +20\log q^2 -39\bigr)\,\bigl(K_{14}+K_{18}\bigr)\notag\\
&\quad{}+W\bigl(\frac{30}{\Delta} +30\log p^2 -60\log q^2 +67\bigr)\,\bigl(K_{15}+K_{19}\bigr)+\bigl(\frac{30}{\Delta} +30\log p^2 -60\log q^2 +137\bigr)\,K_{22}\Bigr]\,,\\[4pt]
&D_5^{\mathrm{soft},(3)} =\frac{\tau^{ab}|q|^3}{576\pi^2N|p|^2}\Bigl[2\bigl(\frac{3}{\Delta} -3\log p^2 -4\bigr)\,\bigl(K_{1}+K_{2}-K_{14}-K_{18}+K_{22}\bigr)\notag\\
&\quad{}+6 W\bigl(-\frac{1}{\Delta} +\log p^2 +1\bigr)\,\bigl(K_{3}-K_{15}\bigr)+2 W\bigl(-\frac{3}{\Delta} +3\log p^2 +1\bigr)\,\bigl(K_{4}-K_{19}\bigr)\notag\\
&\quad{}+\bigl(\frac{6 \left(W^{2} - 1\right)}{\Delta} +6 \left(1 - W^{2}\right)\log p^2 +5 - 3 W^{2}\bigr)\,K_{5}+6\bigl(\frac{1}{\Delta} -\log p^2 -1\bigr)\,\bigl(K_{7}-K_{17}\bigr)\notag\\
&\quad{}+2\bigl(\frac{3}{\Delta} -3\log p^2 -1\bigr)\,\bigl(K_{8}-K_{21}\bigr)+6 W\bigl(-\frac{2}{\Delta} +2\log p^2 +1\bigr)\,K_{9}+6\bigl(\frac{2}{\Delta} -2\log p^2 -1\bigr)\,K_{13}\Bigr]\,,\\[4pt]
&D_6^{\mathrm{soft},(3)}=\left.D_5^{\mathrm{soft},(3)}\right|_{\mu\leftrightarrow\nu}\,,\\[4pt]
&D_7^{\mathrm{soft},(3)} =\frac{\tau^{ab}|q|^3}{3600\pi^2N|p|^2}\Bigl[\bigl(\frac{15}{\Delta} +30\log p^2 -60\log q^2 +20 W^{2} + 182\bigr)\,K_{1}\notag\\
&\quad{}+3\bigl(\frac{5}{\Delta} +10\log p^2 -20\log q^2 +34\bigr)\,K_{2}\notag\\
&\quad{}+W\bigl(-\frac{15}{\Delta} -30\log p^2 +60\log q^2 -112\bigr)\,\bigl(K_{3}-K_{15}-K_{19}\bigr)+W\bigl(-\frac{15}{\Delta} -30\log p^2 +60\log q^2 -52\bigr)\,K_{4}\notag\\
&\quad{}+\bigl(\frac{15 \left(W^{2} - 1\right)}{\Delta} +30 \left(W^{2} - 1\right)\log p^2 +60 \left(1 - W^{2}\right)\log q^2 +117 W^{2} - 107\bigr)\,K_{5}-40 W K_{6}\notag\\
&\quad{}+\bigl(\frac{15}{\Delta} +30\log p^2 -60\log q^2 +112\bigr)\,\bigl(K_{7}+K_{8}\bigr)+6 W\bigl(-\frac{5}{\Delta} -10\log p^2 +20\log q^2 -39\bigr)\,K_{9}+40\,K_{10}\notag\\
&\quad{}+2\bigl(\frac{15}{\Delta} +30\log p^2 -60\log q^2 +137\bigr)\,K_{13}+3\bigl(-\frac{5}{\Delta} -10\log p^2 +20\log q^2 -54\bigr)\,\bigl(K_{14}+K_{18}\bigr)\notag\\
&\quad{}+3\bigl(-\frac{5}{\Delta} -10\log p^2 +20\log q^2 -44\bigr)\,\bigl(K_{17}+K_{21}\bigr)+\bigl(\frac{15}{\Delta} +30\log p^2 -60\log q^2 +182\bigr)\,K_{22}\Bigr]\,,\\[4pt]
&D_8^{\mathrm{soft},(3)} =\frac{\tau^{ab}|q|^3}{1152\pi^2N|p|^2}\Bigl[-\bigl(- 9 \pi^{2} W^{2} + 88 W^{2} - 9 \pi^{2} + 96\bigr)\,K_{1}-8\,\bigl(K_{2}-K_{7}-K_{8}-K_{14}+K_{17}\bigr)\notag\\
&\quad{}+8\,\bigl(K_{18}-K_{21}-K_{22}\bigr)-8 W\,\bigl(K_{3}+K_{4}-K_{15}-K_{19}\bigr)+\bigl(- 9 \pi^{2} W^{2} + 88 W^{2} - 9 \pi^{2} + 72\bigr)\,K_{5}\notag\\
&\quad{}+\bigl(- 18 \pi^{2} W + 176 W\bigr)\,\bigl(K_{6}-K_{9}\bigr)-\bigl(176 - 18 \pi^{2}\bigr)\,\bigl(K_{10}-K_{13}\bigr)\Bigr]\,,\\[4pt]
&D_9^{\mathrm{soft},(3)} =\frac{\tau^{ab}|q|^3}{48\pi^2N|p|^2} \Bigl[-K_{1}+K_{2}-3WK_{3}-WK_{4}+(1+3W^2)K_{5}\Bigr]\,,\\[4pt]
&D_{10}^{\mathrm{soft},(3)} =-\frac{\tau^{ab}|q|^3}{240\pi^2N|p|^2} \Bigl[-K_{1}+K_{2}-3WK_{3}-WK_{4}+(1+3W^2)K_{5}\Bigr]\,,\\[4pt]
&D_{11}^{\mathrm{soft},(3)} =-\frac{\tau^{ab}|q|^3}{120\pi^2N|p|^2} \Bigl[-K_{1}+K_{2}-3WK_{3}-WK_{4}+(1+3W^2)K_{5}\Bigr]\,.
\end{align}
\end{adjustwidth}
\endgroup
The linear and quadratic contributions cancel in the sum of the diagrams.
Adding the stress-tensor counterterm to the cubic contributions gives eq.~\eqref{eq:on-tjj-soft}.

\subsubsection{Contact diagrams}
\label{app:contact-results}

The results for $C_1$ and $C_3$ include both self-energy placements.

The current diagrams carry flavor indices $a,b$ and spacetime indices $\mu,\nu$.
With these indices suppressed on $C_i$, their contributions are
\begin{align}
C_0
&=-\frac{\tau^{ab}|p|}{32}
\left(\delta_{\mu\nu}-\frac{p_\mu p_\nu}{p^2}\right)\,,\\
C_1
&=\frac{\tau^{ab}|p|}{12\pi^2N}
\left[\left(\frac1\Delta-\log p^2+\frac{13}{6}\right)\delta_{\mu\nu}
-\left(\frac1\Delta-\log p^2+\frac76\right)\frac{p_\mu p_\nu}{p^2}\right]\,,\\
C_2
&=\frac{\tau^{ab}|p|}{12\pi^2N}
\left[-\left(\frac1\Delta-\log p^2-\frac12\right)\delta_{\mu\nu}
+\left(\frac1\Delta-\log p^2-\frac32\right)\frac{p_\mu p_\nu}{p^2}\right]\,.
\label{eq:contact-current-diagrams}
\end{align}
The poles and logarithms cancel between $C_1$ and $C_2$.
Their sum is transverse. Adding $C_0$ gives
\begin{equation}
\langle J_\mu^a(p)J_\nu^b(-p)\rangle
=-\frac{\tau^{ab}|p|}{32}
\left(1-\frac{64}{9\pi^2N}\right)
\left(\delta_{\mu\nu}-\frac{p_\mu p_\nu}{p^2}\right)
+O(1/N^2)\,.
\end{equation}
The second mixed contact uses these results with $p\mapsto p+q$.

The insertion $\phi_i\phi_i$ includes the sum over all $N$ scalar components.
We include this flavor trace in $C_3,C_4,C_5$, so their first nonzero terms are of order $N^0$.
With the stress-tensor indices $\alpha,\beta$ suppressed on $C_i$, these terms are
\begin{align}
C_3
&=-\frac{1}{24\pi^2|q|}
\left(q_\alpha q_\beta-\frac{q^2}{3}\delta_{\alpha\beta}\right)\,,\\
C_4
&=\frac{1}{120\pi^2|q|}
\left(q_\alpha q_\beta-\frac{q^2}{3}\delta_{\alpha\beta}\right)\,,\\
C_5
&=\frac{1}{60\pi^2|q|}
\left(q_\alpha q_\beta-\frac{q^2}{3}\delta_{\alpha\beta}\right)\,.
\label{eq:contact-scalar-diagrams}
\end{align}
These terms are finite and traceless in $\alpha,\beta$.
Their sum is
\begin{equation}
C_3+C_4+C_5
=-\frac{1}{60\pi^2|q|}
\left(q_\alpha q_\beta-\frac{q^2}{3}\delta_{\alpha\beta}\right)\,.
\end{equation}
The diagrams $C_6,C_7,C_8$ in the lower row of
figure~\ref{fig:tjj-contacts} complete this partial sum.
For $O=\phi_i\phi_i$, the leading free bilinear bubble is
\begin{equation}
\langle O(q)O(-q)\rangle_0'=\frac{N}{4|q|}.
\end{equation}
To obtain $C_{i+3}$ from $C_i$, $i=3,4,5$, replace its external $O$
insertion by a $\sigma\phi\phi$ vertex and connect this vertex through
one auxiliary-field propagator to a free matter bubble carrying $O$.
The factor $-1/2$ replaces an $O$ insertion by the interaction vertex
from $\exp(-S_{\ON})$. Using the propagator in eq.~\eqref{eq:on-rules},
the finite nonlocal terms at the displayed order give
\begin{equation}
 C_{i+3}
 =\left(-\frac12 C_i\right)
  \left(-\frac{16|q|}{N}\right)
  \left(-\frac{N}{8|q|}\right)
 =-C_i,\qquad i=3,4,5.
 \label{eq:on-tjj-mixed-contact-pairs}
\end{equation}
The last factor is the free bilinear bubble with one $O$ insertion
replaced by the interaction vertex. The self-energy placements included
in $C_3$ are also included in $C_6$.
The local contributions of the mixed subgraphs are kept until the
complete graphs are assembled and the regulator is removed.
Consequently,
\begin{equation}
 \sum_{i=3}^{8}C_i=0,\qquad
 -\frac{\tau^{ab}}{N}\delta_{\mu\nu}\sum_{i=3}^{8}C_i=0.
 \label{eq:on-tjj-mixed-contact-cancellation}
\end{equation}
These equalities concern the nonlocal terms at order $N^0$ in the
mixed kernels, or order $1/N$ after the singlet contact attachment.
Local contact terms follow the common prescription of
section~\ref{sec:background}.

\subsection{\texorpdfstring{$O(N)$ model: $\TTT$}{O(N) model: TTT}}
\label{app:ttt-results}
The formulas below give the individual contributions.
At order $q^0$, they include $D_0,\ldots,D_{12}$ and the additional cyclic
assignments of $D_7,D_8$.
The contributions from soft momentum regions are listed through cubic order.
At cubic order in the soft contribution, the two additional cyclic
assignments of $D_7$ and $D_8$ are given as a combined sum.
\subsubsection{Terms independent of the soft momentum}

The terms independent of $q$ depend only on $V$ and use $L_1,\ldots,L_{11}$.
The results for
Fig.~\ref{fig:ttt} are
\begingroup\allowdisplaybreaks
\begin{adjustwidth}{-5mm}{-5mm}
\small
\begin{align}
&D_{0}^{(0)}=\frac{N|p|^3}{512}\Bigl[L_{1}+4L_{2}-6L_{4}+2L_{6}-8L_{9}+8L_{10}+2L_{11}\Bigr]\,.\\[3pt]
&D_{1}^{(0)}=D_3^{(0)}=-\frac{|p|^3}{23040 \pi^2}\Biggl[\Bigl(\frac{60}{\Delta}-60\log p^2+39\Bigr)L_{1}+\Bigl(\frac{240}{\Delta}-240\log p^2-316\Bigr)L_{2}-16L_{3}\notag\\
&\quad{}+\Bigl(-\frac{360}{\Delta}+360\log p^2-318\Bigr)L_{4}+288L_{5}+\Bigl(\frac{120}{\Delta}-120\log p^2+102\Bigr)L_{6}-16L_{7}+288L_{8}\notag\\
&\quad{}+\Bigl(-\frac{480}{\Delta}+480\log p^2-1032\Bigr)L_{9}+\Bigl(\frac{480}{\Delta}-480\log p^2+936\Bigr)L_{10}+\Bigl(\frac{120}{\Delta}-120\log p^2+102\Bigr)L_{11}\Biggr]\,.\\[3pt]
&D_{2}^{(0)}=-\frac{|p|^3}{23040 \pi^2}\Biggl[\Bigl(\frac{60}{\Delta}-60\log p^2+251\Bigr)L_{1}+\Bigl(\frac{240}{\Delta}-240\log p^2+276\Bigr)L_{2}+16L_{3}\notag\\
&\quad{}+\Bigl(-\frac{360}{\Delta}+360\log p^2-822\Bigr)L_{4}-288L_{5}+\Bigl(\frac{120}{\Delta}-120\log p^2+238\Bigr)L_{6}+16L_{7}-288L_{8}\notag\\
&\quad{}+\Bigl(-\frac{480}{\Delta}+480\log p^2-808\Bigr)L_{9}+\Bigl(\frac{480}{\Delta}-480\log p^2+1224\Bigr)L_{10}+\Bigl(\frac{120}{\Delta}-120\log p^2+238\Bigr)L_{11}\Biggr]\,.\\[3pt]
&D_{4}^{(0)}=\frac{|p|^3}{115200 \pi^2}\Biggl[\Bigl(\frac{60}{\Delta}-60\log p^2+3\Bigr)L_{1}+\Bigl(\frac{240}{\Delta}-240\log p^2-460\Bigr)L_{2}-16L_{3}\notag\\
&\quad{}+\Bigl(-\frac{360}{\Delta}+360\log p^2-102\Bigr)L_{4}+288L_{5}+\Bigl(\frac{120}{\Delta}-120\log p^2+30\Bigr)L_{6}-16L_{7}+288L_{8}\notag\\
&\quad{}+\Bigl(-\frac{480}{\Delta}+480\log p^2-744\Bigr)L_{9}+\Bigl(\frac{480}{\Delta}-480\log p^2+648\Bigr)L_{10}+\Bigl(\frac{120}{\Delta}-120\log p^2+30\Bigr)L_{11}\Biggr]\,.\\[3pt]
&D_{5}^{(0)}=\frac{|p|^3}{5644800 \pi^2}\Biggl[\Bigl(\frac{2940}{\Delta}-2940\log p^2+7819\Bigr)L_{1}+\Bigl(\frac{11760}{\Delta}-11760\log p^2-1036\Bigr)L_{2}-5936L_{3}\notag\\
&\quad{}+\Bigl(-\frac{17640}{\Delta}+17640\log p^2-23478\Bigr)L_{4}+6048L_{5}+\Bigl(\frac{5880}{\Delta}-5880\log p^2+8078\Bigr)L_{6}\notag\\
&\quad{}+\Bigl(\frac{6720}{\Delta}-6720\log p^2-1664\Bigr)L_{7}+\Bigl(-\frac{13440}{\Delta}+13440\log p^2-23552\Bigr)L_{8}\notag\\
&\quad{}+\Bigl(-\frac{36960}{\Delta}+36960\log p^2-58440\Bigr)L_{9}+\Bigl(\frac{36960}{\Delta}-36960\log p^2+69416\Bigr)L_{10}\notag\\
&\quad{}+\Bigl(\frac{12600}{\Delta}-12600\log p^2+16382\Bigr)L_{11}\Biggr]\,.\\[3pt]
&D_{7}^{(0)}=\frac{|p|^3}{28800 \pi^2}\Biggl[\Bigl(\frac{30}{\Delta}-60\log p^2+93\Bigr)L_{1}+\Bigl(\frac{120}{\Delta}-240\log p^2-100\Bigr)L_{2}-16L_{3}\notag\\
&\quad{}+\Bigl(-\frac{180}{\Delta}+360\log p^2-642\Bigr)L_{4}+288L_{5}+\Bigl(\frac{60}{\Delta}-120\log p^2+210\Bigr)L_{6}-16L_{7}+288L_{8}\notag\\
&\quad{}+\Bigl(-\frac{240}{\Delta}+480\log p^2-1464\Bigr)L_{9}+\Bigl(\frac{240}{\Delta}-480\log p^2+1368\Bigr)L_{10}+\Bigl(\frac{60}{\Delta}-120\log p^2+210\Bigr)L_{11}\Biggr]\,.\\[3pt]
&D_{8}^{(0)}=-\frac{|p|^3}{705600 \pi^2}\Biggl[301L_{1}+476L_{2}+\Bigl(\frac{420}{\Delta}-840\log p^2+730\Bigr)L_{3}\notag\\
&\quad{}+378L_{4}+\Bigl(-\frac{840}{\Delta}+1680\log p^2-5548\Bigr)L_{5}\notag\\
&\quad{}+\Bigl(\frac{420}{\Delta}-840\log p^2+2900\Bigr)L_{6}+\Bigl(\frac{420}{\Delta}-840\log p^2+730\Bigr)L_{7}+\Bigl(-\frac{840}{\Delta}+1680\log p^2-5548\Bigr)L_{8}\notag\\
&\quad{}+\Bigl(-\frac{1680}{\Delta}+3360\log p^2-11040\Bigr)L_{9}+\Bigl(\frac{1680}{\Delta}-3360\log p^2+11824\Bigr)L_{10}\notag\\
&\quad{}+\Bigl(\frac{420}{\Delta}-840\log p^2+2900\Bigr)L_{11}\Biggr]\,.
\end{align}
\end{adjustwidth}
\endgroup

The remaining reflected vertex correction is
\begin{adjustwidth}{-5mm}{-5mm}
\small
\begin{equation}
D_6^{(0)}
=\left.D_5^{(0)}\right|_{\mu\leftrightarrow\rho,\,\nu\leftrightarrow\sigma}\,.
\end{equation}
\end{adjustwidth}
The formulas for $D_7$ and $D_8$ refer to the displayed assignment with $T(q)$ on the triangle.
Their other cyclic assignments are given below.

The bridge diagrams have the finite hard coefficients
\begin{equation}
 D_9^{(0)}=-\frac{|p|^3}{96\pi^2}(L_6+L_{11}),\qquad
 D_{10}^{(0)}=-\frac15D_9^{(0)},\qquad
 D_{11}^{(0)}=-\frac25D_9^{(0)}.
 \label{eq:on-ttt-hard-bridges}
\end{equation}
These expressions include all external assignments.
The assignment with the corrected mixed two-point function on the soft
stress-tensor leg has no term independent of $q$ from the hard momentum region.
The two hard-leg assignments give the displayed coefficients.
Their relative factors follow from the three mixed two-point diagrams in
eq.~\eqref{eq:on-contact-mixed-diagrams}.

The diagram $D_{12}$, which contains three scalar loops, gives
\begingroup\small\allowdisplaybreaks[0]
\begin{align}
 D_{12}^{(0)}&=\frac{|p|^3}{705600\pi^2}\Bigl[\notag\\
 &\quad{}-2807L_{1}-9772L_{2}\notag\\
 &\quad{}+\Bigl(-\frac{560}{\Delta }+1680 \log p^2+2320\Bigr)\bigl(L_{3}+L_{7}\bigr)+12474L_{4}\notag\\
 &\quad{}+\Bigl(\frac{1120}{\Delta }-3360 \log p^2+3536\Bigr)\bigl(L_{5}+L_{8}\bigr)+\Bigl(-\frac{560}{\Delta }+1680 \log p^2-6430\Bigr)\bigl(L_{6}+L_{11}\bigr)\notag\\
 &\quad{}+\Bigl(\frac{2240}{\Delta }-6720 \log p^2+24600\Bigr)L_{9}+\Bigl(-\frac{2240}{\Delta }+6720 \log p^2-26168\Bigr)L_{10}\Bigr].
 \label{eq:on-ttt-hard-ring}
\end{align}
\endgroup

\begin{samepage}
The additional cyclic assignments place the hard stress tensor on the
triangle of $D_7$ or $D_8$.
Expanding these assignments before taking $q\to0$ gives
\begingroup\small\allowdisplaybreaks[0]
\begin{align}
 (D_7)_{\mathrm{cyc}}^{(0)}&=-\frac{|p|^3}{705600\pi^2}\Bigl[\notag\\
 &\quad{}\Bigl(-\frac{1470}{\Delta }+2940 \log p^2-12229\Bigr)L_{1}+\Bigl(-\frac{5880}{\Delta }+11760 \log p^2-16604\Bigr)L_{2}\notag\\
 &\quad{}+\Bigl(-\frac{1680}{\Delta }+3360 \log p^2+3800\Bigr)\bigl(L_{3}+L_{7}\bigr)+\Bigl(\frac{8820}{\Delta }-17640 \log p^2+49938\Bigr)L_{4}\notag\\
 &\quad{}+\Bigl(\frac{3360}{\Delta }-6720 \log p^2+8752\Bigr)\bigl(L_{5}+L_{8}\bigr)+\Bigl(-\frac{4620}{\Delta }+9240 \log p^2-23990\Bigr)\bigl(L_{6}+L_{11}\bigr)\notag\\
 &\quad{}+\Bigl(\frac{18480}{\Delta }-36960 \log p^2+93720\Bigr)L_{9}+\Bigl(-\frac{18480}{\Delta }+36960 \log p^2-104696\Bigr)L_{10}\Bigr].
 \label{eq:on-ttt-hard-cyclic-seven}
\end{align}
\begin{align}
 (D_8)_{\mathrm{cyc}}^{(0)}&=\frac{|p|^3}{352800\pi^2}\Bigl[\notag\\
 &\quad{}-301L_{1}-476L_{2}\notag\\
 &\quad{}+\Bigl(-\frac{420}{\Delta }+840 \log p^2-100\Bigr)\bigl(L_{3}+L_{7}\bigr)-378L_{4}\notag\\
 &\quad{}+\Bigl(\frac{840}{\Delta }-1680 \log p^2+4288\Bigr)\bigl(L_{5}+L_{8}\bigr)+\Bigl(-\frac{420}{\Delta }+840 \log p^2-2270\Bigr)\bigl(L_{6}+L_{11}\bigr)\notag\\
 &\quad{}+\Bigl(\frac{1680}{\Delta }-3360 \log p^2+8520\Bigr)L_{9}+\Bigl(-\frac{1680}{\Delta }+3360 \log p^2-9304\Bigr)L_{10}\Bigr].
 \label{eq:on-ttt-hard-cyclic-eight}
\end{align}
\endgroup
\end{samepage}
Each cyclic coefficient includes the two assignments beyond the one
displayed in figure~\ref{fig:ttt}.
The formulas give the pole and finite terms as $\Delta\to0$.
The finite terms include the products of regulator poles with the
regulator-dependent factors from the auxiliary-field propagators.
Together with the preceding diagrams, the contact terms in
appendix~\ref{app:on-ttt-contacts}, and the source correction in
eq.~\eqref{eq:scalar-source-correction}, these coefficients reproduce
eq.~\eqref{eq:on-ttt-hard}.
All regulator poles and logarithms cancel in that sum.

\subsubsection{Contributions from soft momentum regions}

The bridge results $D_9,D_{10},D_{11}$ include all three external assignments.
The results for $D_7,D_8$ initially refer to the displayed assignment with $T(q)$ on the triangle.
Their additional cyclic terms are given together below.

The linear terms are
\begingroup\allowdisplaybreaks
\begin{adjustwidth}{-5mm}{-5mm}
\small
\begin{align}
&D_{1}^{\mathrm{soft},(1)}=-\frac{|q||p|^2}{768 \pi^2}L_{15}\,.\\[3pt]
&D_{3}^{\mathrm{soft},(1)}=-\frac{|q||p|^2}{768 \pi^2}L_{15}\,.\\[3pt]
&D_{4}^{\mathrm{soft},(1)}=\frac{|q||p|^2}{1920 \pi^2}L_{15}\,.\\[3pt]
&D_{7}^{\mathrm{soft},(1)}=\frac{|q||p|^2}{960 \pi^2}L_{15}\,.\\[3pt]
&D_{9}^{\mathrm{soft},(1)}=\frac{|q||p|^2}{384 \pi^2}L_{15}\,.\\[3pt]
&D_i^{\mathrm{soft},(1)}=0\,,\qquad i=0,2,5,6,8,12\,.
\end{align}
\end{adjustwidth}
\endgroup
\begin{adjustwidth}{-5mm}{-5mm}
\small
\begin{equation}
D_{10}^{\mathrm{soft},(1)}=-\frac15D_9^{\mathrm{soft},(1)}\,,\qquad
D_{11}^{\mathrm{soft},(1)}=-\frac25D_9^{\mathrm{soft},(1)}\,.
\end{equation}
\end{adjustwidth}

The quadratic terms are
\begingroup\allowdisplaybreaks
\begin{adjustwidth}{-5mm}{-5mm}
\small
\begin{align}
&D_{1}^{\mathrm{soft},(2)}=-\frac{|q|^2|p|}{2304 \pi^2}\Bigl[-4L_{12}+WL_{15}-4L_{16}-22L_{18}+16L_{28}+16L_{35}\Bigr]\,.\\[3pt]
&D_{3}^{\mathrm{soft},(2)}=\frac{|q|^2|p|}{2304 \pi^2}\Bigl[-4L_{12}+7 WL_{15}-28L_{16}-10L_{18}+16L_{28}+16L_{35}\Bigr]\,.\\[3pt]
&D_{4}^{\mathrm{soft},(2)}=-\frac{|q|^2|p|}{1920 \pi^2}\Bigl[WL_{15}-4L_{16}+2L_{18}\Bigr]\,.\\[3pt]
&D_{7}^{\mathrm{soft},(2)}=-\frac{|q|^2|p|}{960 \pi^2}\Bigl[WL_{15}-4L_{16}+2L_{18}\Bigr]\,.\\[3pt]
&D_{9}^{\mathrm{soft},(2)}=-\frac{|q|^2|p|}{384 \pi^2}\Bigl[WL_{15}-4L_{16}+2L_{18}\Bigr]\,.\\[3pt]
&D_i^{\mathrm{soft},(2)}=0\,,\qquad i=0,2,5,6,8,12\,.
\end{align}
\end{adjustwidth}
\endgroup
\begin{adjustwidth}{-5mm}{-5mm}
\small
\begin{equation}
D_{10}^{\mathrm{soft},(2)}=-\frac15D_9^{\mathrm{soft},(2)}\,,\qquad
D_{11}^{\mathrm{soft},(2)}=-\frac25D_9^{\mathrm{soft},(2)}\,.
\end{equation}
\end{adjustwidth}

The linear and quadratic terms cancel in the diagram sum:
\begin{equation}
\sum_{i=0}^{12}D_i^{\mathrm{soft},(1)}
=\sum_{i=0}^{12}D_i^{\mathrm{soft},(2)}=0\,.
\end{equation}

The cubic terms are
\begingroup\allowdisplaybreaks
\begin{adjustwidth}{-5mm}{-5mm}
\small
\begin{align}
&D_{0}^{\mathrm{soft},(3)}=\frac{N|q|^3}{2048}\Bigl[2L_{1}-8L_{3}+4L_{6}-8L_{7}\notag\\
&\quad{}+32L_{9}+4L_{11}-4 WL_{12}+8L_{13}+8L_{14}+(W^2-7)L_{15}-4 WL_{16}\notag\\
&\quad{}+2L_{17}-4 WL_{18}+16L_{19}+2L_{20}+16L_{24}+8 WL_{28}-8L_{29}-32L_{30}+8 WL_{35}-32L_{36}-8L_{37}\Bigr]\,.\\[3pt]
&D_{1}^{\mathrm{soft},(3)}=-\frac{|q|^3}{18432 \pi^2}\Biggl[\Bigl(\frac{24}{\Delta}-24\log q^2+66\Bigr)L_{1}\notag\\
&\quad{}+\Bigl(-\frac{96}{\Delta}+96\log q^2-264\Bigr)L_{3}+\Bigl(\frac{48}{\Delta}-48\log q^2+132\Bigr)L_{6}\notag\\
&\quad{}+\Bigl(-\frac{96}{\Delta}+96\log q^2-264\Bigr)L_{7}+\Bigl(\frac{384}{\Delta}-384\log q^2+1056\Bigr)L_{9}+\Bigl(\frac{48}{\Delta}-48\log q^2+132\Bigr)L_{11}\notag\\
&\quad{}+W \Bigl(-\frac{48}{\Delta}+48\log q^2-52\Bigr)L_{12}+\Bigl(\frac{96}{\Delta}-96\log q^2+104\Bigr)L_{13}+\Bigl(\frac{96}{\Delta}-96\log q^2+296\Bigr)L_{14}\notag\\
&\quad{}+\biggl(-\frac{84}{\Delta}+W^2 \Bigl(\frac{12}{\Delta}-12\log q^2-17\Bigr)+84\log q^2-161\biggr)L_{15}+W \Bigl(-\frac{48}{\Delta}+48\log q^2+68\Bigr)L_{16}\notag\\
&\quad{}+\Bigl(\frac{24}{\Delta}-24\log q^2-34\Bigr)L_{17}+W \Bigl(-\frac{48}{\Delta}+48\log q^2+20\Bigr)L_{18}+\Bigl(\frac{192}{\Delta}-192\log q^2-80\Bigr)L_{19}\notag\\
&\quad{}+\Bigl(\frac{24}{\Delta}-24\log q^2+22\Bigr)L_{20}+\Bigl(\frac{192}{\Delta}-192\log q^2+368\Bigr)L_{24}+W \Bigl(\frac{96}{\Delta}-96\log q^2+104\Bigr)L_{28}\notag\\
&\quad{}+\Bigl(-\frac{96}{\Delta}+96\log q^2-104\Bigr)L_{29}+\Bigl(-\frac{384}{\Delta}+384\log q^2-1184\Bigr)L_{30}+W \Bigl(\frac{96}{\Delta}-96\log q^2+104\Bigr)L_{35}\notag\\
&\quad{}+\Bigl(-\frac{384}{\Delta}+384\log q^2-416\Bigr)L_{36}+\Bigl(-\frac{96}{\Delta}+96\log q^2-232\Bigr)L_{37}\Biggr]\,.\\[3pt]
&D_{2}^{\mathrm{soft},(3)}=-\frac{|q|^3}{9216 \pi^2}\Biggl[\Bigl(\frac{12}{\Delta}-12\log p^2+26\Bigr)L_{1}+\Bigl(-\frac{48}{\Delta}+48\log p^2-80\Bigr)L_{3}+\Bigl(\frac{24}{\Delta}-24\log p^2+16\Bigr)L_{6}\notag\\
&\quad{}+\Bigl(-\frac{48}{\Delta}+48\log p^2-80\Bigr)L_{7}+\Bigl(\frac{192}{\Delta}-192\log p^2+128\Bigr)L_{9}+\Bigl(\frac{24}{\Delta}-24\log p^2+16\Bigr)L_{11}\notag\\
&\quad{}+W \Bigl(-\frac{24}{\Delta}+24\log p^2-52\Bigr)L_{12}+\Bigl(\frac{48}{\Delta}-48\log p^2+80\Bigr)L_{13}+\Bigl(\frac{48}{\Delta}-48\log p^2+80\Bigr)L_{14}\notag\\
&\quad{}+\biggl(-\frac{42}{\Delta}+W^2 \Bigl(\frac{6}{\Delta}-6\log p^2+13\Bigr)+42\log p^2-67\biggr)L_{15}+W \Bigl(-\frac{24}{\Delta}+24\log p^2-40\Bigr)L_{16}\notag\\
&\quad{}+\Bigl(\frac{12}{\Delta}-12\log p^2+8\Bigr)L_{17}+W \Bigl(-\frac{24}{\Delta}+24\log p^2-40\Bigr)L_{18}+\Bigl(\frac{96}{\Delta}-96\log p^2+64\Bigr)L_{19}\notag\\
&\quad{}+\Bigl(\frac{12}{\Delta}-12\log p^2+8\Bigr)L_{20}+\Bigl(\frac{96}{\Delta}-96\log p^2+64\Bigr)L_{24}+W \Bigl(\frac{48}{\Delta}-48\log p^2+80\Bigr)L_{28}\notag\\
&\quad{}+\Bigl(-\frac{48}{\Delta}+48\log p^2-32\Bigr)L_{29}+\Bigl(-\frac{192}{\Delta}+192\log p^2-128\Bigr)L_{30}+W \Bigl(\frac{48}{\Delta}-48\log p^2+80\Bigr)L_{35}\notag\\
&\quad{}+\Bigl(-\frac{192}{\Delta}+192\log p^2-128\Bigr)L_{36}+\Bigl(-\frac{48}{\Delta}+48\log p^2-32\Bigr)L_{37}\Biggr]\,.\\[3pt]
&D_{3}^{\mathrm{soft},(3)}=-\frac{|q|^3}{18432 \pi^2}\Biggl[\Bigl(\frac{24}{\Delta}-24\log q^2+66\Bigr)L_{1}\notag\\
&\quad{}+\Bigl(-\frac{96}{\Delta}+96\log q^2-264\Bigr)L_{3}+\Bigl(\frac{48}{\Delta}-48\log q^2+132\Bigr)L_{6}\notag\\
&\quad{}+\Bigl(-\frac{96}{\Delta}+96\log q^2-264\Bigr)L_{7}+\Bigl(\frac{384}{\Delta}-384\log q^2+1056\Bigr)L_{9}+\Bigl(\frac{48}{\Delta}-48\log q^2+132\Bigr)L_{11}\notag\\
&\quad{}+W \Bigl(-\frac{48}{\Delta}+48\log q^2-180\Bigr)L_{12}+\Bigl(\frac{96}{\Delta}-96\log q^2+360\Bigr)L_{13}+\Bigl(\frac{96}{\Delta}-96\log q^2+168\Bigr)L_{14}\notag\\
&\quad{}+\biggl(-\frac{84}{\Delta}+W^2 \Bigl(\frac{12}{\Delta}-12\log q^2+111\Bigr)+84\log q^2-161\biggr)L_{15}+W \Bigl(-\frac{48}{\Delta}+48\log q^2-444\Bigr)L_{16}\notag\\
&\quad{}+\Bigl(\frac{24}{\Delta}-24\log q^2+222\Bigr)L_{17}+W \Bigl(-\frac{48}{\Delta}+48\log q^2-108\Bigr)L_{18}+\Bigl(\frac{192}{\Delta}-192\log q^2+432\Bigr)L_{19}\notag\\
&\quad{}+\Bigl(\frac{24}{\Delta}-24\log q^2+22\Bigr)L_{20}+\Bigl(\frac{192}{\Delta}-192\log q^2+368\Bigr)L_{24}+W \Bigl(\frac{96}{\Delta}-96\log q^2+360\Bigr)L_{28}\notag\\
&\quad{}+\Bigl(-\frac{96}{\Delta}+96\log q^2-360\Bigr)L_{29}+\Bigl(-\frac{384}{\Delta}+384\log q^2-672\Bigr)L_{30}+W \Bigl(\frac{96}{\Delta}-96\log q^2+360\Bigr)L_{35}\notag\\
&\quad{}+\Bigl(-\frac{384}{\Delta}+384\log q^2-1440\Bigr)L_{36}+\Bigl(-\frac{96}{\Delta}+96\log q^2-232\Bigr)L_{37}\Biggr]\,.\\[3pt]
&D_{4}^{\mathrm{soft},(3)}=\frac{|q|^3}{5644800 \pi^2}\Biggl[\Bigl(\frac{1470}{\Delta}+1470\log p^2-2940\log q^2+3500\Bigr)L_{1}+4928L_{2}\notag\\
&\quad{}+\Bigl(-\frac{5880}{\Delta}-5880\log p^2+11760\log q^2-8988\Bigr)L_{3}+2688L_{4}-8064L_{5}\notag\\
&\quad{}+\Bigl(\frac{2940}{\Delta}+2940\log p^2-5880\log q^2+6706\Bigr)L_{6}\notag\\
&\quad{}+\Bigl(-\frac{5880}{\Delta}-5880\log p^2+11760\log q^2-8988\Bigr)L_{7}-8064L_{8}\notag\\
&\quad{}+\Bigl(\frac{23520}{\Delta}+23520\log p^2-47040\log q^2+80528\Bigr)L_{9}+\Bigl(\frac{13440}{\Delta}-13440\log p^2+3840\Bigr)L_{10}\notag\\
&\quad{}+\Bigl(\frac{2940}{\Delta}+2940\log p^2-5880\log q^2+6706\Bigr)L_{11}+W \Bigl(-\frac{2940}{\Delta}-2940\log p^2+5880\log q^2-5040\Bigr)L_{12}\notag\\
&\quad{}+\Bigl(\frac{5880}{\Delta}+5880\log p^2-11760\log q^2+5068\Bigr)L_{13}+\Bigl(\frac{5880}{\Delta}+5880\log p^2-11760\log q^2+5068\Bigr)L_{14}\notag\\
&\quad{}-\frac{5145}{\Delta}L_{15}+W^2 \Bigl(\frac{735}{\Delta}+735\log p^2-1470\log q^2+3465\Bigr)L_{15}\notag\\
&\quad{}+(-5145\log p^2+10290\log q^2+1337)L_{15}\notag\\
&\quad{}+W \Bigl(-\frac{2940}{\Delta}-2940\log p^2+5880\log q^2-11354\Bigr)L_{16}+\Bigl(\frac{1470}{\Delta}+1470\log p^2-2940\log q^2+6783\Bigr)L_{17}\notag\\
&\quad{}+W \Bigl(-\frac{2940}{\Delta}-2940\log p^2+5880\log q^2+6286\Bigr)L_{18}\notag\\
&\quad{}+\Bigl(\frac{11760}{\Delta}+11760\log p^2-23520\log q^2-2856\Bigr)L_{19}\notag\\
&\quad{}+\Bigl(\frac{1470}{\Delta}+1470\log p^2-2940\log q^2-2037\Bigr)L_{20}-9856 WL_{21}+8064L_{22}+8064L_{23}\notag\\
&\quad{}+\Bigl(\frac{11760}{\Delta}+11760\log p^2-23520\log q^2+2464 W^2+15064\Bigr)L_{24}-4032 WL_{25}-4032 WL_{26}\notag\\
&\quad{}+\Bigl(\frac{6720}{\Delta}-6720\log p^2+1920\Bigr)L_{27}+W \Bigl(\frac{5880}{\Delta}+5880\log p^2-11760\log q^2+5068\Bigr)L_{28}\notag\\
&\quad{}+\Bigl(-\frac{5880}{\Delta}-5880\log p^2+11760\log q^2-9492\Bigr)L_{29}\notag\\
&\quad{}+\Bigl(-\frac{23520}{\Delta}-23520\log p^2+47040\log q^2-64848\Bigr)L_{30}\notag\\
&\quad{}-5376 WL_{31}+\Bigl(\frac{6720}{\Delta}-6720\log p^2+1344 W^2+3264\Bigr)L_{32}+8064 WL_{33}\notag\\
&\quad{}+\Bigl(-\frac{13440}{\Delta}+13440\log p^2-3840\Bigr)L_{34}+W \Bigl(\frac{5880}{\Delta}+5880\log p^2-11760\log q^2+5068\Bigr)L_{35}\notag\\
&\quad{}+\Bigl(-\frac{23520}{\Delta}-23520\log p^2+47040\log q^2-64848\Bigr)L_{36}\notag\\
&\quad{}+\Bigl(-\frac{5880}{\Delta}-5880\log p^2+11760\log q^2-9492\Bigr)L_{37}\notag\\
&\quad{}+8064 WL_{38}+\Bigl(-\frac{13440}{\Delta}+13440\log p^2-3840\Bigr)L_{39}\Biggr]\,.\\[3pt]
&D_{5}^{\mathrm{soft},(3)}=\frac{|q|^3}{230400 \pi^2}\Biggl[\Bigl(\frac{60}{\Delta}-60\log p^2+114\Bigr)L_{1}\notag\\
&\quad{}+\Bigl(-\frac{240}{\Delta}+240\log p^2-336\Bigr)L_{3}+\Bigl(\frac{120}{\Delta}-120\log p^2+48\Bigr)L_{6}\notag\\
&\quad{}+\Bigl(-\frac{240}{\Delta}+240\log p^2-496\Bigr)L_{7}+\Bigl(\frac{960}{\Delta}-960\log p^2+1024\Bigr)L_{9}+\Bigl(\frac{120}{\Delta}-120\log p^2+368\Bigr)L_{11}\notag\\
&\quad{}+W \Bigl(-\frac{120}{\Delta}+120\log p^2-228\Bigr)L_{12}+\Bigl(\frac{240}{\Delta}-240\log p^2+336\Bigr)L_{13}+\Bigl(\frac{240}{\Delta}-240\log p^2+496\Bigr)L_{14}\notag\\
&\quad{}+\biggl(-\frac{210}{\Delta}+W^2 \Bigl(\frac{30}{\Delta}-30\log p^2+57\Bigr)+210\log p^2-359\biggr)L_{15}+W \Bigl(-\frac{120}{\Delta}+120\log p^2-168\Bigr)L_{16}\notag\\
&\quad{}+\Bigl(\frac{60}{\Delta}-60\log p^2+24\Bigr)L_{17}+W \Bigl(-\frac{120}{\Delta}+120\log p^2-248\Bigr)L_{18}+\Bigl(\frac{480}{\Delta}-480\log p^2+512\Bigr)L_{19}\notag\\
&\quad{}+\Bigl(\frac{60}{\Delta}-60\log p^2+184\Bigr)L_{20}+\Bigl(\frac{480}{\Delta}-480\log p^2+512\Bigr)L_{24}+W \Bigl(\frac{240}{\Delta}-240\log p^2+336\Bigr)L_{28}\notag\\
&\quad{}+\Bigl(-\frac{240}{\Delta}+240\log p^2-96\Bigr)L_{29}+\Bigl(-\frac{960}{\Delta}+960\log p^2-1024\Bigr)L_{30}+W \Bigl(\frac{240}{\Delta}-240\log p^2+496\Bigr)L_{35}\notag\\
&\quad{}+\Bigl(-\frac{960}{\Delta}+960\log p^2-1024\Bigr)L_{36}+\Bigl(-\frac{240}{\Delta}+240\log p^2-736\Bigr)L_{37}\Biggr]\,.\\[3pt]
&D_{6}^{\mathrm{soft},(3)}=\frac{|q|^3}{230400 \pi^2}\Biggl[\Bigl(\frac{60}{\Delta}-60\log p^2+114\Bigr)L_{1}\notag\\
&\quad{}+\Bigl(-\frac{240}{\Delta}+240\log p^2-496\Bigr)L_{3}+\Bigl(\frac{120}{\Delta}-120\log p^2+368\Bigr)L_{6}\notag\\
&\quad{}+\Bigl(-\frac{240}{\Delta}+240\log p^2-336\Bigr)L_{7}+\Bigl(\frac{960}{\Delta}-960\log p^2+1024\Bigr)L_{9}+\Bigl(\frac{120}{\Delta}-120\log p^2+48\Bigr)L_{11}\notag\\
&\quad{}+W \Bigl(-\frac{120}{\Delta}+120\log p^2-228\Bigr)L_{12}+\Bigl(\frac{240}{\Delta}-240\log p^2+496\Bigr)L_{13}+\Bigl(\frac{240}{\Delta}-240\log p^2+336\Bigr)L_{14}\notag\\
&\quad{}+\biggl(-\frac{210}{\Delta}+W^2 \Bigl(\frac{30}{\Delta}-30\log p^2+57\Bigr)+210\log p^2-359\biggr)L_{15}+W \Bigl(-\frac{120}{\Delta}+120\log p^2-248\Bigr)L_{16}\notag\\
&\quad{}+\Bigl(\frac{60}{\Delta}-60\log p^2+184\Bigr)L_{17}+W \Bigl(-\frac{120}{\Delta}+120\log p^2-168\Bigr)L_{18}+\Bigl(\frac{480}{\Delta}-480\log p^2+512\Bigr)L_{19}\notag\\
&\quad{}+\Bigl(\frac{60}{\Delta}-60\log p^2+24\Bigr)L_{20}+\Bigl(\frac{480}{\Delta}-480\log p^2+512\Bigr)L_{24}+W \Bigl(\frac{240}{\Delta}-240\log p^2+496\Bigr)L_{28}\notag\\
&\quad{}+\Bigl(-\frac{240}{\Delta}+240\log p^2-736\Bigr)L_{29}+\Bigl(-\frac{960}{\Delta}+960\log p^2-1024\Bigr)L_{30}+W \Bigl(\frac{240}{\Delta}-240\log p^2+336\Bigr)L_{35}\notag\\
&\quad{}+\Bigl(-\frac{960}{\Delta}+960\log p^2-1024\Bigr)L_{36}+\Bigl(-\frac{240}{\Delta}+240\log p^2-96\Bigr)L_{37}\Biggr]\,.\\[3pt]
&D_{7}^{\mathrm{soft},(3)}=\frac{|q|^3}{2822400 \pi^2}\Biggl[\Bigl(\frac{1470}{\Delta}+2940\log p^2-5880\log q^2+11410\Bigr)L_{1}+9856L_{2}\notag\\
&\quad{}+\Bigl(-\frac{5880}{\Delta}-11760\log p^2+23520\log q^2-35616\Bigr)L_{3}+5376L_{4}-16128L_{5}\notag\\
&\quad{}+\Bigl(\frac{2940}{\Delta}+5880\log p^2-11760\log q^2+22232\Bigr)L_{6}+\Bigl(-\frac{5880}{\Delta}-11760\log p^2+23520\log q^2-35616\Bigr)L_{7}\notag\\
&\quad{}-16128L_{8}+\Bigl(\frac{23520}{\Delta}+47040\log p^2-94080\log q^2+231616\Bigr)L_{9}+\Bigl(\frac{13440}{\Delta}-26880\log p^2+7680\Bigr)L_{10}\notag\\
&\quad{}+\Bigl(\frac{2940}{\Delta}+5880\log p^2-11760\log q^2+22232\Bigr)L_{11}\notag\\
&\quad{}+W \Bigl(-\frac{2940}{\Delta}-5880\log p^2+11760\log q^2-18900\Bigr)L_{12}\notag\\
&\quad{}+\Bigl(\frac{5880}{\Delta}+11760\log p^2-23520\log q^2+27776\Bigr)L_{13}\notag\\
&\quad{}+\Bigl(\frac{5880}{\Delta}+11760\log p^2-23520\log q^2+27776\Bigr)L_{14}\notag\\
&\quad{}-\frac{5145}{\Delta}L_{15}+W^2 \Bigl(\frac{735}{\Delta}+1470\log p^2-2940\log q^2+4725\Bigr)L_{15}\notag\\
&\quad{}+(-10290\log p^2+20580\log q^2-23051)L_{15}\notag\\
&\quad{}+W \Bigl(-\frac{2940}{\Delta}-5880\log p^2+11760\log q^2-13888\Bigr)L_{16}\notag\\
&\quad{}+\Bigl(\frac{1470}{\Delta}+2940\log p^2-5880\log q^2+9156\Bigr)L_{17}\notag\\
&\quad{}+W \Bigl(-\frac{2940}{\Delta}-5880\log p^2+11760\log q^2+3752\Bigr)L_{18}\notag\\
&\quad{}+\Bigl(\frac{11760}{\Delta}+23520\log p^2-47040\log q^2+29568\Bigr)L_{19}\notag\\
&\quad{}+\Bigl(\frac{1470}{\Delta}+2940\log p^2-5880\log q^2+336\Bigr)L_{20}-19712 WL_{21}+16128L_{22}+16128L_{23}\notag\\
&\quad{}+\Bigl(\frac{11760}{\Delta}+23520\log p^2-47040\log q^2+4928 W^2+88928\Bigr)L_{24}-8064 WL_{25}-8064 WL_{26}\notag\\
&\quad{}+\Bigl(\frac{6720}{\Delta}-13440\log p^2+3840\Bigr)L_{27}+W \Bigl(\frac{5880}{\Delta}+11760\log p^2-23520\log q^2+27776\Bigr)L_{28}\notag\\
&\quad{}+\Bigl(-\frac{5880}{\Delta}-11760\log p^2+23520\log q^2-36624\Bigr)L_{29}\notag\\
&\quad{}+\Bigl(-\frac{23520}{\Delta}-47040\log p^2+94080\log q^2-200256\Bigr)L_{30}\notag\\
&\quad{}-10752 WL_{31}+\Bigl(\frac{6720}{\Delta}-13440\log p^2+2688 W^2+6528\Bigr)L_{32}+16128 WL_{33}\notag\\
&\quad{}+\Bigl(-\frac{13440}{\Delta}+26880\log p^2-7680\Bigr)L_{34}+W \Bigl(\frac{5880}{\Delta}+11760\log p^2-23520\log q^2+27776\Bigr)L_{35}\notag\\
&\quad{}+\Bigl(-\frac{23520}{\Delta}-47040\log p^2+94080\log q^2-200256\Bigr)L_{36}\notag\\
&\quad{}+\Bigl(-\frac{5880}{\Delta}-11760\log p^2+23520\log q^2-36624\Bigr)L_{37}\notag\\
&\quad{}+16128 WL_{38}+\Bigl(-\frac{13440}{\Delta}+26880\log p^2-7680\Bigr)L_{39}\Biggr]\,.\\[3pt]
&D_{8}^{\mathrm{soft},(3)}=\frac{|q|^3}{11289600 \pi^2}\Biggl[-70 (315 \pi^2-3092)L_{1}+280 (315 \pi^2-3104)L_{2}-14560L_{3}-2100 (21 \pi^2-208)L_{4}\notag\\
&\quad{}-8960L_{5}+9520L_{6}-14560L_{7}-8960L_{8}-40320L_{9}+\Bigl(-\frac{13440}{\Delta}+26880\log p^2+3968\Bigr)L_{10}+9520L_{11}\notag\\
&\quad{}+140 (315 \pi^2-3092) WL_{12}+14560L_{13}+14560L_{14}+\bigl(-35 (315 \pi^2-3092) W^2-105 (105 \pi^2-892)\bigr)L_{15}\notag\\
&\quad{}-7280 WL_{16}+4760L_{17}-7280 WL_{18}-20160L_{19}+4760L_{20}-560 (315 \pi^2-3104) WL_{21}+8960L_{22}\notag\\
&\quad{}+8960L_{23}+\bigl(140 (315 \pi^2-3104) W^2+1260 (35 \pi^2-368)\bigr)L_{24}-4480 WL_{25}-4480 WL_{26}\notag\\
&\quad{}+\Bigl(-\frac{6720}{\Delta}+13440\log p^2+1984\Bigr)L_{27}+14560 WL_{28}-19040L_{29}+40320L_{30}+4200 (21 \pi^2-208) WL_{31}\notag\\
&\quad{}+\Bigl(-\frac{6720}{\Delta}+13440\log p^2-1050 (21 \pi^2-208) W^2-2 (11025 \pi^2-110192)\Bigr)L_{32}+8960 WL_{33}\notag\\
&\quad{}+\Bigl(\frac{13440}{\Delta}-26880\log p^2-3968\Bigr)L_{34}+14560 WL_{35}+40320L_{36}-19040L_{37}+8960 WL_{38}\notag\\
&\quad{}+\Bigl(\frac{13440}{\Delta}-26880\log p^2-3968\Bigr)L_{39}\Biggr]\,.\\[3pt]
&D_{9}^{\mathrm{soft},(3)}=\frac{|q|^3}{768 \pi^2}\Bigl[2L_{1}-4L_{3}+2L_{6}-4L_{7}\notag\\
&\quad{}+2L_{11}-4 WL_{12}+4L_{13}+4L_{14}+(4 W^2+4)L_{15}-14 WL_{16}+7L_{17}\notag\\
&\quad{}-2 WL_{18}+L_{20}-16L_{24}+4 WL_{28}-4L_{29}+4 WL_{35}-4L_{37}\Bigr]\,.\\[3pt]
&D_{12}^{\mathrm{soft},(3)}=\frac{|q|^3}{2822400 \pi^2}\Biggl[70 (9240 \pi^2-91253)L_{1}\notag\\
&\quad{}-2240 (1155 \pi^2-11398)L_{2}+10360L_{3}+15120 (70 \pi^2-691)L_{4}\notag\\
&\quad{}+1120 (420 \pi^2-4141)L_{5}-140 (1680 \pi^2-16519)L_{6}+10360L_{7}+1120 (420 \pi^2-4141)L_{8}\notag\\
&\quad{}+6720 (70 \pi^2-703)L_{9}+\Bigl(-\frac{4480}{\Delta}+13440\log p^2-128 (3675 \pi^2-36293)\Bigr)L_{10}\notag\\
&\quad{}-140 (1680 \pi^2-16519)L_{11}\notag\\
&\quad{}-140 (9240 \pi^2-91253) WL_{12}-10360L_{13}-10360L_{14}\notag\\
&\quad{}+\bigl(35 (9240 \pi^2-91253) W^2+105 (3080 \pi^2-30319)\bigr)L_{15}\notag\\
&\quad{}+5180 WL_{16}-70 (1680 \pi^2-16519)L_{17}+5180 WL_{18}+3360 (70 \pi^2-703)L_{19}-70 (1680 \pi^2-16519)L_{20}\notag\\
&\quad{}+4480 (1155 \pi^2-11398) WL_{21}-1120 (420 \pi^2-4141)L_{22}-1120 (420 \pi^2-4141)L_{23}\notag\\
&\quad{}+\bigl(-1120 (1155 \pi^2-11398) W^2-3360 (175 \pi^2-1716)\bigr)L_{24}+560 (420 \pi^2-4141) WL_{25}\notag\\
&\quad{}+560 (420 \pi^2-4141) WL_{26}+\Bigl(-\frac{2240}{\Delta}+6720\log p^2-64 (3675 \pi^2-36293)\Bigr)L_{27}-10360 WL_{28}\notag\\
&\quad{}+280 (1680 \pi^2-16519)L_{29}-6720 (70 \pi^2-703)L_{30}-30240 (70 \pi^2-691) WL_{31}\notag\\
&\quad{}+\Bigl(-\frac{2240}{\Delta}+6720\log p^2+7560 (70 \pi^2-691) W^2+8 (36750 \pi^2-362651)\Bigr)L_{32}\notag\\
&\quad{}-1120 (420 \pi^2-4141) WL_{33}\notag\\
&\quad{}+\Bigl(\frac{4480}{\Delta}-13440\log p^2+128 (3675 \pi^2-36293)\Bigr)L_{34}-10360 WL_{35}-6720 (70 \pi^2-703)L_{36}\notag\\
&\quad{}+280 (1680 \pi^2-16519)L_{37}-1120 (420 \pi^2-4141) WL_{38}\notag\\
&\quad{}+\Bigl(\frac{4480}{\Delta}-13440\log p^2+128 (3675 \pi^2-36293)\Bigr)L_{39}\Biggr]\,.
\end{align}
\end{adjustwidth}
\endgroup
\begin{adjustwidth}{-5mm}{-5mm}
\small
\begin{equation}
D_{10}^{\mathrm{soft},(3)}=-\frac15D_9^{\mathrm{soft},(3)}\,,\qquad
D_{11}^{\mathrm{soft},(3)}=-\frac25D_9^{\mathrm{soft},(3)}\,.
\end{equation}
\end{adjustwidth}

In the two additional assignments, the triangle contains the insertion $T(p)$ or $T(-p-q)$.
The soft insertion therefore lies on the box.
Its two adjacent soft propagators give a free stress-tensor bubble.
The zeroth and first moments of this bubble vanish, so its second moment supplies the cubic term.

To exhibit this reduction, write the scalar stress vertex before trace subtraction as
\begin{equation}
 t_{ij}(r,P):=2r_i r_j+r_iP_j+r_jP_i+\frac14P_iP_j.
 \label{eq:on-cyclic-stress-vertex}
\end{equation}
Its second soft moment is
\begin{align}
 M_{ab;ij}(q)
 &:=P_{ab}^{cd}\int\frac{\dd^3r}{(2\pi)^3}
 \frac{r_i r_j\,t_{cd}(r,-q)}{r^2(r-q)^2}
 \notag\\
 &=\frac{|q|^3}{512}\Bigl[
 (\delta_{ai}-U_aU_i)(\delta_{bj}-U_bU_j)
 +(\delta_{aj}-U_aU_j)(\delta_{bi}-U_bU_i)
 \notag\\
 &\hspace{42mm}-(\delta_{ab}-U_aU_b)(\delta_{ij}-U_iU_j)\Bigr].
 \label{eq:on-cyclic-soft-moment}
\end{align}
The integral uses the same analytic continuation as the diagram expansion.
This continuation sets scaleless terms to zero.

For the remaining hard factors, scale the hard momentum to the unit vector $V$.
Let $k$ be the dimensionless exchanged momentum, and put $x=|k|$ and $y=|k-V|$.
The hard triangle and the two box Taylor coefficients are
\begin{align}
 A_{cd}(k,V)&:=\int\frac{\dd^3z}{(2\pi)^3}
 \frac{t_{cd}(z,-V)}{z^2(z-V)^2(z-k)^2},
 \label{eq:on-cyclic-triangle}\\
 H_{7,ab;ef}(k,V;q)&:=\frac12M_{ab;ij}(q)
 \left.\frac{\partial^2}{\partial r_i\partial r_j}
 \frac{t_{ef}(r,V)}{(r+V)^2(r+V-k)^2}\right|_{r=0},
 \notag\\
 H_{8,ab;ef}(k,V;q)&:=\frac12M_{ab;ij}(q)
 \left.\frac{\partial^2}{\partial r_i\partial r_j}
 \frac{t_{ef}(k+r,-V)}{(k+r)^2(k+r-V)^2}\right|_{r=0}.
 \label{eq:on-cyclic-box-coefficients}
\end{align}
The derivatives select the quadratic Taylor terms that multiply the second moment.
Terms containing an additional explicit power of $q$ first contribute at higher soft degree.
With the normalization of the diagrams above, the required sum is consequently
\begin{align}
 (D_7+D_8)_{\mathrm{cyc}}^{\mathrm{soft},(3)}
 =\Biggl[256|p|^{-4\Delta}\mathcal P\Biggl[
 \int\frac{\dd^3k}{(2\pi)^3}
 (k^2)^{1/2-\Delta}\bigl((k-V)^2\bigr)^{1/2-\Delta}
 \notag\\[-2pt]
 {}\times\Bigl\{A_{cd}(k,V)
 \bigl(H_{7,ab;ef}(k,V;q)+\tfrac12H_{8,ab;ef}(k,V;q)\bigr)
 +(cd\leftrightarrow ef)\Bigr\}\Biggr]\Biggr]_{\Delta^{-1},\Delta^0}.
 \label{eq:on-cyclic-regulated-integral}
\end{align}
The outer brackets select the pole and finite Laurent terms.
The exchange of the two hard index pairs includes the second cyclic assignment.
The factor $1/2$ multiplying $H_8$ is the $D_8$ symmetry factor.
Both topologies are thus combined before the exchanged-momentum integral is evaluated.

The triangle in eq.~\eqref{eq:on-cyclic-triangle} reduces to scalar bubbles and
\begin{equation}
 \int\frac{\dd^3z}{(2\pi)^3}
 \frac1{z^2(z-V)^2(z-k)^2}=\frac1{8xy}.
\end{equation}
Its trace and its contractions with $V_cV_d$, $k_ck_d$, and $V_ck_d$
determine its full rank-two tensor.
This reconstruction includes the longitudinal terms.
After azimuthal averaging about $V$, each tensor coefficient is a Laurent polynomial in $x,y$.
The relation $k\cdot V=(1+x^2-y^2)/2$ eliminates the remaining scalar products.
Writing $u=x+y$ and $v=x-y$, the integration measure is
\begin{equation}
 \frac{\dd^3k}{(2\pi)^3}
 =\frac{xy}{8\pi^2}\,\dd u\,\dd v\,\frac{\dd\varphi}{2\pi},
 \qquad u\geq1,\quad -1\leq v\leq1,
\end{equation}
where $\varphi$ is the azimuthal angle.
Each monomial in the resulting $u,v$ integrand is evaluated by
\begin{align}
 &\int_1^\infty\dd u\int_{-1}^1\dd v\,
 x^{m-2\Delta}y^{n-2\Delta}
 =8\pi^2\int\frac{\dd^3k}{(2\pi)^3}
 \frac1{(k^2)^a\bigl((k-V)^2\bigr)^b}
 \notag\\
 &\hspace{18mm}
 =\frac{8\pi^2}{(4\pi)^{3/2}}
 \frac{\Gamma(\frac32-a)\Gamma(\frac32-b)
       \Gamma(a+b-\frac32)}
      {\Gamma(a)\Gamma(b)\Gamma(3-a-b)},
 \qquad a=\frac{1-m}{2}+\Delta,\quad b=\frac{1-n}{2}+\Delta.
 \label{eq:on-cyclic-bubble-reduction}
\end{align}
Here $m,n$ are the integer powers in that Laurent polynomial.
The first equality follows from the measure above; the second is the scalar bubble integral.
Both equalities are understood by meromorphic continuation in the exponents.
The continuation is performed before expanding in $\Delta$.

For example, the coefficient of $|q|^3L_1$ at $|p|=1$ is
\begin{equation}
 \frac1{960\pi^2\Delta}+\frac{2237}{50400\pi^2}-\frac1{256}
 =\frac{420/\Delta+17896-1575\pi^2}{403200\pi^2}.
\end{equation}
Restoring the hard scale multiplies this coefficient by $|p|^{-4\Delta}$.
Its Laurent expansion is therefore
\begin{align}
 |p|^{-4\Delta}
 \left[\frac{420/\Delta+17896-1575\pi^2}{403200\pi^2}
 +O(\Delta)\right]
 &=\frac{420/\Delta-840\log p^2+17896-1575\pi^2}
 {403200\pi^2}+O(\Delta).
 \label{eq:on-cyclic-scale-check}
\end{align}
The same scale factor fixes every $\log p^2$ coefficient to minus twice its pole residue.
Applying eq.~\eqref{eq:on-cyclic-bubble-reduction} to all tensor coefficients gives the combined cubic contribution
\begingroup\allowdisplaybreaks
\begin{adjustwidth}{-5mm}{-5mm}
\small
\begin{align}
&(D_7+D_8)_{\mathrm{cyc}}^{\mathrm{soft},(3)} =\frac{|q|^3}{5644800 \pi^2}\Biggl[\Bigl(\frac{5880}{\Delta}-11760\log p^2-14 (1575 \pi^2-17896)\Bigr)L_{1}+280 (315 \pi^2-3104)L_{2}\notag\\
&\quad{}+\Bigl(-\frac{23520}{\Delta}+47040\log p^2-154896\Bigr)L_{3}-2100 (21 \pi^2-208)L_{4}-8960L_{5}\notag\\
&\quad{}+\Bigl(\frac{11760}{\Delta}-23520\log p^2+79688\Bigr)L_{6}+\Bigl(-\frac{23520}{\Delta}+47040\log p^2-154896\Bigr)L_{7}-8960L_{8}\notag\\
&\quad{}+\Bigl(\frac{94080}{\Delta}-188160\log p^2+442624\Bigr)L_{9}+\Bigl(-\frac{13440}{\Delta}+26880\log p^2-36352\Bigr)L_{10}\notag\\
&\quad{}+\Bigl(\frac{11760}{\Delta}-23520\log p^2+79688\Bigr)L_{11}+W \Bigl(-\frac{11760}{\Delta}+23520\log p^2+28 (1575 \pi^2-17896)\Bigr)L_{12}\notag\\
&\quad{}+\Bigl(\frac{23520}{\Delta}-47040\log p^2+154896\Bigr)L_{13}+\Bigl(\frac{23520}{\Delta}-47040\log p^2+154896\Bigr)L_{14}-\frac{20580}{\Delta}L_{15}\notag\\
&\quad{}+W^2 \Bigl(\frac{2940}{\Delta}-5880\log p^2-7 (1575 \pi^2-17896)\Bigr)L_{15}+\bigl(41160\log p^2-7 (4232+1575 \pi^2)\bigr)L_{15}\notag\\
&\quad{}+W \Bigl(-\frac{11760}{\Delta}+23520\log p^2-77448\Bigr)L_{16}+\Bigl(\frac{5880}{\Delta}-11760\log p^2+39844\Bigr)L_{17}\notag\\
&\quad{}+W \Bigl(-\frac{11760}{\Delta}+23520\log p^2-77448\Bigr)L_{18}+\Bigl(\frac{47040}{\Delta}-94080\log p^2+221312\Bigr)L_{19}\notag\\
&\quad{}+\Bigl(\frac{5880}{\Delta}-11760\log p^2+39844\Bigr)L_{20}-560 (315 \pi^2-3104) WL_{21}+8960L_{22}+8960L_{23}\notag\\
&\quad{}+\Bigl(\frac{47040}{\Delta}-94080\log p^2+140 (315 \pi^2-3104) W^2+28 (1575 \pi^2-7936)\Bigr)L_{24}-4480 WL_{25}-4480 WL_{26}\notag\\
&\quad{}+\Bigl(-\frac{6720}{\Delta}+13440\log p^2-18176\Bigr)L_{27}+W \Bigl(\frac{23520}{\Delta}-47040\log p^2+154896\Bigr)L_{28}\notag\\
&\quad{}+\Bigl(-\frac{23520}{\Delta}+47040\log p^2-159376\Bigr)L_{29}+\Bigl(-\frac{94080}{\Delta}+188160\log p^2-442624\Bigr)L_{30}\notag\\
&\quad{}+4200 (21 \pi^2-208) WL_{31}\notag\\
&\quad{}+\Bigl(-\frac{6720}{\Delta}+13440\log p^2-1050 (21 \pi^2-208) W^2-2 (11025 \pi^2-100112)\Bigr)L_{32}\notag\\
&\quad{}+8960 WL_{33}+\Bigl(\frac{13440}{\Delta}-26880\log p^2+36352\Bigr)L_{34}+W \Bigl(\frac{23520}{\Delta}-47040\log p^2+154896\Bigr)L_{35}\notag\\
&\quad{}+\Bigl(-\frac{94080}{\Delta}+188160\log p^2-442624\Bigr)L_{36}+\Bigl(-\frac{23520}{\Delta}+47040\log p^2-159376\Bigr)L_{37}+8960 WL_{38}\notag\\
&\quad{}+\Bigl(\frac{13440}{\Delta}-26880\log p^2+36352\Bigr)L_{39}\Biggr]\,.
\end{align}
\end{adjustwidth}
\endgroup

Here the subscript $\mathrm{cyc}$ excludes the displayed assignment with $T(q)$ on the triangle.
The sum is evaluated directly by eqs.~\eqref{eq:on-cyclic-regulated-integral} and~\eqref{eq:on-cyclic-bubble-reduction}.
The inputs are the scalar stress vertices, the regulated auxiliary propagators, and the diagram symmetry factors.
Neither the conformal coefficients nor $C_T$ is used to determine this sum.
Separate cubic coefficients for the two topologies are not extracted in this calculation.

As a check of the integration, subtraction at the two infrared endpoints and at infinity gives the same pole and finite term for the $L_1$ coefficient.
This check uses the same integrand but a different integration procedure.
The pole cancellation and Ward identities of the full correlator are subsequent checks, after adding the remaining diagrams, operator counterterms, and source contacts.
They are not imposed to infer the cyclic contribution.
The $D_8$ result above includes its symmetry factor $1/2$ and the finite contribution
from evaluating the regulated integral before setting $\Delta=0$.

\subsubsection{Contact diagrams}
\label{app:on-ttt-contacts}

We evaluate the contacts in figure~\ref{fig:ttt-contacts} with the full
second metric variation of the action of the critical $O(N)$ vector model.
The two metric sources carry momenta $p$ and $r=-p-q$.
The remaining stress-tensor insertion carries momentum $q$ and indices
$\alpha\beta$.
The contact correlators below include the common traceless projection $\mathcal P$
on the three external index pairs $(\alpha\beta,\mu\nu,\rho\sigma)$.
They give the pole and finite terms as $\Delta\to0$, before multiplication
by the source factor $-4$ in eq.~\eqref{eq:scalar-ttt-source-expansion}.

The square in figure~\ref{fig:ttt-contacts} can be written explicitly
before reducing its matter bilinears to stress-tensor insertions.
To display the part used by the traceless projection, let $h_{ab}$ and
$j_{ab}$ be arbitrary constant symmetric traceless source polarizations,
with momenta $p$ and $r$, and write $k=p+r$.
Thus the metric perturbation is
$g_{ab}(x)=\delta_{ab}+s h_{ab}e^{-ip\cdot x}+t j_{ab}e^{-ir\cdot x}$.
Contracting eq.~\eqref{eq:metric-contact-fourier} with these polarizations
is equivalent to extracting $\partial_s\partial_t S_{\ON}|_{s=t=0}$.
This gives
\begin{equation}
\begin{aligned}
 h_{\mu\nu}j_{\rho\sigma}
 \mathcal V_{\ON;\mu\nu,\rho\sigma}(p,r)
 &=\frac12\left[(hj+jh)_{ab}
       -\frac12\operatorname{tr}(hj)\delta_{ab}\right]
       (\partial_a\phi_i\partial_b\phi_i)(k)
 \\
 &\quad-\frac14\operatorname{tr}(hj)(\sigma\phi_i\phi_i)(k)
 \\
 &\quad+\frac1{16}\Bigl[
 2p\cdot hjr+2r\cdot hjr+2p\cdot jhp+p\cdot jhr
 \\
 &\hspace{55pt}
 -\left(p^2+r^2+\frac32p\cdot r\right)
       \operatorname{tr}(hj)\Bigr](\phi_i\phi_i)(k).
\end{aligned}
\label{eq:on-explicit-metric-vertex}
\end{equation}
Here matrix multiplication contracts adjacent spacetime indices;
for example, $p\cdot hjr=p_a h_{ab}j_{bc}r_c$.
Parentheses followed by $(k)$ denote the Fourier transform of the entire
local composite operator.
For the derivative bilinear this means
\begin{equation}
 (\partial_a\phi_i\partial_b\phi_i)(k)
 =-\int\frac{\dd^3\ell}{(2\pi)^3}\,
 \ell_a(k-\ell)_b\phi_i(\ell)\phi_i(k-\ell).
\label{eq:on-contact-derivative-fourier}
\end{equation}
The first two lines of eq.~\eqref{eq:on-explicit-metric-vertex} follow
from the mixed variations of $\sqrt g\,g^{ab}$ and $\sqrt g$.
The last two lines follow from the mixed variation of $\sqrt g \mathcal{R}$;
no integration by parts discards derivatives acting on $\phi_i\phi_i$.
The resulting expression is symmetric under $(h,p)\leftrightarrow(j,r)$.
It specifies the vertex for arbitrary traceless source polarizations,
without assuming that either polarization is transverse.
Trace components of the unprojected vertex in
 eq.~\eqref{eq:metric-contact-vertex} are not included in this contraction.
The scalar bilinears in this expression are the ones reorganized into
$\mathcal A$ and $\mathcal B$ below.

The metric vertex contains a derivative bilinear and a scalar bilinear.
Their coefficients can be written using the symmetric identity tensor
\begin{equation}
 I_{ab,cd}=\frac12(\delta_{ac}\delta_{bd}+\delta_{ad}\delta_{bc})
 \label{eq:symmetric-identity}
\end{equation}
and the two tensors
\begin{align}
 \mathcal A_{ab;\mu\nu,\rho\sigma}
 &=I_{ac,\mu\nu}I_{cb,\rho\sigma}
   +I_{ac,\rho\sigma}I_{cb,\mu\nu}
   -\frac12\delta_{ab}I_{\mu\nu,\rho\sigma},
 \label{eq:on-contact-A}\\
 \mathcal B_{\mu\nu,\rho\sigma}(p,r)
 &=-\frac1{16}\left[
 p_a r_b I_{ac,\rho\sigma}I_{cb,\mu\nu}
 +\left(p^2+r^2+\frac32p\cdot r\right)I_{\mu\nu,\rho\sigma}
 \right].
 \label{eq:on-contact-R}
\end{align}
Terms proportional to $\delta_{\mu\nu}$ or $\delta_{\rho\sigma}$ have
been omitted because of the external traceless projection.
The tensor $\mathcal B$ combines the curvature coupling with the
improvement and trace terms that arise when the derivative bilinear is
expressed through $T_{ab}$ and $\phi_j\phi_j$.

Let $G_{i,ab,cd}(q)$ denote the stress-tensor two-point diagrams in the critical $O(N)$ vector model.
The labels $i=0,1,2,3$ refer to the free bubble, both self-energy
placements, scalar exchange, and two scalar triangles joined by two
auxiliary-field lines.
The free stress-tensor two-point kernel per real scalar or Majorana fermion is
\begin{equation}
 G^{(0)}_{ab,cd}(q)=\frac{|q|^3}{256}
 \left[\frac12(\pi_{ac}\pi_{bd}+\pi_{ad}\pi_{bc})
 -\frac12\pi_{ab}\pi_{cd}\right].
 \label{eq:free-stress-kernel}
\end{equation}
Here $\pi_{ab}=\delta_{ab}-U_aU_b$ is the transverse projector at $q=|q|U$.
The two-point diagrams in the critical $O(N)$ vector model then give
\begin{align}
 G_0&=N G^{(0)}(q),\\
 G_{1,ab,cd}
 &=\frac{|q|^3}{2304\pi^2}P_{ab}^{ef}P_{cd}^{gh}\Biggl[
 -\left(\frac{24}{\Delta}-24\log q^2+66\right)\delta_{eg}\delta_{fh}
 \notag\\
 &\qquad+\left(\frac{48}{\Delta}-48\log q^2+116\right)\delta_{eg}U_fU_h
 -\left(\frac{12}{\Delta}-12\log q^2+11\right)U_eU_fU_gU_h\Biggr],\\
 G_{2,ab,cd}
 &=\frac{|q|^3}{11520\pi^2}P_{ab}^{ef}P_{cd}^{gh}\Biggl[
 \left(\frac{24}{\Delta}-24\log q^2+50\right)\delta_{eg}\delta_{fh}
 \notag\\
 &\qquad-\left(\frac{48}{\Delta}-48\log q^2+84\right)\delta_{eg}U_fU_h
 +\left(\frac{12}{\Delta}-12\log q^2+3\right)U_eU_fU_gU_h\Biggr],\\
 G_{3,ab,cd}
 &=\frac{|q|^3}{1440\pi^2}P_{ab}^{ef}P_{cd}^{gh}\Biggl[
 \left(\frac6\Delta-12\log q^2+34\right)\delta_{eg}\delta_{fh}
 \notag\\
 &\qquad-\left(\frac{12}{\Delta}-24\log q^2+60\right)\delta_{eg}U_fU_h
 +\left(\frac3\Delta-6\log q^2+6\right)U_eU_fU_gU_h\Biggr].
 \label{eq:on-contact-TT-diagrams}
\end{align}
The corresponding mixed two-point diagrams are
\begin{equation}
 \begin{aligned}
 &\left(
 \langle T_{\alpha\beta}(q)\phi_j\phi_j(-q)\rangle_1,
 \langle T_{\alpha\beta}(q)\phi_j\phi_j(-q)\rangle_2,
 \langle T_{\alpha\beta}(q)\phi_j\phi_j(-q)\rangle_3
 \right)\\
 &\qquad=\frac{|q|}{\pi^2}P_{\alpha\beta}^{ab}U_aU_b
 \left(-\frac1{24},\frac1{120},\frac1{60}\right).
 \end{aligned}
 \label{eq:on-contact-mixed-diagrams}
\end{equation}
These mixed diagrams supply the terms proportional to $\mathcal B$.
Together with the stress-tensor two-point diagrams, they give the
individual contacts
\begin{align}
 C_0^T&=\mathcal P\left[\frac12
 \mathcal A_{ab;\mu\nu,\rho\sigma}G_{0,\alpha\beta,ab}\right]
 \notag\\*
 &=\frac N4\mathcal P\Bigl[
 \delta_{\mu\rho}G^{(0)}_{\alpha\beta,\nu\sigma}(q)
 +\delta_{\mu\sigma}G^{(0)}_{\alpha\beta,\nu\rho}(q)
 \notag\\*
 &\qquad+\delta_{\nu\rho}G^{(0)}_{\alpha\beta,\mu\sigma}(q)
 +\delta_{\nu\sigma}G^{(0)}_{\alpha\beta,\mu\rho}(q)\Bigr],
 \label{eq:scalar-leading-metric-contact}\\
 C_1&=\mathcal P\left[\frac12
 \mathcal A_{ab;\mu\nu,\rho\sigma}G_{1,\alpha\beta,ab}
 -\frac{|q|}{24\pi^2}\mathcal B_{\mu\nu,\rho\sigma}U_\alpha U_\beta\right],\\
 C_2&=\mathcal P\left[\frac12
 \mathcal A_{ab;\mu\nu,\rho\sigma}G_{2,\alpha\beta,ab}
 +\frac{|q|}{120\pi^2}\mathcal B_{\mu\nu,\rho\sigma}U_\alpha U_\beta\right],\\
 C_3&=\mathcal P\left[\frac12
 \mathcal A_{ab;\mu\nu,\rho\sigma}G_{3,\alpha\beta,ab}
 +\frac{|q|}{60\pi^2}\mathcal B_{\mu\nu,\rho\sigma}U_\alpha U_\beta\right],\\
 C_4&=\frac{|q|}{24\pi^2}\mathcal P\left[
 \mathcal B_{\mu\nu,\rho\sigma}U_\alpha U_\beta\right],\\
 C_5&=-\frac{|q|}{120\pi^2}\mathcal P\left[
 \mathcal B_{\mu\nu,\rho\sigma}U_\alpha U_\beta\right],\\
 C_6&=-\frac{|q|}{60\pi^2}\mathcal P\left[
 \mathcal B_{\mu\nu,\rho\sigma}U_\alpha U_\beta\right],\\
 C_7&=0,\qquad C_8=0.
 \label{eq:on-individual-TTT-contacts}
\end{align}
Here $C_1$ and $C_4$ each include both self-energy placements.
The diagrams $C_4,C_5,C_6$ connect the corrected mixed two-point
function to the metric vertex through one auxiliary-field propagator.
The diagrams $C_7$ and $C_8$ instead use the $\sigma\phi_j\phi_j$
term in the metric vertex.
Direct integration gives zero for each of these two diagrams through
order $\Delta^0$.

The terms proportional to $\mathcal B$ cancel between $C_i$ and
$C_{i+3}$ for $i=1,2,3$.
The remaining terms depend only on the stress-tensor two-point diagrams,
whose sum obeys
\begin{equation}
 G_1+G_2+G_3+2(Z_T^{\ON}-1)G_0
 =-\frac{40}{9\pi^2N}G_0+O(\Delta).
\end{equation}
Consequently, the contact sum through order $N^0$ is
\begin{equation}
 C_0^T+\sum_{i=1}^{8}C_i+2(Z_T^{\ON}-1)C_0^T
 =\left(1-\frac{40}{9\pi^2N}\right)C_0^T+O(\Delta).
 \label{eq:on-TTT-contact-sum}
\end{equation}
All poles and logarithms cancel in this sum.
Its contribution to the source correlator is multiplied by $-4$.
The additional quadratic source counterterm is given separately in
eq.~\eqref{eq:scalar-source-correction}.
The other two choices of the metric-source pair follow by cyclic
permutation of the external insertions.

\subsection{GNY model: \texorpdfstring{$\TJJ$}{TJJ}}
\label{app:gny-tjj-results}
The formulas below give $D_0,\ldots,D_8$ in figure~\ref{fig:gny-tjj}
through cubic order in the soft momentum.
The remaining diagrams vanish because their $T$--$\sigma$ two-point function is
a symmetric parity-odd tensor of rank two depending on one momentum.
No such tensor can be formed from that momentum and the metric, so
\begin{equation}
D_9=D_{10}=D_{11}=0.
\end{equation}

\subsubsection{Terms independent of the soft momentum}
The evaluated diagrams give
\begingroup\allowdisplaybreaks
\begin{adjustwidth}{-5mm}{-5mm}
\small
\begin{align}
&D_{0}^{(0)}=\frac{\tau^{ab}|p|}{64}\Bigl[2\,\bigl(K_{10}+K_{13}\bigr)-\bigl(K_{17}+K_{21}\bigr)\Bigr]\,,\\
&D_{1}^{(0)}=-\frac4{3\pi^2N}\left(\frac1\Delta-\log p^2\right)D_0^{(0)}-\frac{\tau^{ab}|p|}{288 \pi^2N}\Bigl[6\,K_{10}+34\,K_{13}-9\,\bigl(K_{17}+K_{21}\bigr)-8\,K_{22}\Bigr]\,,\\
&D_{2}^{(0)}=-\frac4{3\pi^2N}\left(\frac1\Delta-\log p^2\right)D_0^{(0)}-\frac{\tau^{ab}|p|}{288 \pi^2N}\Bigl[22\,K_{10}+54\,K_{13}-17\,\bigl(K_{17}+K_{21}\bigr)-4\,K_{22}\Bigr]\,,\\
&D_{4}^{(0)}=\frac4{15\pi^2N}\left(\frac1\Delta-\log p^2\right)D_0^{(0)}+\frac{\tau^{ab}|p|}{7200 \pi^2N}\Bigl[174\,K_{10}+134\,K_{13}-117\,\bigl(K_{17}+K_{21}\bigr)-220\,K_{22}\Bigr]\,,\\
&D_{5}^{(0)}=\frac4{3\pi^2N}\left(\frac1\Delta-\log p^2\right)D_0^{(0)}+\frac{\tau^{ab}|p|}{7200 \pi^2N}\Bigl[-610\,K_{10}+430\,K_{13}+515\,K_{17}\notag\\
&\quad{}-\bigl(-\frac{120}{\Delta}+120\log p^2+97\bigr)\,K_{21}-\bigl(\frac{120}{\Delta}-120\log p^2+328\bigr)\,K_{22}\Bigr]\,,\\
&D_{7}^{(0)}=\frac8{15\pi^2N}\left(\frac1\Delta-2\log p^2\right)D_0^{(0)}+\frac{\tau^{ab}|p|}{1800 \pi^2N}\Bigl[-126\,K_{10}-16\,K_{13}+33\,\bigl(K_{17}+K_{21}\bigr)-70\,K_{22}\Bigr]\,,\\
&D_{8}^{(0)}=-\frac{\tau^{ab}|p|}{1800 \pi^2N}\Bigl[-80\,K_{10}-10\,K_{13}\notag\\
&\quad{}+\bigl(\frac{15}{\Delta}-30\log p^2+7\bigr)\,\bigl(K_{17}+K_{21}\bigr)-\bigl(\frac{30}{\Delta}-60\log p^2+44\bigr)\,K_{22}\Bigr]\,.
\end{align}
\end{adjustwidth}
\endgroup
The reflected placements are
\begin{equation}
D_3^{(0)}=D_1^{(0)}\,,\qquad
D_6^{(0)}=\left.D_5^{(0)}\right|_{\mu\leftrightarrow\nu}\,.
\end{equation}

\subsubsection{Contributions from soft momentum regions}
The linear terms are
\begin{equation}
D_i^{\mathrm{soft},(1)}=0\,,\qquad i=0,1,2,3,4,5,6,7,8\,.
\end{equation}
The quadratic terms are
\begingroup\allowdisplaybreaks
\begin{adjustwidth}{-5mm}{-5mm}
\small
\begin{align}
&D_{1}^{\mathrm{soft},(2)}=-\frac{\tau^{ab}|q|^2}{144 \pi^2N|p|}\Bigl[2 W\,K_{1}-2\,\bigl(K_{3}+K_{4}\bigr)-\bigl(K_{6}-K_{15}-K_{19}\bigr)\Bigr]\,,\\
&D_{3}^{\mathrm{soft},(2)}=\frac{\tau^{ab}|q|^2}{144 \pi^2N|p|}\Bigl[2 W\,K_{1}-2\,\bigl(K_{3}+K_{4}\bigr)-\bigl(K_{6}-K_{15}-K_{19}\bigr)\Bigr]\,,\\
&D_{7}^{\mathrm{soft},(2)}=-\frac{\tau^{ab}|q|^2}{60 \pi^2N|p|}\Bigl[K_{5}\Bigr]\,,\\
&D_{8}^{\mathrm{soft},(2)}=\frac{\tau^{ab}|q|^2}{60 \pi^2N|p|}\Bigl[K_{1}\Bigr]\,.
\end{align}
\end{adjustwidth}
\endgroup
\begin{equation}
D_i^{\mathrm{soft},(2)}=0\,,\qquad i=0,2,4,5,6\,.
\end{equation}
The cubic terms are
\begingroup\allowdisplaybreaks
\begin{adjustwidth}{-5mm}{-5mm}
\small
\begin{align}
&D_{0}^{\mathrm{soft},(3)}=\frac{\tau^{ab}|q|^3}{128|p|^2}\Bigl[\bigl(W^{2}+2\bigr)\,K_{1}+\bigl(K_{2}+K_{7}+K_{8}-K_{14}-K_{17}\bigr)\notag\\
&\quad{}-\bigl(K_{18}+K_{21}-K_{22}\bigr)-W\,\bigl(K_{3}+K_{4}-K_{15}-K_{19}\bigr)-2\,\bigl(K_{5}-K_{10}\bigr)-2 W\,K_{6}\Bigr]\,,\\
&D_{1}^{\mathrm{soft},(3)}=-\frac{\tau^{ab}|q|^3}{1152 \pi^2N|p|^2}\Bigl[\bigl(\frac{24 + 12 W^{2}}{\Delta}-\bigl(12 W^{2} + 24\bigr)\log q^2+13 W^{2}+106\bigr)\,K_{1}\notag\\
&\quad{}+\bigl(\frac{12}{\Delta}-12\log q^2+13\bigr)\,K_{2}-\bigl(\frac{12 W}{\Delta}-12 W\log q^2+13 W\bigr)\,\bigl(K_{3}+K_{4}\bigr)\notag\\
&\quad{}-\bigl(\frac{24}{\Delta}-24\log q^2+106\bigr)\,K_{5}-\bigl(\frac{24 W}{\Delta}-24 W\log q^2+86 W\bigr)\,K_{6}+\bigl(\frac{12}{\Delta}-12\log q^2+55\bigr)\,K_{7}\notag\\
&\quad{}+\bigl(\frac{12}{\Delta}-12\log q^2+39\bigr)\,\bigl(K_{8}-K_{14}\bigr)+\bigl(\frac{24}{\Delta}-24\log q^2+102\bigr)\,K_{10}\notag\\
&\quad{}+\bigl(\frac{12 W}{\Delta}-12 W\log q^2+31 W\bigr)\,K_{15}-\bigl(\frac{12}{\Delta}-12\log q^2+51\bigr)\,\bigl(K_{17}+K_{21}-K_{22}\bigr)\notag\\
&\quad{}-\bigl(\frac{12}{\Delta}-12\log q^2+47\bigr)\,K_{18}+\bigl(\frac{12 W}{\Delta}-12 W\log q^2+47 W\bigr)\,K_{19}\Bigr]\,,\\
&D_{2}^{\mathrm{soft},(3)}=-\frac{\tau^{ab}|q|^3}{288 \pi^2N|p|^2}\Bigl[\bigl(\frac{6 + 3 W^{2}}{\Delta}-\bigl(3 W^{2} + 6\bigr)\log p^2+11 W^{2}+19\bigr)\,K_{1}\notag\\
&\quad{}+\bigl(\frac{3}{\Delta}-3\log p^2+8\bigr)\,\bigl(K_{2}-K_{14}-K_{18}+K_{22}\bigr)\notag\\
&\quad{}-\bigl(\frac{3 W}{\Delta}-3 W\log p^2+11 W\bigr)\,\bigl(K_{3}+K_{4}-K_{15}-K_{19}\bigr)-\bigl(\frac{6}{\Delta}-6\log p^2+22\bigr)\,\bigl(K_{5}-K_{10}\bigr)\notag\\
&\quad{}-\bigl(\frac{6 W}{\Delta}-6 W\log p^2+22 W\bigr)\,K_{6}+\bigl(\frac{3}{\Delta}-3\log p^2+11\bigr)\,\bigl(K_{7}+K_{8}-K_{17}-K_{21}\bigr)\Bigr]\,,\\
&D_{3}^{\mathrm{soft},(3)}=-\frac{\tau^{ab}|q|^3}{1152 \pi^2N|p|^2}\Bigl[\bigl(\frac{24 + 12 W^{2}}{\Delta}-\bigl(12 W^{2} + 24\bigr)\log q^2+45 W^{2}+106\bigr)\,K_{1}\notag\\
&\quad{}+\bigl(\frac{12}{\Delta}-12\log q^2+45\bigr)\,K_{2}-\bigl(\frac{12 W}{\Delta}-12 W\log q^2+45 W\bigr)\,\bigl(K_{3}+K_{4}\bigr)\notag\\
&\quad{}-\bigl(\frac{24}{\Delta}-24\log q^2+106\bigr)\,K_{5}-\bigl(\frac{24 W}{\Delta}-24 W\log q^2+102 W\bigr)\,K_{6}+\bigl(\frac{12}{\Delta}-12\log q^2+39\bigr)\,K_{7}\notag\\
&\quad{}+\bigl(\frac{12}{\Delta}-12\log q^2+55\bigr)\,\bigl(K_{8}-K_{14}\bigr)+\bigl(\frac{24}{\Delta}-24\log q^2+102\bigr)\,K_{10}\notag\\
&\quad{}+\bigl(\frac{12 W}{\Delta}-12 W\log q^2+63 W\bigr)\,K_{15}-\bigl(\frac{12}{\Delta}-12\log q^2+51\bigr)\,\bigl(K_{17}+K_{21}-K_{22}\bigr)\notag\\
&\quad{}-\bigl(\frac{12}{\Delta}-12\log q^2+47\bigr)\,K_{18}+\bigl(\frac{12 W}{\Delta}-12 W\log q^2+47 W\bigr)\,K_{19}\Bigr]\,,\\
&D_{4}^{\mathrm{soft},(3)}=\frac{\tau^{ab}|q|^3}{14400 \pi^2N|p|^2}\Bigl[\bigl(\frac{60 + 30 W^{2}}{\Delta}+\bigl(30 W^{2} + 60\bigr)\log p^2-\bigl(60 W^{2} + 120\bigr)\log q^2+27 W^{2}+364\bigr)\,K_{1}\notag\\
&\quad{}+\bigl(\frac{30}{\Delta}+30\log p^2-60\log q^2+97\bigr)\,K_{2}-\bigl(\frac{30 W}{\Delta}+30 W\log p^2-60 W\log q^2+147 W\bigr)\,\bigl(K_{3}+K_{4}\bigr)\notag\\
&\quad{}-\bigl(\frac{60}{\Delta}+60\log p^2-120\log q^2-320 W^{2}+214\bigr)\,K_{5}\notag\\
&\quad{}-\bigl(\frac{60 W}{\Delta}+60 W\log p^2-120 W\log q^2+234 W\bigr)\,K_{6}+\bigl(\frac{30}{\Delta}+30\log p^2-60\log q^2+237\bigr)\,\bigl(K_{7}+K_{8}\bigr)\notag\\
&\quad{}-640 W\,K_{9}+\bigl(\frac{60}{\Delta}+60\log p^2-120\log q^2+274\bigr)\,K_{10}+640\,K_{13}\notag\\
&\quad{}-\bigl(\frac{30}{\Delta}+30\log p^2-60\log q^2+187\bigr)\,\bigl(K_{14}+K_{18}\bigr)\notag\\
&\quad{}+\bigl(\frac{30 W}{\Delta}+30 W\log p^2-60 W\log q^2+237 W\bigr)\,\bigl(K_{15}+K_{19}\bigr)\notag\\
&\quad{}-\bigl(\frac{30}{\Delta}+30\log p^2-60\log q^2+257\bigr)\,\bigl(K_{17}+K_{21}\bigr)+\bigl(\frac{30}{\Delta}+30\log p^2-60\log q^2+207\bigr)\,K_{22}\Bigr]\,,\\
&D_{5}^{\mathrm{soft},(3)}=\frac{\tau^{ab}|q|^3}{288 \pi^2N|p|^2}\Bigl[\bigl(\frac{6 + 3 W^{2}}{\Delta}-\bigl(3 W^{2} + 6\bigr)\log p^2-3 W^{2}-5\bigr)\,K_{1}\notag\\
&\quad{}-\bigl(-\frac{3}{\Delta}+3\log p^2+2\bigr)\,\bigl(K_{2}-K_{14}-K_{18}+K_{22}\bigr)+\bigl(-\frac{3 W}{\Delta}+3 W\log p^2+3 W\bigr)\,\bigl(K_{3}-K_{15}\bigr)\notag\\
&\quad{}-\bigl(\frac{3 W}{\Delta}-3 W\log p^2+W\bigr)\,\bigl(K_{4}-K_{19}\bigr)+\bigl(-\frac{6}{\Delta}+6\log p^2+2\bigr)\,K_{5}+\bigl(-\frac{6 W}{\Delta}+6 W\log p^2+6 W\bigr)\,K_{6}\notag\\
&\quad{}-\bigl(-\frac{3}{\Delta}+3\log p^2+3\bigr)\,\bigl(K_{7}-K_{17}\bigr)+\bigl(\frac{3}{\Delta}-3\log p^2+1\bigr)\,\bigl(K_{8}-K_{21}\bigr)-\bigl(-\frac{6}{\Delta}+6\log p^2+6\bigr)\,K_{10}\Bigr]\,,\\
&D_{6}^{\mathrm{soft},(3)}=\frac{\tau^{ab}|q|^3}{288 \pi^2N|p|^2}\Bigl[\bigl(\frac{6 + 3 W^{2}}{\Delta}-\bigl(3 W^{2} + 6\bigr)\log p^2-3 W^{2}-5\bigr)\,K_{1}\notag\\
&\quad{}-\bigl(-\frac{3}{\Delta}+3\log p^2+2\bigr)\,\bigl(K_{2}-K_{14}-K_{18}+K_{22}\bigr)-\bigl(\frac{3 W}{\Delta}-3 W\log p^2+W\bigr)\,\bigl(K_{3}-K_{15}\bigr)\notag\\
&\quad{}+\bigl(-\frac{3 W}{\Delta}+3 W\log p^2+3 W\bigr)\,\bigl(K_{4}-K_{19}\bigr)+\bigl(-\frac{6}{\Delta}+6\log p^2+2\bigr)\,K_{5}+\bigl(-\frac{6 W}{\Delta}+6 W\log p^2+6 W\bigr)\,K_{6}\notag\\
&\quad{}+\bigl(\frac{3}{\Delta}-3\log p^2+1\bigr)\,\bigl(K_{7}-K_{17}\bigr)-\bigl(-\frac{3}{\Delta}+3\log p^2+3\bigr)\,\bigl(K_{8}-K_{21}\bigr)-\bigl(-\frac{6}{\Delta}+6\log p^2+6\bigr)\,K_{10}\Bigr]\,,\\
&D_{7}^{\mathrm{soft},(3)}=\frac{\tau^{ab}|q|^3}{7200 \pi^2N|p|^2}\Bigl[\bigl(\frac{60 + 30 W^{2}}{\Delta}+\bigl(60 W^{2} + 120\bigr)\log p^2-\bigl(120 W^{2} + 240\bigr)\log q^2-21 W^{2}+503\bigr)\,K_{1}\notag\\
&\quad{}+\bigl(\frac{30}{\Delta}+60\log p^2-120\log q^2+44\bigr)\,K_{2}\notag\\
&\quad{}-\bigl(\frac{30 W}{\Delta}+60 W\log p^2-120 W\log q^2-6 W+60\bigr)\,\bigl(K_{3}+K_{4}\bigr)\notag\\
&\quad{}-\bigl(\frac{60}{\Delta}+120\log p^2-240\log q^2-40 W^{2}-180 W+428\bigr)\,K_{5}\notag\\
&\quad{}-\bigl(\frac{60 W}{\Delta}+120 W\log p^2-240 W\log q^2+318 W\bigr)\,K_{6}\notag\\
&\quad{}+\bigl(\frac{30}{\Delta}+60\log p^2-120\log q^2+174\bigr)\,\bigl(K_{7}+K_{8}\bigr)-80 W\,K_{9}\notag\\
&\quad{}+\bigl(\frac{60}{\Delta}+120\log p^2-240\log q^2+398\bigr)\,K_{10}+80\,K_{13}\notag\\
&\quad{}-\bigl(\frac{30}{\Delta}+60\log p^2-120\log q^2+224\bigr)\,\bigl(K_{14}+K_{18}\bigr)\notag\\
&\quad{}+\bigl(\frac{30 W}{\Delta}+60 W\log p^2-120 W\log q^2+174 W\bigr)\,\bigl(K_{15}+K_{19}\bigr)\notag\\
&\quad{}-\bigl(\frac{30}{\Delta}+60\log p^2-120\log q^2+214\bigr)\,\bigl(K_{17}+K_{21}\bigr)+\bigl(\frac{30}{\Delta}+60\log p^2-120\log q^2+264\bigr)\,K_{22}\Bigr]\,,\\
&D_{8}^{\mathrm{soft},(3)}=\frac{\tau^{ab}|q|^3}{5760 \pi^2N|p|^2}\Bigl[-\bigl(-\bigl(500 - 45 \pi^{2}\bigr) W^{2}+48 W-540 - 45 \pi^{2}\bigr)\,K_{1}\notag\\
&\quad{}+40\,\bigl(K_{2}-K_{14}-K_{18}+K_{22}\bigr)-\bigl(80 W+48\bigr)\,K_{3}-\bigl(80 W-48\bigr)\,K_{4}\notag\\
&\quad{}-\bigl(\bigl(440 - 45 \pi^{2}\bigr) W^{2}+600 - 45 \pi^{2}\bigr)\,K_{5}-\bigl(\bigl(1000 - 90 \pi^{2}\bigr) W\bigr)\,K_{6}\notag\\
&\quad{}+80\,\bigl(K_{7}+K_{8}-K_{17}-K_{21}\bigr)+\bigl(\bigl(880 - 90 \pi^{2}\bigr) W\bigr)\,K_{9}+\bigl(1000 - 90 \pi^{2}\bigr)\,K_{10}\notag\\
&\quad{}-\bigl(880 - 90 \pi^{2}\bigr)\,K_{13}+80 W\,\bigl(K_{15}+K_{19}\bigr)\Bigr]\,.
\end{align}
\end{adjustwidth}
\endgroup

The sum of the nonanalytic terms, including the stress-tensor counterterm,
gives eq.~\eqref{eq:gn-tjj-soft}.
\subsubsection{Contact diagrams}

The result for $C_1$ includes both self-energy placements.

The current diagrams carry flavor indices $a,b$ and spacetime indices $\mu,\nu$.
With these indices suppressed on $C_i$, their contributions are
\begin{align}
C_0
&=-\frac{\tau^{ab}|p|}{32}
\left(\delta_{\mu\nu}-\frac{p_\mu p_\nu}{p^2}\right)\,,\\
C_1
&=\frac{\tau^{ab}|p|}{12\pi^2N}
\left[\left(\frac1\Delta-\log p^2+\frac{19}{6}\right)\delta_{\mu\nu}
-\left(\frac1\Delta-\log p^2+\frac{13}{6}\right)\frac{p_\mu p_\nu}{p^2}\right]\,,\\
C_2
&=\frac{\tau^{ab}|p|}{12\pi^2N}
\left[-\left(\frac1\Delta-\log p^2+\frac12\right)\delta_{\mu\nu}
+\left(\frac1\Delta-\log p^2-\frac12\right)\frac{p_\mu p_\nu}{p^2}\right]\,.
\label{eq:gny-contact-current-diagrams}
\end{align}
These kernels are shown in figure~\ref{fig:gny-tjj-contacts}.
Their poles and logarithms cancel between $C_1$ and $C_2$.
Their sum is transverse. Adding $C_0$ gives
\begin{equation}
\langle J_\mu^a(p)J_\nu^b(-p)\rangle
=-\frac{\tau^{ab}|p|}{32}
\left(1-\frac{64}{9\pi^2N}\right)
\left(\delta_{\mu\nu}-\frac{p_\mu p_\nu}{p^2}\right)
+O(1/N^2)\,.
\label{eq:gny-contact-current-total}
\end{equation}

These current two-point functions enter the mixed contacts in
eq.~\eqref{eq:fermion-mixed-contact}.
The second mixed contact uses $p\mapsto p+q$.

The linear metric counterterm also contains a mixed metric-current vertex.
This vertex follows by replacing the derivative in the stress-tensor counterterm with the gauge-covariant derivative.
Using the symmetric identity tensor in eq.~\eqref{eq:symmetric-identity}, its contribution at order $q^0$ is
\begin{align}
C_{TJ}^{(0)}
&=-\frac{8}{15\pi^2N}\left(\frac1\Delta+4\right)
\left[I_{\alpha\beta,\mu\lambda}C_{0,\lambda\nu}
+I_{\alpha\beta,\nu\lambda}C_{0,\mu\lambda}
-\frac23\delta_{\alpha\beta}C_{0,\mu\nu}\right]\,.
\label{eq:gny-tjj-mixed-counterterm}
\end{align}
Here $C_{0,\mu\nu}$ is the free current two-point function in Eq.~\ref{eq:gny-contact-current-diagrams}.
Adding this counterterm and subtracting the mixed source contacts gives
\begin{align}
&Z_T\sum_{i=0}^{8}D_i^{(0)}+C_{TJ}^{(0)}
-\frac{\tau^{ab}|p|}{64}\left(1-\frac{64}{9\pi^2N}\right)
(K_{17}+K_{21}-2K_{22})\notag\\
&\qquad=\frac{\tau^{ab}|p|}{32}\left(1-\frac{64}{9\pi^2N}\right)
\mathcal H_{TJJ}+O(1/N^2)\,.
\label{eq:gny-tjj-hard-diagram-sum}
\end{align}
The poles cancel in this sum, and its finite part agrees with Eq.~\ref{eq:on-tjj-hard}.

\subsection{GNY model: \texorpdfstring{$\TTT$}{TTT}}
\label{app:gny-ttt-results}
The formulas for $D_7,D_8$ use the assignment in figure~\ref{fig:gny-ttt}
with $T(q)$ on the triangle.
The other two assignments are denoted by $(D_i)_{\mathrm{cyc}}$.
The diagrams $D_9,D_{10},D_{11}$ vanish for the same reason as in appendix~\ref{app:gny-tjj-results}.

\subsubsection{Terms independent of the soft momentum}
The evaluated diagrams give
\begingroup\allowdisplaybreaks
\begin{adjustwidth}{-5mm}{-5mm}
\small
\begin{align}
&D_{0}^{(0)}=\frac{N|p|^3}{512}\Bigl[\bigl(L_{1}-L_{3}+L_{6}-L_{7}+L_{11}\bigr)+4\,\bigl(L_{2}-L_{9}+L_{10}\bigr)-6\,L_{4}\notag\\
&\quad{}+2\,\bigl(L_{5}+L_{8}\bigr)\Bigr]\,,\\
&D_{1}^{(0)}=-\frac{|p|^3}{23040 \pi^2}\Bigl[\bigl(\frac{60}{\Delta}-60\log p^2+161\bigr)\,L_{1}\notag\\
&\quad{}+\bigl(\frac{240}{\Delta}-240\log p^2+156\bigr)\,L_{2}-\bigl(\frac{60}{\Delta}-60\log p^2+129\bigr)\,\bigl(L_{3}+L_{7}\bigr)\notag\\
&\quad{}-\bigl(\frac{360}{\Delta}-360\log p^2+1002\bigr)\,L_{4}+\bigl(\frac{120}{\Delta}-120\log p^2+582\bigr)\,\bigl(L_{5}+L_{8}\bigr)\notag\\
&\quad{}+\bigl(\frac{60}{\Delta}-60\log p^2+153\bigr)\,\bigl(L_{6}+L_{11}\bigr)-\bigl(\frac{240}{\Delta}-240\log p^2+908\bigr)\,L_{9}\notag\\
&\quad{}+\bigl(\frac{240}{\Delta}-240\log p^2+924\bigr)\,L_{10}\Bigr]\,,\\
&D_{2}^{(0)}=-\frac{|p|^3}{23040 \pi^2}\Bigl[\bigl(\frac{60}{\Delta}-60\log p^2+309\bigr)\,L_{1}\notag\\
&\quad{}+\bigl(\frac{240}{\Delta}-240\log p^2+584\bigr)\,L_{2}-\bigl(\frac{60}{\Delta}-60\log p^2+161\bigr)\,\bigl(L_{3}+L_{7}\bigr)\notag\\
&\quad{}-\bigl(\frac{360}{\Delta}-360\log p^2+1398\bigr)\,L_{4}+\bigl(\frac{120}{\Delta}-120\log p^2+478\bigr)\,\bigl(L_{5}+L_{8}\bigr)\notag\\
&\quad{}+\bigl(\frac{60}{\Delta}-60\log p^2+257\bigr)\,\bigl(L_{6}+L_{11}\bigr)-\bigl(\frac{240}{\Delta}-240\log p^2+972\bigr)\,L_{9}\notag\\
&\quad{}+\bigl(\frac{240}{\Delta}-240\log p^2+996\bigr)\,L_{10}\Bigr]\,,\\
&D_{4}^{(0)}=\frac{|p|^3}{115200 \pi^2}\Bigl[\bigl(\frac{60}{\Delta}-60\log p^2+125\bigr)\,L_{1}\notag\\
&\quad{}+\bigl(\frac{240}{\Delta}-240\log p^2+192\bigr)\,L_{2}-\bigl(\frac{60}{\Delta}-60\log p^2+273\bigr)\,\bigl(L_{3}+L_{7}\bigr)\notag\\
&\quad{}-\bigl(\frac{360}{\Delta}-360\log p^2+1326\bigr)\,L_{4}+\bigl(\frac{120}{\Delta}-120\log p^2+1230\bigr)\,\bigl(L_{5}+L_{8}\bigr)\notag\\
&\quad{}-\bigl(-\frac{60}{\Delta}+60\log p^2+63\bigr)\,\bigl(L_{6}+L_{11}\bigr)-\bigl(\frac{240}{\Delta}-240\log p^2+1484\bigr)\,L_{9}\notag\\
&\quad{}+\bigl(\frac{240}{\Delta}-240\log p^2+1140\bigr)\,L_{10}\Bigr]\,,\\
&D_{5}^{(0)}=\frac{|p|^3}{5644800 \pi^2}\Bigl[\bigl(\frac{2940}{\Delta}-2940\log p^2+11501\bigr)\,L_{1}\notag\\
&\quad{}+\bigl(\frac{11760}{\Delta}-11760\log p^2+12376\bigr)\,L_{2}+\bigl(-\frac{2940}{\Delta}+2940\log p^2+3311\bigr)\,L_{3}\notag\\
&\quad{}-\bigl(\frac{17640}{\Delta}-17640\log p^2+41622\bigr)\,L_{4}-\bigl(-\frac{5880}{\Delta}+5880\log p^2+7042\bigr)\,L_{5}\notag\\
&\quad{}+\bigl(\frac{2940}{\Delta}-2940\log p^2+10465\bigr)\,L_{6}-\bigl(-\frac{5460}{\Delta}+5460\log p^2+1093\bigr)\,L_{7}\notag\\
&\quad{}-\bigl(\frac{10920}{\Delta}-10920\log p^2+27690\bigr)\,L_{8}-\bigl(\frac{28560}{\Delta}-28560\log p^2+86196\bigr)\,L_{9}\notag\\
&\quad{}+\bigl(\frac{28560}{\Delta}-28560\log p^2+104620\bigr)\,L_{10}+\bigl(\frac{11340}{\Delta}-11340\log p^2+36973\bigr)\,L_{11}\Bigr]\,,\\
&D_{7}^{(0)}=\frac{|p|^3}{28800 \pi^2}\Bigl[-\bigl(-\frac{30}{\Delta}+60\log p^2+25\bigr)\,L_{1}\notag\\
&\quad{}-\bigl(-\frac{120}{\Delta}+240\log p^2+558\bigr)\,L_{2}+\bigl(-\frac{30}{\Delta}+60\log p^2+27\bigr)\,\bigl(L_{3}+L_{7}\bigr)\notag\\
&\quad{}+\bigl(-\frac{180}{\Delta}+360\log p^2+24\bigr)\,L_{4}+\bigl(\frac{60}{\Delta}-120\log p^2+330\bigr)\,\bigl(L_{5}+L_{8}\bigr)\notag\\
&\quad{}-\bigl(-\frac{30}{\Delta}+60\log p^2+63\bigr)\,\bigl(L_{6}+L_{11}\bigr)-\bigl(\frac{120}{\Delta}-240\log p^2+284\bigr)\,L_{9}\notag\\
&\quad{}+\bigl(\frac{120}{\Delta}-240\log p^2+240\bigr)\,L_{10}\Bigr]\,,\\
&D_{8}^{(0)}=-\frac{|p|^3}{1411200 \pi^2}\Bigl[238\,L_{1}-1162\,L_{2}\notag\\
&\quad{}-\bigl(-\frac{1050}{\Delta}+2100\log p^2+1073\bigr)\,\bigl(L_{3}+L_{7}\bigr)+4914\,L_{4}\notag\\
&\quad{}-\bigl(\frac{2100}{\Delta}-4200\log p^2+9838\bigr)\,\bigl(L_{5}+L_{8}\bigr)+\bigl(\frac{1050}{\Delta}-2100\log p^2+2147\bigr)\,\bigl(L_{6}+L_{11}\bigr)\notag\\
&\quad{}-\bigl(\frac{4200}{\Delta}-8400\log p^2+7188\bigr)\,L_{9}+\bigl(\frac{4200}{\Delta}-8400\log p^2+14440\bigr)\,L_{10}\Bigr]\,,\\
&D_{12}^{(0)}=\frac{|p|^3}{705600 \pi^2}\Bigl[2807\,L_{1}+9772\,L_{2}\notag\\
&\quad{}-\bigl(-\frac{560}{\Delta}+1680\log p^2+2320\bigr)\,\bigl(L_{3}+L_{7}\bigr)-12474\,L_{4}\notag\\
&\quad{}-\bigl(\frac{1120}{\Delta}-3360\log p^2+3536\bigr)\,\bigl(L_{5}+L_{8}\bigr)+\bigl(\frac{560}{\Delta}-1680\log p^2+6430\bigr)\,\bigl(L_{6}+L_{11}\bigr)\notag\\
&\quad{}-\bigl(\frac{2240}{\Delta}-6720\log p^2+24600\bigr)\,L_{9}+\bigl(\frac{2240}{\Delta}-6720\log p^2+26168\bigr)\,L_{10}\Bigr]\,,\\
&(D_{7})_{\mathrm{cyc}}^{(0)}=\frac{|p|^3}{2822400 \pi^2}\Bigl[\bigl(\frac{5880}{\Delta}-11760\log p^2+25844\bigr)\,L_{1}\notag\\
&\quad{}-\bigl(-\frac{23520}{\Delta}+47040\log p^2+16016\bigr)\,L_{2}-\bigl(\frac{3150}{\Delta}-6300\log p^2+2605\bigr)\,\bigl(L_{3}+L_{7}\bigr)\notag\\
&\quad{}-\bigl(\frac{35280}{\Delta}-70560\log p^2+90888\bigr)\,L_{4}+\bigl(\frac{6300}{\Delta}-12600\log p^2+19798\bigr)\,\bigl(L_{5}+L_{8}\bigr)\notag\\
&\quad{}+\bigl(\frac{8610}{\Delta}-17220\log p^2+35755\bigr)\,\bigl(L_{6}+L_{11}\bigr)-\bigl(\frac{34440}{\Delta}-68880\log p^2+139380\bigr)\,L_{9}\notag\\
&\quad{}+\bigl(\frac{34440}{\Delta}-68880\log p^2+154276\bigr)\,L_{10}\Bigr]\,,\\
&(D_{8})_{\mathrm{cyc}}^{(0)}=-\frac{|p|^3}{2822400 \pi^2}\Bigl[1792\,L_{1}-448\,L_{2}\notag\\
&\quad{}+\bigl(\frac{6090}{\Delta}-12180\log p^2+6595\bigr)\,\bigl(L_{3}+L_{7}\bigr)+12096\,L_{4}\notag\\
&\quad{}-\bigl(\frac{12180}{\Delta}-24360\log p^2+66586\bigr)\,\bigl(L_{5}+L_{8}\bigr)\notag\\
&\quad{}+\bigl(\frac{6090}{\Delta}-12180\log p^2+32915\bigr)\,\bigl(L_{6}+L_{11}\bigr)-\bigl(\frac{24360}{\Delta}-48720\log p^2+147060\bigr)\,L_{9}\notag\\
&\quad{}+\bigl(\frac{24360}{\Delta}-48720\log p^2+152548\bigr)\,L_{10}\Bigr]\,.
\end{align}
\end{adjustwidth}
\endgroup
The reflected placements are
\begin{equation}
D_3^{(0)}=D_1^{(0)}\,,\qquad
D_6^{(0)}=\left.D_5^{(0)}\right|_{\mu\leftrightarrow\rho,\,\nu\leftrightarrow\sigma}\,.
\end{equation}

\subsubsection{Contributions from soft momentum regions}
The linear terms are
\begin{equation}
D_i^{\mathrm{soft},(1)}=0\,,\qquad i=0,1,2,3,4,5,6,7,8,12\,.
\end{equation}
\begin{equation}
(D_i)_{\mathrm{cyc}}^{\mathrm{soft},(1)}=0\,,\qquad i=7,8\,.
\end{equation}
The quadratic terms are
\begingroup\allowdisplaybreaks
\begin{adjustwidth}{-5mm}{-5mm}
\small
\begin{align}
&D_{1}^{\mathrm{soft},(2)}=-\frac{|q|^2|p|}{288 \pi^2}\Bigl[-\bigl(L_{16}+L_{18}+L_{21}-L_{28}-L_{35}\bigr)+W\,L_{24}\Bigr]\,,\\
&D_{3}^{\mathrm{soft},(2)}=\frac{|q|^2|p|}{288 \pi^2}\Bigl[-\bigl(L_{16}+L_{18}+L_{21}-L_{28}-L_{35}\bigr)+W\,L_{24}\Bigr]\,,\\
&D_{7}^{\mathrm{soft},(2)}=-\frac{|q|^2|p|}{480 \pi^2}\Bigl[\bigl(L_{15}+L_{24}-L_{32}\bigr)\Bigr]\,,\\
&D_{8}^{\mathrm{soft},(2)}=\frac{|q|^2|p|}{480 \pi^2}\Bigl[\bigl(L_{24}-L_{32}\bigr)\Bigr]\,,\\
&D_{12}^{\mathrm{soft},(2)}=\frac{|q|^2|p|}{480 \pi^2}\Bigl[L_{15}\Bigr]\,.
\end{align}
\end{adjustwidth}
\endgroup
\begin{equation}
D_i^{\mathrm{soft},(2)}=0\,,\qquad i=0,2,4,5,6\,.
\end{equation}
\begin{equation}
(D_i)_{\mathrm{cyc}}^{\mathrm{soft},(2)}=0\,,\qquad i=7,8\,.
\end{equation}
The cubic terms are
\begingroup\allowdisplaybreaks
\begin{adjustwidth}{-5mm}{-5mm}
\small
\begin{align}
&D_{0}^{\mathrm{soft},(3)}=\frac{N|q|^3}{512}\Bigl[2\,\bigl(L_{2}-L_{3}-L_{5}+L_{6}-L_{7}\bigr)\notag\\
&\quad{}-2\,\bigl(L_{8}-L_{11}-L_{13}-L_{14}+L_{15}\bigr)+2\,\bigl(L_{22}+L_{23}\bigr)+6\,\bigl(L_{9}-L_{30}-L_{36}\bigr)\notag\\
&\quad{}-W\,\bigl(L_{16}+L_{18}+L_{25}+L_{26}\bigr)+\bigl(L_{17}+L_{20}\bigr)+3\,L_{19}-4 W\,L_{21}+\bigl(W^{2}+2\bigr)\,L_{24}\notag\\
&\quad{}+2 W\,\bigl(L_{28}+L_{33}+L_{35}+L_{38}\bigr)-4\,\bigl(L_{29}+L_{37}\bigr)\Bigr]\,,\\
&D_{1}^{\mathrm{soft},(3)}=-\frac{|q|^3}{4608 \pi^2}\Bigl[\bigl(\frac{24}{\Delta}-24\log q^2+102\bigr)\,\bigl(L_{2}-L_{3}-L_{5}+L_{6}-L_{7}\bigr)\notag\\
&\quad{}-\bigl(\frac{24}{\Delta}-24\log q^2+102\bigr)\,\bigl(L_{8}-L_{11}\bigr)+\bigl(\frac{72}{\Delta}-72\log q^2+306\bigr)\,L_{9}\notag\\
&\quad{}+\bigl(\frac{24}{\Delta}-24\log q^2+110\bigr)\,\bigl(L_{13}+L_{23}\bigr)+\bigl(\frac{24}{\Delta}-24\log q^2+78\bigr)\,L_{14}\notag\\
&\quad{}-\bigl(\frac{24}{\Delta}-24\log q^2+106\bigr)\,L_{15}-\bigl(\frac{12 W}{\Delta}-12 W\log q^2+13 W\bigr)\,\bigl(L_{16}+L_{18}+L_{25}\bigr)\notag\\
&\quad{}+\bigl(\frac{12}{\Delta}-12\log q^2+13\bigr)\,L_{17}+\bigl(\frac{36}{\Delta}-36\log q^2+55\bigr)\,L_{19}+\bigl(\frac{12}{\Delta}-12\log q^2+29\bigr)\,L_{20}\notag\\
&\quad{}-\bigl(\frac{48 W}{\Delta}-48 W\log q^2+172 W\bigr)\,L_{21}+\bigl(\frac{24}{\Delta}-24\log q^2+62\bigr)\,L_{22}\notag\\
&\quad{}+\bigl(\frac{24 + 12 W^{2}}{\Delta}-\bigl(12 W^{2} + 24\bigr)\log q^2+13 W^{2}+106\bigr)\,L_{24}-\bigl(\frac{12 W}{\Delta}-12 W\log q^2+29 W\bigr)\,L_{26}\notag\\
&\quad{}+\bigl(\frac{24 W}{\Delta}-24 W\log q^2+62 W\bigr)\,L_{28}-\bigl(\frac{48}{\Delta}-48\log q^2+172\bigr)\,L_{29}\notag\\
&\quad{}-\bigl(\frac{72}{\Delta}-72\log q^2+298\bigr)\,L_{30}+\bigl(\frac{24 W}{\Delta}-24 W\log q^2+110 W\bigr)\,L_{33}\notag\\
&\quad{}+\bigl(\frac{24 W}{\Delta}-24 W\log q^2+94 W\bigr)\,L_{35}-\bigl(\frac{72}{\Delta}-72\log q^2+234\bigr)\,L_{36}\notag\\
&\quad{}-\bigl(\frac{48}{\Delta}-48\log q^2+188\bigr)\,L_{37}+\bigl(\frac{24 W}{\Delta}-24 W\log q^2+78 W\bigr)\,L_{38}\Bigr]\,,\\
&D_{2}^{\mathrm{soft},(3)}=-\frac{|q|^3}{1152 \pi^2}\Bigl[\bigl(\frac{6}{\Delta}-6\log p^2+22\bigr)\,\bigl(L_{2}-L_{3}-L_{7}+L_{13}+L_{14}\bigr)\notag\\
&\quad{}-\bigl(\frac{6}{\Delta}-6\log p^2+22\bigr)\,L_{15}-\bigl(\frac{6}{\Delta}-6\log p^2+16\bigr)\,\bigl(L_{5}-L_{6}+L_{8}-L_{11}-L_{22}\bigr)\notag\\
&\quad{}+\bigl(\frac{6}{\Delta}-6\log p^2+16\bigr)\,L_{23}+\bigl(\frac{18}{\Delta}-18\log p^2+48\bigr)\,\bigl(L_{9}-L_{30}-L_{36}\bigr)\notag\\
&\quad{}-\bigl(\frac{3 W}{\Delta}-3 W\log p^2+11 W\bigr)\,\bigl(L_{16}+L_{18}\bigr)+\bigl(\frac{3}{\Delta}-3\log p^2+8\bigr)\,\bigl(L_{17}+L_{20}\bigr)\notag\\
&\quad{}+\bigl(\frac{9}{\Delta}-9\log p^2+24\bigr)\,L_{19}-\bigl(\frac{12 W}{\Delta}-12 W\log p^2+44 W\bigr)\,L_{21}\notag\\
&\quad{}+\bigl(\frac{6 + 3 W^{2}}{\Delta}-\bigl(3 W^{2} + 6\bigr)\log p^2+11 W^{2}+19\bigr)\,L_{24}-\bigl(\frac{3 W}{\Delta}-3 W\log p^2+8 W\bigr)\,\bigl(L_{25}+L_{26}\bigr)\notag\\
&\quad{}+\bigl(\frac{6 W}{\Delta}-6 W\log p^2+22 W\bigr)\,\bigl(L_{28}+L_{35}\bigr)-\bigl(\frac{12}{\Delta}-12\log p^2+32\bigr)\,\bigl(L_{29}+L_{37}\bigr)\notag\\
&\quad{}+\bigl(\frac{6 W}{\Delta}-6 W\log p^2+16 W\bigr)\,\bigl(L_{33}+L_{38}\bigr)\Bigr]\,,\\
&D_{3}^{\mathrm{soft},(3)}=-\frac{|q|^3}{4608 \pi^2}\Bigl[\bigl(\frac{24}{\Delta}-24\log q^2+102\bigr)\,\bigl(L_{2}-L_{3}-L_{5}+L_{6}-L_{7}\bigr)\notag\\
&\quad{}-\bigl(\frac{24}{\Delta}-24\log q^2+102\bigr)\,\bigl(L_{8}-L_{11}\bigr)+\bigl(\frac{72}{\Delta}-72\log q^2+306\bigr)\,L_{9}\notag\\
&\quad{}+\bigl(\frac{24}{\Delta}-24\log q^2+78\bigr)\,\bigl(L_{13}+L_{23}\bigr)+\bigl(\frac{24}{\Delta}-24\log q^2+110\bigr)\,L_{14}\notag\\
&\quad{}-\bigl(\frac{24}{\Delta}-24\log q^2+106\bigr)\,L_{15}-\bigl(\frac{12 W}{\Delta}-12 W\log q^2+45 W\bigr)\,\bigl(L_{16}+L_{18}+L_{25}\bigr)\notag\\
&\quad{}+\bigl(\frac{12}{\Delta}-12\log q^2+45\bigr)\,L_{17}+\bigl(\frac{36}{\Delta}-36\log q^2+119\bigr)\,L_{19}+\bigl(\frac{12}{\Delta}-12\log q^2+29\bigr)\,L_{20}\notag\\
&\quad{}-\bigl(\frac{48 W}{\Delta}-48 W\log q^2+204 W\bigr)\,L_{21}+\bigl(\frac{24}{\Delta}-24\log q^2+126\bigr)\,L_{22}\notag\\
&\quad{}+\bigl(\frac{24 + 12 W^{2}}{\Delta}-\bigl(12 W^{2} + 24\bigr)\log q^2+45 W^{2}+106\bigr)\,L_{24}-\bigl(\frac{12 W}{\Delta}-12 W\log q^2+29 W\bigr)\,L_{26}\notag\\
&\quad{}+\bigl(\frac{24 W}{\Delta}-24 W\log q^2+126 W\bigr)\,L_{28}-\bigl(\frac{48}{\Delta}-48\log q^2+204\bigr)\,L_{29}\notag\\
&\quad{}-\bigl(\frac{72}{\Delta}-72\log q^2+266\bigr)\,L_{30}+\bigl(\frac{24 W}{\Delta}-24 W\log q^2+78 W\bigr)\,L_{33}\notag\\
&\quad{}+\bigl(\frac{24 W}{\Delta}-24 W\log q^2+94 W\bigr)\,L_{35}-\bigl(\frac{72}{\Delta}-72\log q^2+330\bigr)\,L_{36}\notag\\
&\quad{}-\bigl(\frac{48}{\Delta}-48\log q^2+188\bigr)\,L_{37}+\bigl(\frac{24 W}{\Delta}-24 W\log q^2+110 W\bigr)\,L_{38}\Bigr]\,,\\
&D_{4}^{\mathrm{soft},(3)}=\frac{|q|^3}{2822400 \pi^2}\Bigl[9856\,L_{1}\notag\\
&\quad{}+\bigl(\frac{2940}{\Delta}+2940\log p^2-5880\log q^2+11410\bigr)\,L_{2}\notag\\
&\quad{}-\bigl(\frac{2940}{\Delta}+2940\log p^2-5880\log q^2+17794\bigr)\,\bigl(L_{3}+L_{7}\bigr)+6384\,L_{4}\notag\\
&\quad{}-\bigl(\frac{2940}{\Delta}+2940\log p^2-5880\log q^2+17878\bigr)\,\bigl(L_{5}+L_{8}\bigr)\notag\\
&\quad{}+\bigl(\frac{2940}{\Delta}+2940\log p^2-5880\log q^2+12166\bigr)\,\bigl(L_{6}+L_{11}\bigr)\notag\\
&\quad{}+\bigl(\frac{8820}{\Delta}+8820\log p^2-17640\log q^2+47978\bigr)\,L_{9}\notag\\
&\quad{}+\bigl(\frac{8400}{\Delta}-8400\log p^2+13824\bigr)\,\bigl(L_{10}-L_{34}-L_{39}\bigr)-19712 W\,L_{12}\notag\\
&\quad{}+\bigl(\frac{2940}{\Delta}+2940\log p^2-5880\log q^2+15834\bigr)\,\bigl(L_{13}+L_{14}\bigr)\notag\\
&\quad{}-\bigl(\frac{2940}{\Delta}+2940\log p^2-5880\log q^2-4928 W^{2}+13846\bigr)\,L_{15}\notag\\
&\quad{}-\bigl(\frac{1470 W}{\Delta}+1470 W\log p^2-2940 W\log q^2+3507 W\bigr)\,\bigl(L_{16}+L_{18}\bigr)\notag\\
&\quad{}+\bigl(\frac{1470}{\Delta}+1470\log p^2-2940\log q^2+693\bigr)\,\bigl(L_{17}+L_{20}\bigr)\notag\\
&\quad{}+\bigl(\frac{4410}{\Delta}+4410\log p^2-8820\log q^2+7819\bigr)\,L_{19}\notag\\
&\quad{}-\bigl(\frac{5880 W}{\Delta}+5880 W\log p^2-11760 W\log q^2+18900 W\bigr)\,L_{21}\notag\\
&\quad{}+\bigl(\frac{2940}{\Delta}+2940\log p^2-5880\log q^2+15918\bigr)\,\bigl(L_{22}+L_{23}\bigr)\notag\\
&\quad{}+\bigl(\frac{2940 + 1470 W^{2}}{\Delta}+\bigl(1470 W^{2} + 2940\bigr)\log p^2-\bigl(2940 W^{2} + 5880\bigr)\log q^2+315 W^{2}+12796\bigr)\,L_{24}\notag\\
&\quad{}-\bigl(\frac{1470 W}{\Delta}+1470 W\log p^2-2940 W\log q^2+3549 W\bigr)\,\bigl(L_{25}+L_{26}\bigr)\notag\\
&\quad{}+\bigl(\frac{4200}{\Delta}-4200\log p^2+6912\bigr)\,L_{27}\notag\\
&\quad{}+\bigl(\frac{2940 W}{\Delta}+2940 W\log p^2-5880 W\log q^2+15834 W\bigr)\,\bigl(L_{28}+L_{35}\bigr)\notag\\
&\quad{}-\bigl(\frac{5880}{\Delta}+5880\log p^2-11760\log q^2+20412\bigr)\,\bigl(L_{29}+L_{37}\bigr)\notag\\
&\quad{}-\bigl(\frac{8820}{\Delta}+8820\log p^2-17640\log q^2+42098\bigr)\,\bigl(L_{30}+L_{36}\bigr)-12768 W\,L_{31}\notag\\
&\quad{}+\bigl(\frac{4200}{\Delta}-4200\log p^2+3192 W^{2}+10104\bigr)\,L_{32}\notag\\
&\quad{}+\bigl(\frac{2940 W}{\Delta}+2940 W\log p^2-5880 W\log q^2+15918 W\bigr)\,\bigl(L_{33}+L_{38}\bigr)\Bigr]\,,\\
&D_{5}^{\mathrm{soft},(3)}=\frac{|q|^3}{28800 \pi^2}\Bigl[\bigl(\frac{30}{\Delta}-30\log p^2+42\bigr)\,\bigl(L_{2}-L_{3}+L_{13}\bigr)\notag\\
&\quad{}-\bigl(\frac{30}{\Delta}-30\log p^2+12\bigr)\,\bigl(L_{5}-L_{6}-L_{22}\bigr)-\bigl(\frac{30}{\Delta}-30\log p^2+82\bigr)\,\bigl(L_{7}-L_{14}\bigr)\notag\\
&\quad{}-\bigl(\frac{30}{\Delta}-30\log p^2+32\bigr)\,\bigl(L_{8}-L_{23}\bigr)+\bigl(\frac{90}{\Delta}-90\log p^2+216\bigr)\,\bigl(L_{9}-L_{30}-L_{36}\bigr)\notag\\
&\quad{}+\bigl(\frac{30}{\Delta}-30\log p^2+152\bigr)\,L_{11}-\bigl(\frac{30}{\Delta}-30\log p^2+62\bigr)\,L_{15}\notag\\
&\quad{}-\bigl(\frac{15 W}{\Delta}-15 W\log p^2+21 W\bigr)\,L_{16}+\bigl(\frac{15}{\Delta}-15\log p^2+6\bigr)\,L_{17}\notag\\
&\quad{}-\bigl(\frac{15 W}{\Delta}-15 W\log p^2+41 W\bigr)\,L_{18}+\bigl(\frac{45}{\Delta}-45\log p^2+108\bigr)\,L_{19}\notag\\
&\quad{}+\bigl(\frac{15}{\Delta}-15\log p^2+76\bigr)\,L_{20}-\bigl(\frac{60 W}{\Delta}-60 W\log p^2+84 W\bigr)\,L_{21}\notag\\
&\quad{}+\bigl(\frac{30 + 15 W^{2}}{\Delta}-\bigl(15 W^{2} + 30\bigr)\log p^2+21 W^{2}+107\bigr)\,L_{24}-\bigl(\frac{15 W}{\Delta}-15 W\log p^2+6 W\bigr)\,L_{25}\notag\\
&\quad{}-\bigl(\frac{15 W}{\Delta}-15 W\log p^2+16 W\bigr)\,L_{26}+\bigl(\frac{30 W}{\Delta}-30 W\log p^2+42 W\bigr)\,L_{28}\notag\\
&\quad{}-\bigl(\frac{60}{\Delta}-60\log p^2+24\bigr)\,L_{29}+\bigl(\frac{30 W}{\Delta}-30 W\log p^2+12 W\bigr)\,L_{33}\notag\\
&\quad{}+\bigl(\frac{30 W}{\Delta}-30 W\log p^2+82 W\bigr)\,L_{35}-\bigl(\frac{60}{\Delta}-60\log p^2+304\bigr)\,L_{37}\notag\\
&\quad{}+\bigl(\frac{30 W}{\Delta}-30 W\log p^2+32 W\bigr)\,L_{38}\Bigr]\,,\\
&D_{6}^{\mathrm{soft},(3)}=\frac{|q|^3}{28800 \pi^2}\Bigl[\bigl(\frac{30}{\Delta}-30\log p^2+42\bigr)\,\bigl(L_{2}-L_{7}+L_{14}\bigr)\notag\\
&\quad{}-\bigl(\frac{30}{\Delta}-30\log p^2+82\bigr)\,\bigl(L_{3}-L_{13}\bigr)-\bigl(\frac{30}{\Delta}-30\log p^2+32\bigr)\,\bigl(L_{5}-L_{22}\bigr)\notag\\
&\quad{}+\bigl(\frac{30}{\Delta}-30\log p^2+152\bigr)\,L_{6}-\bigl(\frac{30}{\Delta}-30\log p^2+12\bigr)\,\bigl(L_{8}-L_{11}-L_{23}\bigr)\notag\\
&\quad{}+\bigl(\frac{90}{\Delta}-90\log p^2+216\bigr)\,\bigl(L_{9}-L_{30}-L_{36}\bigr)-\bigl(\frac{30}{\Delta}-30\log p^2+62\bigr)\,L_{15}\notag\\
&\quad{}-\bigl(\frac{15 W}{\Delta}-15 W\log p^2+41 W\bigr)\,L_{16}+\bigl(\frac{15}{\Delta}-15\log p^2+76\bigr)\,L_{17}\notag\\
&\quad{}-\bigl(\frac{15 W}{\Delta}-15 W\log p^2+21 W\bigr)\,L_{18}+\bigl(\frac{45}{\Delta}-45\log p^2+108\bigr)\,L_{19}+\bigl(\frac{15}{\Delta}-15\log p^2+6\bigr)\,L_{20}\notag\\
&\quad{}-\bigl(\frac{60 W}{\Delta}-60 W\log p^2+84 W\bigr)\,L_{21}+\bigl(\frac{30 + 15 W^{2}}{\Delta}-\bigl(15 W^{2} + 30\bigr)\log p^2+21 W^{2}+107\bigr)\,L_{24}\notag\\
&\quad{}-\bigl(\frac{15 W}{\Delta}-15 W\log p^2+16 W\bigr)\,L_{25}-\bigl(\frac{15 W}{\Delta}-15 W\log p^2+6 W\bigr)\,L_{26}\notag\\
&\quad{}+\bigl(\frac{30 W}{\Delta}-30 W\log p^2+82 W\bigr)\,L_{28}-\bigl(\frac{60}{\Delta}-60\log p^2+304\bigr)\,L_{29}\notag\\
&\quad{}+\bigl(\frac{30 W}{\Delta}-30 W\log p^2+32 W\bigr)\,L_{33}+\bigl(\frac{30 W}{\Delta}-30 W\log p^2+42 W\bigr)\,L_{35}\notag\\
&\quad{}-\bigl(\frac{60}{\Delta}-60\log p^2+24\bigr)\,L_{37}+\bigl(\frac{30 W}{\Delta}-30 W\log p^2+12 W\bigr)\,L_{38}\Bigr]\,,\\
&D_{7}^{\mathrm{soft},(3)}=\frac{|q|^3}{1411200 \pi^2}\Bigl[1232\,L_{1}\notag\\
&\quad{}+\bigl(\frac{2940}{\Delta}+5880\log p^2-11760\log q^2+10430\bigr)\,L_{2}\notag\\
&\quad{}-\bigl(\frac{2940}{\Delta}+5880\log p^2-11760\log q^2+11228\bigr)\,\bigl(L_{3}+L_{7}\bigr)+5208\,L_{4}\notag\\
&\quad{}-\bigl(\frac{2940}{\Delta}+5880\log p^2-11760\log q^2+21896\bigr)\,\bigl(L_{5}+L_{8}\bigr)\notag\\
&\quad{}+\bigl(\frac{2940}{\Delta}+5880\log p^2-11760\log q^2+16772\bigr)\,\bigl(L_{6}+L_{11}\bigr)\notag\\
&\quad{}+\bigl(\frac{8820}{\Delta}+17640\log p^2-35280\log q^2+51016\bigr)\,L_{9}\notag\\
&\quad{}+\bigl(\frac{2730}{\Delta}-5460\log p^2+10422\bigr)\,\bigl(L_{10}-L_{34}-L_{39}\bigr)-2464 W\,L_{12}\notag\\
&\quad{}+\bigl(\frac{2940}{\Delta}+5880\log p^2-11760\log q^2+7308\bigr)\,\bigl(L_{13}+L_{14}\bigr)\notag\\
&\quad{}-\bigl(\frac{2940}{\Delta}+5880\log p^2-11760\log q^2-616 W^{2}-4410 W+12572\bigr)\,L_{15}\notag\\
&\quad{}-\bigl(\frac{1470 W}{\Delta}+2940 W\log p^2-5880 W\log q^2-5166 W+4410\bigr)\,L_{16}\notag\\
&\quad{}+\bigl(\frac{1470}{\Delta}+2940\log p^2-5880\log q^2-2394\bigr)\,\bigl(L_{17}+L_{20}\bigr)\notag\\
&\quad{}-\bigl(\frac{1470 W}{\Delta}+2940 W\log p^2-5880 W\log q^2-5166 W+1470\bigr)\,L_{18}\notag\\
&\quad{}+\bigl(\frac{4410}{\Delta}+8820\log p^2-17640\log q^2-6832\bigr)\,L_{19}\notag\\
&\quad{}-\bigl(\frac{5880 W}{\Delta}+11760 W\log p^2-23520 W\log q^2+13020 W\bigr)\,L_{21}\notag\\
&\quad{}+\bigl(\frac{2940}{\Delta}+5880\log p^2-11760\log q^2+17976\bigr)\,\bigl(L_{22}+L_{23}\bigr)\notag\\
&\quad{}+\bigl(\frac{2940 + 1470 W^{2}}{\Delta}+\bigl(2940 W^{2} + 5880\bigr)\log p^2\notag\\
&\quad{}-\bigl(5880 W^{2} + 11760\bigr)\log q^2-5565 W^{2}+1470 W+10787\bigr)\,L_{24}\notag\\
&\quad{}-\bigl(\frac{1470 W}{\Delta}+2940 W\log p^2-5880 W\log q^2+168 W+1470\bigr)\,\bigl(L_{25}+L_{26}\bigr)\notag\\
&\quad{}+\bigl(\frac{1365}{\Delta}-2730\log p^2+5211\bigr)\,L_{27}\notag\\
&\quad{}+\bigl(\frac{2940 W}{\Delta}+5880 W\log p^2-11760 W\log q^2+7308 W\bigr)\,\bigl(L_{28}+L_{35}\bigr)\notag\\
&\quad{}-\bigl(\frac{5880}{\Delta}+11760\log p^2-23520\log q^2+25704\bigr)\,\bigl(L_{29}+L_{37}\bigr)\notag\\
&\quad{}-\bigl(\frac{8820}{\Delta}+17640\log p^2-35280\log q^2+39256\bigr)\,\bigl(L_{30}+L_{36}\bigr)-10416 W\,L_{31}\notag\\
&\quad{}+\bigl(\frac{1365}{\Delta}-2730\log p^2+2604 W^{2}+1470 W+7815\bigr)\,L_{32}\notag\\
&\quad{}+\bigl(\frac{2940 W}{\Delta}+5880 W\log p^2-11760 W\log q^2+17976 W\bigr)\,\bigl(L_{33}+L_{38}\bigr)\Bigr]\,,\\
&D_{8}^{\mathrm{soft},(3)}=\frac{|q|^3}{11289600 \pi^2}\Bigl[-\bigl(217280 - 22050 \pi^{2}\bigr)\,L_{1}\notag\\
&\quad{}+\bigl(887600 - 88200 \pi^{2}\bigr)\,L_{2}-25760\,\bigl(L_{3}+L_{7}-L_{13}-L_{14}\bigr)-\bigl(203280 - 22050 \pi^{2}\bigr)\,L_{4}\notag\\
&\quad{}-\bigl(468160 - 44100 \pi^{2}\bigr)\,\bigl(L_{5}+L_{8}-L_{22}-L_{23}\bigr)+\bigl(224000 - 22050 \pi^{2}\bigr)\,\bigl(L_{6}+L_{11}\bigr)\notag\\
&\quad{}-\bigl(436800 - 44100 \pi^{2}\bigr)\,\bigl(L_{9}-L_{30}-L_{36}\bigr)\notag\\
&\quad{}+\bigl(-\frac{31920}{\Delta}+63840\log p^2+284944 - 44100 \pi^{2}\bigr)\,\bigl(L_{10}-L_{34}-L_{39}\bigr)\notag\\
&\quad{}+\bigl(\bigl(434560 - 44100 \pi^{2}\bigr) W\bigr)\,L_{12}\notag\\
&\quad{}-\bigl(\bigl(108640 - 11025 \pi^{2}\bigr) W^{2}+134400 - 11025 \pi^{2}\bigr)\,L_{15}-\bigl(12880 W+11760\bigr)\,L_{16}\notag\\
&\quad{}+\bigl(112000 - 11025 \pi^{2}\bigr)\,\bigl(L_{17}+L_{20}\bigr)-\bigl(12880 W-11760\bigr)\,L_{18}\notag\\
&\quad{}-\bigl(218400 - 22050 \pi^{2}\bigr)\,L_{19}-\bigl(\bigl(1775200 - 176400 \pi^{2}\bigr) W\bigr)\,L_{21}\notag\\
&\quad{}-\bigl(-\bigl(443800 - 44100 \pi^{2}\bigr) W^{2}+11760 W+242760 - 22050 \pi^{2}\bigr)\,L_{24}\notag\\
&\quad{}+\bigl(-\bigl(234080 - 22050 \pi^{2}\bigr) W+11760\bigr)\,\bigl(L_{25}+L_{26}\bigr)\notag\\
&\quad{}+\bigl(-\frac{15960}{\Delta}+31920\log p^2+142472 - 22050 \pi^{2}\bigr)\,L_{27}+25760 W\,\bigl(L_{28}+L_{35}\bigr)\notag\\
&\quad{}-\bigl(448000 - 44100 \pi^{2}\bigr)\,\bigl(L_{29}+L_{37}\bigr)+\bigl(\bigl(406560 - 44100 \pi^{2}\bigr) W\bigr)\,L_{31}\notag\\
&\quad{}-\bigl(\frac{15960}{\Delta}-31920\log p^2+\bigl(101640 - 11025 \pi^{2}\bigr) W^{2}+11760 W-40832 - 11025 \pi^{2}\bigr)\,L_{32}\notag\\
&\quad{}+\bigl(\bigl(468160 - 44100 \pi^{2}\bigr) W\bigr)\,\bigl(L_{33}+L_{38}\bigr)\Bigr]\,,\\
&D_{12}^{\mathrm{soft},(3)}=\frac{|q|^3}{2822400 \pi^2}\Bigl[\bigl(6387710 - 646800 \pi^{2}\bigr)\,L_{1}\notag\\
&\quad{}-\bigl(25531520 - 2587200 \pi^{2}\bigr)\,L_{2}-10360\,\bigl(L_{3}+L_{7}-L_{13}-L_{14}\bigr)\notag\\
&\quad{}+\bigl(10447920 - 1058400 \pi^{2}\bigr)\,L_{4}+\bigl(4637920 - 470400 \pi^{2}\bigr)\,\bigl(L_{5}+L_{8}-L_{22}-L_{23}\bigr)\notag\\
&\quad{}-\bigl(2312660 - 235200 \pi^{2}\bigr)\,\bigl(L_{6}+L_{11}\bigr)+\bigl(4724160 - 470400 \pi^{2}\bigr)\,\bigl(L_{9}-L_{30}-L_{36}\bigr)\notag\\
&\quad{}-\bigl(-\frac{4480}{\Delta}+13440\log p^2+4645504 - 470400 \pi^{2}\bigr)\,\bigl(L_{10}-L_{34}-L_{39}\bigr)\notag\\
&\quad{}-\bigl(\bigl(12775420 - 1293600 \pi^{2}\bigr) W\bigr)\,L_{12}\notag\\
&\quad{}-\bigl(-\bigl(3193855 - 323400 \pi^{2}\bigr) W^{2}+8820 W-3183495 - 323400 \pi^{2}\bigr)\,L_{15}\notag\\
&\quad{}-\bigl(5180 W-11760\bigr)\,L_{16}-\bigl(1156330 - 117600 \pi^{2}\bigr)\,\bigl(L_{17}+L_{20}\bigr)-5180 W\,L_{18}\notag\\
&\quad{}+\bigl(2362080 - 235200 \pi^{2}\bigr)\,L_{19}+\bigl(\bigl(51063040 - 5174400 \pi^{2}\bigr) W\bigr)\,L_{21}\notag\\
&\quad{}-\bigl(\bigl(12765760 - 1293600 \pi^{2}\bigr) W^{2}+5765760 - 588000 \pi^{2}\bigr)\,L_{24}\notag\\
&\quad{}+\bigl(\bigl(2318960 - 235200 \pi^{2}\bigr) W\bigr)\,\bigl(L_{25}+L_{26}\bigr)\notag\\
&\quad{}-\bigl(-\frac{2240}{\Delta}+6720\log p^2+2322752 - 235200 \pi^{2}\bigr)\,L_{27}+10360 W\,\bigl(L_{28}+L_{35}\bigr)\notag\\
&\quad{}+\bigl(4625320 - 470400 \pi^{2}\bigr)\,\bigl(L_{29}+L_{37}\bigr)-\bigl(\bigl(20895840 - 2116800 \pi^{2}\bigr) W\bigr)\,L_{31}\notag\\
&\quad{}-\bigl(-\frac{2240}{\Delta}+6720\log p^2-\bigl(5223960 - 529200 \pi^{2}\bigr) W^{2}-2901208 - 294000 \pi^{2}\bigr)\,L_{32}\notag\\
&\quad{}-\bigl(\bigl(4637920 - 470400 \pi^{2}\bigr) W\bigr)\,\bigl(L_{33}+L_{38}\bigr)\Bigr]\,,\\
&(D_{7})_{\mathrm{cyc}}^{\mathrm{soft},(3)}=\frac{|q|^3}{28800 \pi^2}\Bigl[\bigl(\frac{120}{\Delta}-240\log p^2+336\bigr)\,L_{2}\notag\\
&\quad{}-\bigl(\frac{120}{\Delta}-240\log p^2+346\bigr)\,\bigl(L_{3}+L_{7}-L_{13}-L_{14}+L_{15}\bigr)\notag\\
&\quad{}-\bigl(\frac{120}{\Delta}-240\log p^2+176\bigr)\,\bigl(L_{5}+L_{8}-L_{22}-L_{23}\bigr)\notag\\
&\quad{}+\bigl(\frac{120}{\Delta}-240\log p^2+206\bigr)\,\bigl(L_{6}+L_{11}\bigr)+\bigl(\frac{360}{\Delta}-720\log p^2+528\bigr)\,\bigl(L_{9}-L_{30}-L_{36}\bigr)\notag\\
&\quad{}-\bigl(\frac{60 W}{\Delta}-120 W\log p^2+173 W\bigr)\,\bigl(L_{16}+L_{18}\bigr)+\bigl(\frac{60}{\Delta}-120\log p^2+103\bigr)\,\bigl(L_{17}+L_{20}\bigr)\notag\\
&\quad{}+\bigl(\frac{180}{\Delta}-360\log p^2+264\bigr)\,L_{19}-\bigl(\frac{240 W}{\Delta}-480 W\log p^2+672 W\bigr)\,L_{21}\notag\\
&\quad{}+\bigl(\frac{120 + 60 W^{2}}{\Delta}-\bigl(120 W^{2} + 240\bigr)\log p^2+168 W^{2}+256\bigr)\,L_{24}\notag\\
&\quad{}-\bigl(\frac{60 W}{\Delta}-120 W\log p^2+88 W\bigr)\,\bigl(L_{25}+L_{26}\bigr)+\bigl(\frac{120 W}{\Delta}-240 W\log p^2+346 W\bigr)\,\bigl(L_{28}+L_{35}\bigr)\notag\\
&\quad{}-\bigl(\frac{240}{\Delta}-480\log p^2+412\bigr)\,\bigl(L_{29}+L_{37}\bigr)+\bigl(\frac{120 W}{\Delta}-240 W\log p^2+176 W\bigr)\,\bigl(L_{33}+L_{38}\bigr)\Bigr]\,,\\
&(D_{8})_{\mathrm{cyc}}^{\mathrm{soft},(3)}=\frac{|q|^3}{1128960 \pi^2}\Bigl[-\bigl(46144 - 4410 \pi^{2}\bigr)\,L_{1}\notag\\
&\quad{}+\bigl(173152 - 17640 \pi^{2}\bigr)\,L_{2}-5992\,\bigl(L_{3}+L_{7}-L_{13}-L_{14}\bigr)-\bigl(43680 - 4410 \pi^{2}\bigr)\,L_{4}\notag\\
&\quad{}-\bigl(91616 - 8820 \pi^{2}\bigr)\,\bigl(L_{5}+L_{8}-L_{22}-L_{23}\bigr)+\bigl(48832 - 4410 \pi^{2}\bigr)\,\bigl(L_{6}+L_{11}\bigr)\notag\\
&\quad{}-\bigl(94752 - 8820 \pi^{2}\bigr)\,\bigl(L_{9}-L_{30}-L_{36}\bigr)\notag\\
&\quad{}+\bigl(-\frac{3360}{\Delta}+6720\log p^2+92384 - 8820 \pi^{2}\bigr)\,\bigl(L_{10}-L_{34}-L_{39}\bigr)\notag\\
&\quad{}+\bigl(\bigl(92288 - 8820 \pi^{2}\bigr) W\bigr)\,L_{12}-\bigl(\bigl(23072 - 2205 \pi^{2}\bigr) W^{2}+29064 - 2205 \pi^{2}\bigr)\,L_{15}\notag\\
&\quad{}-2996 W\,\bigl(L_{16}+L_{18}\bigr)+\bigl(24416 - 2205 \pi^{2}\bigr)\,\bigl(L_{17}+L_{20}\bigr)-\bigl(47376 - 4410 \pi^{2}\bigr)\,L_{19}\notag\\
&\quad{}-\bigl(\bigl(346304 - 35280 \pi^{2}\bigr) W\bigr)\,L_{21}+\bigl(\bigl(86576 - 8820 \pi^{2}\bigr) W^{2}-52416 - 4410 \pi^{2}\bigr)\,L_{24}\notag\\
&\quad{}-\bigl(\bigl(45808 - 4410 \pi^{2}\bigr) W\bigr)\,\bigl(L_{25}+L_{26}\bigr)\notag\\
&\quad{}+\bigl(-\frac{1680}{\Delta}+3360\log p^2+46192 - 4410 \pi^{2}\bigr)\,L_{27}+5992 W\,\bigl(L_{28}+L_{35}\bigr)\notag\\
&\quad{}-\bigl(97664 - 8820 \pi^{2}\bigr)\,\bigl(L_{29}+L_{37}\bigr)+\bigl(\bigl(87360 - 8820 \pi^{2}\bigr) W\bigr)\,L_{31}\notag\\
&\quad{}+\bigl(-\frac{1680}{\Delta}+3360\log p^2-\bigl(21840 - 2205 \pi^{2}\bigr) W^{2}+24352 - 2205 \pi^{2}\bigr)\,L_{32}\notag\\
&\quad{}+\bigl(\bigl(91616 - 8820 \pi^{2}\bigr) W\bigr)\,\bigl(L_{33}+L_{38}\bigr)\Bigr]\,.
\end{align}
\end{adjustwidth}
\endgroup
The tensors obey the three-dimensional identity
\begin{align}
&0=2\,\bigl(L_{4}+L_{6}+L_{11}-L_{19}+L_{27}\bigr)\notag\\
&\quad{}-4\,\bigl(L_{5}+L_{8}+L_{9}-L_{10}-L_{22}\bigr)+4\,\bigl(L_{23}-L_{29}+L_{30}-L_{34}+L_{36}\bigr)-4\,\bigl(L_{37}+L_{39}\bigr)\notag\\
&\quad{}+\bigl(L_{17}+L_{20}\bigr)-6\,L_{24}-2 W\,\bigl(L_{25}+L_{26}\bigr)-4 W\,\bigl(L_{31}-L_{33}-L_{38}\bigr)+\bigl(W^{2}+3\bigr)\,L_{32}\,.
\end{align}

\subsubsection{Contact diagrams}
\label{app:gny-full-contacts}

The GNY contact diagrams follow from the full second metric variation of
 eq.~\eqref{eq:fermion-curved}.
This variation includes the fermion derivative term, the spin connection,
and the Yukawa interaction.
We evaluate each term with the regulator in eq.~\eqref{eq:regulated-propagators}.
The results below include the pole and finite terms as $\Delta\to0$.

To display the vertex, use traceless symmetric variations $h_{ab}(x)$ and
$j_{ab}(x)$ in $g_{ab}=\delta_{ab}+s h_{ab}+t j_{ab}$.
The traceless condition implements the external projection used throughout
this appendix.
In the symmetric vielbein convention, the derivative bilinear is
\begin{equation}
 B_{ab}=\frac18\bar\psi_i
 \left(\gamma_a\overleftrightarrow{\partial}_b
       +\gamma_b\overleftrightarrow{\partial}_a\right)\psi_i,
 \qquad
 \bar\psi\overleftrightarrow{\partial}_a\psi
 =\bar\psi\partial_a\psi-(\partial_a\bar\psi)\psi.
 \label{eq:gny-derivative-bilinear}
\end{equation}
Its traceless projection is the stress-tensor insertion used in the
loop integrals.
Expanding $\sqrt g\,g^{-1/2}$, $\sqrt g$, and the spin connection gives
\begin{align}
 \left.\partial_s\partial_t S_{\GN}\right|_{s=t=0}
 &=\int\dd^3x\,\Biggl\{
 \left[\frac38(hj+jh)_{ab}
       -\frac12\operatorname{tr}(hj)\delta_{ab}\right]B_{ab}
 \notag\\
 &\qquad-\frac14\operatorname{tr}(hj)\,
           \sigma\bar\psi_i\psi_i
       -\frac{N}{4g}\operatorname{tr}(hj)\,\sigma^2
 \notag\\
 &\qquad-\frac1{32}
 \left(h_{b\ell}\partial_a j_{\ell c}
       +j_{b\ell}\partial_a h_{\ell c}\right)
 \bar\psi_i\gamma^{abc}\psi_i\Biggr\}.
 \label{eq:gny-full-second-variation}
\end{align}
Here $\gamma^{abc}=\gamma^{[a}\gamma^b\gamma^{c]}$, with antisymmetrization
of unit weight.
The last line comes from the spin connection after writing the fermion
action with derivatives acting on both fermion fields, as in
 eq.~\eqref{eq:gny-derivative-bilinear}.
For the momentum-space square in figure~\ref{fig:gny-ttt-contacts},
insert $h_{ab}(x)=h_{ab}e^{-ip\cdot x}$ and
$j_{ab}(x)=j_{ab}e^{-ir\cdot x}$ into
 eq.~\eqref{eq:gny-full-second-variation}, with constant symmetric
traceless polarizations and $k=p+r$.
Using the Fourier convention of section~\ref{sec:source-momentum-identities}
then gives the explicit contraction
\begin{equation}
\begin{aligned}
 h_{\mu\nu}j_{\rho\sigma}
 \mathcal V_{\GN;\mu\nu,\rho\sigma}(p,r)
 &=\left[\frac38(hj+jh)_{ab}
       -\frac12\operatorname{tr}(hj)\delta_{ab}\right]B_{ab}(k)
 \\
 &\quad-\frac14\operatorname{tr}(hj)
       (\sigma\bar\psi_i\psi_i)(k)
       -\frac{N}{4g}\operatorname{tr}(hj)(\sigma^2)(k)
 \\
 &\quad+\frac{\mathrm{i}}{32}
       \left[r_a(hj)_{bc}+p_a(jh)_{bc}\right]
       (\bar\psi_i\gamma^{abc}\psi_i)(k).
\end{aligned}
\label{eq:gny-explicit-metric-vertex}
\end{equation}
Here $B_{ab}(k)$ is the Fourier transform of
 eq.~\eqref{eq:gny-derivative-bilinear}, and the other composite operators
are also Fourier transformed as a whole.
The factor $+\mathrm{i}$ in the last line follows from
$\partial_a e^{-ir\cdot x}=-\mathrm{i}r_a e^{-ir\cdot x}$ and the
minus sign in the spin-connection term of
 eq.~\eqref{eq:gny-full-second-variation}.
This formula includes every term of that variation for traceless source
polarizations; it does not impose transversality.
The reductions to the contributing fermion bilinear in the figure are
established below.

At the critical point, the renormalized local coefficient of $\sigma^2$
is zero.
The momentum-dependent inverse $\sigma$ propagator is generated by
fermion loops and is included in the diagrams below.

In three dimensions, $\bar\psi_i\gamma^{abc}\psi_i$ is proportional to
$\epsilon^{abc}\bar\psi_i\psi_i$.
The bilinear $\bar\psi_i\psi_i$ is parity odd, whereas the stress tensor
is parity even.
Their mixed two-point function would require a symmetric parity-odd
rank-two tensor constructed from a single momentum.
No such tensor exists, so
\begin{equation}
 \langle T_{ab}(q)\bar\psi_i\psi_i(-q)\rangle=0,
 \qquad
 \langle T_{ab}(q)\bar\psi_i\gamma^{cde}\psi_i(-q)\rangle=0.
 \label{eq:gny-spin-contact-zero}
\end{equation}
The regulator preserves this symmetry.
Thus every spin-connection contact vanishes separately.
A diagram in which a single auxiliary-field line separates the metric
vertex from the stress-tensor insertion also vanishes.
Cutting that line leaves a mixed two-point function of the stress tensor
with a parity-odd scalar insertion.

The Yukawa term requires a separate integral calculation because
$\sigma\bar\psi_i\psi_i$ is parity even.
At order $N^0$, it gives two diagrams: one with both Yukawa insertions on
the fermion loop containing $T$, and one with an additional fermion bubble.
Both contain the same fermion triangle with one stress-tensor insertion
and two scalar insertions.
Write this triangle per flavor as $\mathcal T_{ab}(q,k)$, with the
stress-tensor pair projected to its traceless part.
For $x=|k|/|q|$ and $y=|k-q|/|q|$, its longitudinal contraction is
\begin{equation}
 \frac32 U^aU^b\mathcal T_{ab}(q,k)
 =\frac{|q|}{64}(x+y)\left[3(x-y)^2-1\right].
 \label{eq:gny-contact-triangle}
\end{equation}
The diagram with one fermion loop contains one regulated auxiliary propagator.
The additional bubble in the second diagram cancels the unregulated
part of a second auxiliary propagator and leaves an extra regulator
factor.
The two required integrals therefore have the forms
\begin{equation}
 \int\frac{\dd^3k}{(2\pi)^3}
 \frac{\mathcal T_{ab}(q,k)}
 {(k^2)^{1/2+\Delta}((k-q)^2)^{m\Delta/2}}
 =O(\Delta),\qquad m=0,2.
 \label{eq:gny-yukawa-integrals-zero}
\end{equation}
To evaluate them, insert eq.~\eqref{eq:gny-contact-triangle} and expand
$(x+y)[3(x-y)^2-1]=-x-y+3x^3-3x^2y-3xy^2+3y^3$.
Each monomial is a scalar convolution of the form
\eqref{eq:scalar-master}.
Their Laurent expansions give zero pole and finite coefficients for
both $m=0$ and $m=2$.
Rotational covariance then fixes the full traceless rank-two integrals
to vanish at this order.
The two Yukawa contact diagrams therefore vanish individually.

The trace $B_{cc}$ in the first line of
 eq.~\eqref{eq:gny-full-second-variation} must also be checked before
replacing $B_{ab}$ by its traceless part.
Direct evaluation gives
\begin{equation}
 \langle T_{ab}(q)B_{cc}(-q)\rangle_i=O(\Delta),
 \qquad i=0,1,2,3.
 \label{eq:gny-derivative-traces-zero}
\end{equation}
Here $i=0,1,2,3$ denotes the free bubble, both self-energy placements,
exchange, and two fermion triangles joined by two auxiliary-field lines.
The first three correlators reduce to scalar convolution integrals.
For the fourth, the trace insertion cancels an adjacent fermion
propagator and gives the $m=2$ integral in
 eq.~\eqref{eq:gny-yukawa-integrals-zero} plus the same integral with
the auxiliary momenta exchanged.
Thus only the term $\tfrac38(hj+jh)_{ab}B_{ab}$ contributes through
order $\Delta^0$.

Its contribution is expressed through the individual stress-tensor
two-point diagrams $G_{i,ab,cd}(q)$.
The free two-point function $G^{(0)}$ is given in
eq.~\eqref{eq:free-stress-kernel}.
With the topology labels just specified, the diagrams are
\begin{align}
G_0
&=N G^{(0)}(q)\,,\\
G_1
&=\frac{|q|^3}{2304\pi^2}
P_{ab}^{ef}P_{cd}^{gh}\Biggl[
-\left(\frac{24}{\Delta}-24\log q^2+102\right)\delta_{eg}\delta_{fh}
\notag\\*
&\qquad+\left(\frac{48}{\Delta}-48\log q^2+188\right)\delta_{eg}U_fU_h
-\left(\frac{12}{\Delta}-12\log q^2+29\right)U_eU_fU_gU_h\Biggr]\,,\\
G_2
&=\frac{|q|^3}{11520\pi^2}
P_{ab}^{ef}P_{cd}^{gh}\Biggl[
\left(\frac{24}{\Delta}-24\log q^2+86\right)\delta_{eg}\delta_{fh}
\notag\\*
&\qquad-\left(\frac{48}{\Delta}-48\log q^2+156\right)\delta_{eg}U_fU_h
+\left(\frac{12}{\Delta}-12\log q^2+21\right)U_eU_fU_gU_h\Biggr]\,,\\
G_3
&=\frac{|q|^3}{1440\pi^2}
P_{ab}^{ef}P_{cd}^{gh}\Biggl[
\left(\frac6\Delta-12\log q^2+34\right)\delta_{eg}\delta_{fh}
\notag\\*
&\qquad-\left(\frac{12}{\Delta}-24\log q^2+60\right)\delta_{eg}U_fU_h
+\left(\frac3\Delta-6\log q^2+6\right)U_eU_fU_gU_h\Biggr]\,.
\label{eq:gny-stress-two-point-diagrams}
\end{align}
The vanishing results above leave only the contraction with the stress-tensor two-point function.
For the free diagram $G_0=N G^{(0)}$, this contraction gives
\begin{equation}
 \begin{aligned}
 C_0^{T,\GN}
 &=\frac{3N}{16}\mathcal P\Bigl[
 \delta_{ce}G^{(0)}_{ab,df}(q)+\delta_{cf}G^{(0)}_{ab,de}(q)\\
 &\qquad+\delta_{de}G^{(0)}_{ab,cf}(q)+\delta_{df}G^{(0)}_{ab,ce}(q)\Bigr].
 \end{aligned}
 \label{eq:fermion-leading-metric-contact}
\end{equation}
The interacting diagrams use the same contraction:
\begin{align}
 C_i^{\mathrm{GNY}}
 &=\frac3{16}\mathcal P\Bigl[
 \delta_{ce}G_{i,ab,df}(q)+\delta_{cf}G_{i,ab,de}(q)
 \notag\\
 &\qquad+\delta_{de}G_{i,ab,cf}(q)+\delta_{df}G_{i,ab,ce}(q)
 \Bigr],\qquad i=1,2,3.
 \label{eq:gny-kinetic-contact-diagrams}
\end{align}
The index $i$ labels the GNY contact diagrams in figure~\ref{fig:gny-ttt-contacts}.
The other contributions from the full vertex are zero.

After stress-tensor renormalization, the two-point diagrams obey
\begin{equation}
 G_1+G_2+G_3+2(Z_T^{\GN}-1)G_0
 =\frac8{9\pi^2N}G_0+O(\Delta).
 \label{eq:gny-stress-two-point-sum}
\end{equation}
The same contraction therefore gives the sum of the contact diagrams
through order $N^0$:
\begin{equation}
 C_0^{T,\GN}+\sum_{i=1}^3 C_i^{\mathrm{GNY}}
 +2(Z_T^{\GN}-1)C_0^{T,\GN}
 =\left(1+\frac8{9\pi^2N}\right)C_0^{T,\GN}+O(\Delta).
 \label{eq:gny-full-contact-sum}
\end{equation}
The poles and logarithms cancel in this sum.
Its contribution to the source correlator has the factor $-4$ in
 eq.~\eqref{eq:scalar-ttt-source-expansion}.
The additional quadratic source counterterm is given separately in
 eq.~\eqref{eq:fermion-source-correction}.
The other two choices of the metric-source pair follow by cyclic
permutation of the external insertions.

\section{Metric-source counterterms}\label{app:counterterms}

In this appendix, we compute the metric-source counterterms.
The aim is to make the two-point functions respond covariantly to two metric variations.
The resulting counterterms multiply the leading contacts in
eqs.~\eqref{eq:scalar-leading-metric-contact} and~\eqref{eq:fermion-leading-metric-contact}.
Both contacts scale as $N|q|^3$, so the $1/N$ source coefficients below
contribute at order $N^0$.

A constant metric isolates the local kinetic terms with no derivatives
of the metric sources.
Their coefficients also apply when the sources carry momentum, since
these coefficients multiply the same local matter bilinears.
The resulting contact is nonlocal in the remaining stress-tensor
insertion: its momentum dependence comes from the two-point function
of that insertion with the bilinear.
For the pair of hard metric sources, the bilinear carries momentum $-q$,
so its stress-tensor component produces the required $|q|^3$ term.
The other source pairs leave a two-point function at hard momentum,
which is analytic near $q=0$.

Constant-metric matching does not determine vertices containing
derivatives of the metric sources.
Their relevance must instead be checked with the full vertices in
appendices~\ref{app:on-ttt-contacts} and~\ref{app:gny-full-contacts}.
Those reductions leave only the kinetic stress-tensor component for the
nonanalytic contact considered here, as detailed below for each model.
Purely background local terms give momentum polynomials and cannot
supply this nonanalytic coefficient.
Since the additional source coefficients are of order $1/N$, their
order-$N^0$ contributions multiply the leading contacts only.

\subsection{\texorpdfstring{$O(N)$}{O(N)} vector model}

Let $g_{ab}=\delta_{ab}+s h_{ab}+t j_{ab}$ be constant. The scalar
segment and induced bubble are
\begin{equation}
 M_g(\ell)=\frac{\sqrt g}{g^{ab}\ell_a\ell_b},\qquad
 R_g(k)=\frac{\sqrt g\,|k|}{\sqrt{g^{ab}k_ak_b}},\qquad \ell=p-k.
\end{equation}
With the regulator kernel fixed,
\begin{equation}
 G_{\sigma,\Delta,g}(k)=-\frac{16}{N}
 \frac{(k^2)^{1/2-\Delta}}{1+(R_g(k)-1)(k^2)^{-\Delta}},\qquad
 \Sigma_{\Delta,g}(p)=\intk G_{\sigma,\Delta,g}(k)M_g(p-k).
\end{equation}
For either $X_g=M_g$ or $X_g=R_g$, use the metric derivatives
\begin{equation}
 X_0=X_g|_0,\qquad
 X_h=\partial_sX_g|_0,\qquad
 X_j=\partial_tX_g|_0,\qquad
 X_{hj}=\partial_s\partial_tX_g|_0,
 \label{eq:on-source-metric-derivatives}
\end{equation}
where $|_0$ means $s=t=0$ at fixed momentum.
The $M$ factors below have argument $\ell=p-k$, and the $R$ factors have argument $k$.
For symmetric traceless $h,j$, matrix inversion and the determinant expansion give
\begin{equation}
\begin{aligned}
 &M_0=\frac1{\ell^2},\qquad
 M_h=\frac{\ell\cdot h\ell}{(\ell^2)^2},\qquad
 R_0=1,\qquad R_h=\frac{k\cdot hk}{2k^2},\\
 &M_{hj}=-\frac{\Tr(hj)}{2\ell^2}
 -\frac{\ell\cdot(hj+jh)\ell}{(\ell^2)^2}
 +\frac{2(\ell\cdot h\ell)(\ell\cdot j\ell)}{(\ell^2)^3},\\
 &R_{hj}=-\frac{\Tr(hj)}2
 -\frac{k\cdot(hj+jh)k}{2k^2}
 +\frac{3(k\cdot hk)(k\cdot jk)}{4(k^2)^2}.
\end{aligned}
 \label{eq:on-source-explicit-variations}
\end{equation}
The expressions for $M_j,R_j$ follow by exchanging $h$ and $j$.
These derivatives enter three groups of regulated integrals:
\begin{align}
 \Sigma_1&=-\frac{16}{N}\intk(k^2)^{1/2-\Delta}M_{hj},\notag\\
 \Sigma_2&=\frac{16}{N}\intk(k^2)^{1/2-2\Delta}
 \bigl(R_hM_j+R_jM_h+R_{hj}M_0\bigr),\notag\\
 \Sigma_3&=-\frac{32}{N}\intk(k^2)^{1/2-3\Delta}R_hR_jM_0.
 \label{eq:on-source-group-integrals}
\end{align}
Applying the product rule to the regulated self-energy gives
\begin{equation}
\begin{aligned}
 \left.\partial_s\partial_t\Sigma_{\Delta,g}(p)\right|_0
 &=-\frac{16}{N}\intk(k^2)^{1/2-\Delta}
 \Bigl\{M_{hj}
 \\
 &\quad{}-(k^2)^{-\Delta}\bigl(R_hM_j+R_jM_h+R_{hj}M_0\bigr)
 +2(k^2)^{-2\Delta}R_hR_jM_0\Bigr\}
 \\
 &=\Sigma_1+\Sigma_2+\Sigma_3.
\end{aligned}
 \label{eq:on-source-group-sum}
\end{equation}
Here $\Sigma_1$ has both derivatives on the scalar segment.
The derivatives on the induced bubble give $\Sigma_2$ when they act once or twice on a single factor $R_g$.
The product of two first derivatives gives $\Sigma_3$.
Its factor $2$ comes from differentiating the inverse denominator twice.
Each differentiated factor $R_g$ also supplies $(k^2)^{-\Delta}$.
These factors produce the three regulator powers in
eq.~\eqref{eq:on-source-group-integrals} and affect the finite terms after integration.

Let $B(p)=p_a(h_{ab}j_{bc}+j_{ab}h_{bc})p_c$ and denote its coefficient by $[X]_B$.
Evaluating the integrals in eq.~\eqref{eq:on-source-group-integrals} gives
\begin{align}
 [\Sigma_1]_B&=\frac1{N\pi^2}
 \left(-\frac4{35\Delta}-\frac{148}{3675}+\frac4{35}\log p^2\right),\\
 [\Sigma_2]_B&=\frac1{N\pi^2}
 \left(-\frac{16}{35\Delta}-\frac{8464}{3675}+\frac{32}{35}\log p^2\right),\\
 [\Sigma_3]_B&=\frac1{N\pi^2}
 \left(-\frac{32}{315\Delta}+\frac{496}{11025}+\frac{32}{105}\log p^2\right).
\end{align}
Here and below $O(\Delta)$ terms are omitted. Covariance compares their
sum with the metric variation of the flat self-energy:
\begin{align}
 \Sigma_{\Delta,\delta}(p)&=\frac{p^2}{N\pi^2}
 \left(-\frac4{3\Delta}-\frac89+\frac43\log p^2\right),\\
 \delta K_{\Delta,g}(p)&=\Sigma_{\Delta,g}(p)
 -\sqrt g\,\Sigma_{\Delta,\delta}\!\left(\sqrt{g^{ab}p_ap_b}\right).
\end{align}
The quadratic kinetic coefficient is
\begin{equation}
 [\partial_s\partial_t\delta K_{\Delta,g}|_0]_B
 =\frac1{N\pi^2}\left(\frac{208}{315\Delta}-\frac{96}{35}\right).
\end{equation}
The coefficient of $B(p)$ in the tree inverse-propagator variation is
one.
Subtracting the part already carried by $Z_T^{\ON}-1$ therefore gives
the coefficient relative to this tree vertex:
\begin{equation}
\begin{aligned}
 \kappa_{\ON}
 &=[\partial_s\partial_t\delta K_{\Delta,g}|_0]_B
   -(Z_T^{\ON}-1)\\
 &=\frac1{N\pi^2}\left[
 \frac{208}{315\Delta}-\frac{96}{35}
 -\frac8{15}\left(\frac1\Delta-4\right)\right]
 =\frac1{N\pi^2}\left(\frac8{63\Delta}-\frac{64}{105}\right).
\end{aligned}
\end{equation}
To relate this coefficient to the momentum-space contact, give the
polarizations $h,j$ total momentum $-q$.
The corresponding kinetic vertex is
$\frac12(hj+jh)_{ab}(\partial_a\phi_i\partial_b\phi_i)(-q)$,
up to the trace term displayed in eq.~\eqref{eq:on-explicit-metric-vertex}.
Its nonlocal contraction with $T_{\alpha\beta}(q)$ reduces to
\begin{equation}
 h_{\mu\nu}j_{\rho\sigma}
 (C_0^T)_{\alpha\beta,\mu\nu,\rho\sigma}(q)
 =\frac12(hj+jh)_{ab}G_{0,\alpha\beta,ab}(q),
\end{equation}
with $G_0=N G^{(0)}$ from eq.~\eqref{eq:free-stress-kernel}.
The improvement and trace terms reduce to the mixed two-point function
of the conformally improved free stress tensor with $\phi_i\phi_i$.
This function has no nonlocal part.
At the interacting order, the scalar terms proportional to $\mathcal B$
cancel between $C_i$ and $C_{i+3}$ for $i=1,2,3$ in
eq.~\eqref{eq:on-individual-TTT-contacts}.
That calculation also gives zero pole and finite terms for the
$\sigma\phi_i\phi_i$ contacts $C_7,C_8$.
These reductions explain why the kinetic coefficient suffices for the
contact correction at the order considered.
Multiplying the leading contact by its relative coefficient
$\kappa_{\ON}$ and the source factor $-4$ gives
\begin{equation}
 \delta\TTT^{\mathrm{soft},(3)}_{\mathrm{source}}
 =-4\kappa_{\ON}C_0^T
 =\frac1{N\pi^2}\left(-\frac{32}{63\Delta}+\frac{256}{105}\right)C_0^T.
 \label{eq:scalar-source-correction}
\end{equation}

\subsection{GNY model}

The fermion background factors are
\begin{equation}
 M_g^{\GN,a}(\ell)=\frac{\sqrt g\,(g^{-1/2})^a{}_b\ell_b}{g^{cd}\ell_c\ell_d},
 \qquad R_g^{\GN}(k)=\frac{\sqrt g\sqrt{g^{ab}k_ak_b}}{|k|}.
\end{equation}
The self-energy vector and covariant comparison are
\begin{align}
 \Sigma_{\Delta,g}^{\GN,a}(p)
 &=\frac{16}{N}\intk\frac{(k^2)^{-1/2-\Delta}M_g^{\GN,a}(p-k)}
 {1+(R_g^{\GN}(k)-1)(k^2)^{-\Delta}},\\
 \delta\Gamma_{\Delta,g}^{\GN,a}(p)
 &=\sqrt g\,\Sigma_{\Delta,\delta}^{\GN,a}(g^{-1/2}p)
 -\Sigma_{\Delta,g}^{\GN,a}(p).
\end{align}
The flat answer is
\begin{equation}
 \Sigma_{\Delta,\delta}^{\GN,a}(p)=\frac{p_a}{N\pi^2}
 \left(\frac4{3\Delta}+\frac{32}{9}-\frac43\log p^2\right).
\end{equation}

The mixed derivative $\partial_s\partial_t\Sigma_{\Delta,g}^{\GN,a}|_{s=t=0}$
has three contributions.
We denote by $\Sigma_1^{\mathrm{GNY},a}$ the contribution with both derivatives on $M_g^{\GN,a}$.
The terms with one or two derivatives on a single factor $R_g^{\GN}$ form $\Sigma_2^{\mathrm{GNY},a}$.
The product of two first derivatives of $R_g^{\GN}$ forms $\Sigma_3^{\mathrm{GNY},a}$.
For traceless $h,j$, their coefficients of $((hj+jh)p)_a$ are
\begin{align}
[\Sigma_1^{\mathrm{GNY}}]_{(hj+jh)p}
&=\frac1{N\pi^2}\left(\frac1{210\Delta}+\frac{2696}{11025}-\frac1{210}\log p^2\right)+O(\Delta)\,,
\\
[\Sigma_2^{\mathrm{GNY}}]_{(hj+jh)p}
&=\frac1{N\pi^2}\left(\frac{34}{105\Delta}+\frac{2236}{11025}-\frac{68}{105}\log p^2\right)+O(\Delta)\,,
\\
[\Sigma_3^{\mathrm{GNY}}]_{(hj+jh)p}
&=\frac1{N\pi^2}\left(-\frac{16}{315\Delta}+\frac{176}{3675}+\frac{16}{105}\log p^2\right)+O(\Delta)\,.
\label{eq:gny-source-individual-diagrams}
\end{align}

The quadratic tree vertex contains $3((hj+jh)p)_a/8$.
The coefficient of this tensor in the quadratic counterterm is
$(2/(9\Delta)+88/105)/(N\pi^2)$.
Subtracting the contribution already carried by $Z_T^{\GN}-1$ gives
\begin{equation}
\begin{aligned}
 \kappa_{\GN}
 &=\frac1{N\pi^2}\left(\frac2{9\Delta}+\frac{88}{105}\right)
   -\frac38(Z_T^{\GN}-1)\\
 &=\frac1{N\pi^2}\left[
 \frac2{9\Delta}+\frac{88}{105}
 -\frac15\left(\frac1\Delta+4\right)\right]
 =\frac1{N\pi^2}\left(\frac1{45\Delta}+\frac4{105}\right).
\end{aligned}
\end{equation}
Here $\kappa_{\GN}$ is an absolute kinetic coefficient.
Its ratio to the tree coefficient $3/8$ is therefore $8\kappa_{\GN}/3$.
The tree coefficient multiplies $(hj+jh)_{ab}B_{ab}$ in
eq.~\eqref{eq:gny-full-second-variation}.
Giving the sources total momentum $-q$ and contracting with the remaining
stress tensor yields
\begin{equation}
 h_{\mu\nu}j_{\rho\sigma}
 (C_0^{T,\GN})_{\alpha\beta,\mu\nu,\rho\sigma}(q)
 =\frac38(hj+jh)_{ab}G_{0,\alpha\beta,ab}(q),
\end{equation}
where again $G_0=N G^{(0)}$.
The trace of $B_{ab}$ has no leading nonlocal contraction with $T$.
The spin-connection bilinear has zero contraction by parity, as shown
in eq.~\eqref{eq:gny-spin-contact-zero}.
The remaining Yukawa and trace contributions are evaluated in
appendix~\ref{app:gny-full-contacts} and vanish through the pole and finite
terms at the order displayed there.
Thus the kinetic coefficient multiplies precisely the leading contact
in eq.~\eqref{eq:fermion-leading-metric-contact}.
Its relative coefficient $8\kappa_{\GN}/3$ and the source factor $-4$ give
\begin{equation}
 \delta\TTT^{\mathrm{soft},(3)}_{\mathrm{source}}
 =-4\frac83\kappa_{\GN}C_0^{T,\GN}
 =-\frac1{N\pi^2}\left(\frac{32}{135\Delta}+\frac{128}{315}\right)C_0^{T,\GN}.
 \label{eq:fermion-source-correction}
\end{equation}